%% file: main.tex
\documentclass[a4paper, 11pt]{article}

\usepackage{jheppub}
\usepackage{amsmath, amsfonts}
\usepackage[normalem]{ulem}
\usepackage{bm}
\usepackage{physics}

\usepackage{comment}
\usepackage{color}

\usepackage[framemethod=default]{mdframed}
\newmdenv[skipabove=7pt,
skipbelow=7pt,
rightline=false,
leftline=false,
topline=false,
bottomline=false,
backgroundcolor=gray!10,
linecolor=gray,
innerleftmargin=5pt,
innerrightmargin=5pt,
innertopmargin=5pt,
innerbottommargin=5pt,
leftmargin=0cm,
rightmargin=0cm,
linewidth=4pt]{eBox}

\makeatletter
\@addtoreset{equation}{section}

\makeatother

\allowdisplaybreaks[4]
\usepackage{cancel}
\usepackage{bbold}
\usepackage{empheq}
\usepackage{framed}
\usepackage[most]{tcolorbox}
\usepackage{xcolor}
\colorlet{shadecolor}{orange!15}
\usepackage{braket}
\usepackage{caption}
\usepackage{subcaption}
\usepackage{tikz}
\definecolor{pyblue}{RGB}{31, 119, 180}
\definecolor{pyorange}{RGB}{255, 127, 14}
\definecolor{pygreen}{RGB}{44, 160, 44}
\definecolor{pyred}{RGB}{214, 39, 40}
\usetikzlibrary{intersections, calc, arrows.meta}
\usetikzlibrary{patterns}

\usepackage{array}
\usepackage{longtable}

\usepackage{adjustbox}
\usepackage{tikz-feynman}
\definecolor{rossocorsa}{rgb}{0.83, 0.0, 0.0}

\definecolor{gbcolor2}{rgb}{.9,.2,.6}
\definecolor{gbcolor3}{rgb}{.3,.2,.6}

\definecolor{verdechiaro}{rgb}{0.6,1,0.6}
\definecolor{giallochiaro}{rgb}{1,1,0.6}
\definecolor{bluscuro}{rgb}{0.15, 0.2, 0.9}
\definecolor{verdes}{rgb}{0.1, 0.5, 0.1}
\definecolor{tangerineyellow}{rgb}{1.0, 0.8, 0.0}
\definecolor{smokyblack}{rgb}{0.06, 0.05, 0.03}
\definecolor{americanrose}{rgb}{1.0, 0.01, 0.24}
\definecolor{cobalt}{rgb}{0.0, 0.28, 0.67}
\definecolor{brandeisblue}{rgb}{0.0, 0.44, 1.0}
\definecolor{mycolor}{rgb}{0.0, 0.0, 0.5}
\definecolor{oxfordblue}{rgb}{0.0, 0.13, 0.28}
\definecolor{azure}{rgb}{0.0, 0.5, 1.0}
\definecolor{turquoiseblue}{rgb}{0.0, 1.0, 0.94}
\definecolor{venetianred}{rgb}{0.78, 0.03, 0.08}
\newtcolorbox{mynamedbox1}[1]{colback=venetianred!5!white,colframe=venetianred!80!black,title=#1}
\usepackage{amssymb}
\usepackage{pifont}

\newcommand{\ctext}[1]{\raise0.2ex\hbox{\textcircled{\scriptsize{#1}}}}

\newtcolorbox{mybox}[3][]
{enhanced, 
breakable,
skin first=enhanced,
skin middle=enhanced,
skin last=enhanced,
before upper={\parindent15pt},
  colframe = #2!25,
  colback  = #2!10,
  coltitle = #2!20!black,  
  title    = {#3},
  #1,
}

\makeatletter
\gdef\@fpheader{}
\g@addto@macro\bfseries{\boldmath}
\makeatother
\DeclareMathAlphabet      {\mathbfit}{OML}{cmm}{b}{it}

\usepackage{savesym}

\savesymbol{Ref}
\savesymbol{erf}

\input{newcommands}

\restoresymbol{phys}{Ref}
\restoresymbol{phys}{erf}

\usepackage{booktabs}

\usepackage{amstext} % for \text macro
\usepackage{array}   % for \newcolumntype macro
\newcolumntype{L}{>{$}l<{$}} % math-mode version of "l" column type

\def\pa{\varphi_{a}}
\def\pr{\varphi_{r}}
\usepackage{latexsym}
\usepackage{graphicx}  % Allows to import images
\usepackage{float} % Allows for control of float positions
\usepackage{ifoddpage}
\graphicspath{ {images/} }

\usepackage{tikz}	
\usepackage{tikz-feynman}
\tikzfeynmanset{compat=1.0.0}

\title{Stochastic inflation from a non-equilibrium renormalization group}

\author[a]{Sebasti\'an C\'espedes}
\author[b]{and Thomas Colas}
\affiliation[a]{Abdus Salam Centre for Theoretical Physics, Imperial College, London, SW7 2AZ, UK}
\affiliation[b]{Department of Applied Mathematics and Theoretical Physics, University of Cambridge, Wilberforce Road, Cambridge, CB3 0WA, UK}
\emailAdd{s.cespedes-castillo@imperial.ac.uk}
\emailAdd{tc683@cam.ac.uk}

\begin{document}
	\sloppy
	
\abstract{Understanding stochastic inflation, and in particular the systematic computation of controlled corrections from first principles, remains an important open problem. In this work, we address this problem from two complementary perspectives.
First, we derive the effective field theory governing long-wavelength modes from the reduced density matrix of a coarse-grained description. In this framework, locality in time follows from the thin-shell approximation, while locality in space is recovered dynamically in the super-Hubble regime. The resulting open effective field theory contains both dissipative and diffusive operators, with diffusion dominating as the coarse-graining scale is pushed into the infrared. This construction reproduces the usual Fokker--Planck equation at leading order and allows us to compute its corrections, including subleading contributions to the stochastic dynamics.
Second, we study the evolution of the density matrix under changes of the coarse-graining scale. We show that this flow is governed by a Polchinski-type renormalisation group equation formulated directly for the density matrix. Dissipative and diffusive operators are generated dynamically along the flow, and the resulting effective action matches the Schwinger--Keldysh description. We derive a generalised Fokker--Planck equation directly from the renormalisation group flow, systematically incorporating subleading corrections and recovering the results obtained in the open effective field theory approach.} 
	
	%\keywords{physics of the early universe, inflation, alternatives to inflation, cosmological perturbation theory}
	
	%\arxivnumber{16XX.XXXXX}
	
	\maketitle
\newpage
    
\section{Introduction}

Understanding quantum field theories in expanding backgrounds is essential for both phenomenological and conceptual reasons. Cosmological observations strongly support the idea that inflation provides the initial conditions for the primordial fluctuations imprinted on the Cosmic Microwave Background and Large Scale Structure of the universe. At the same time, time-dependent geometries provide a setting in which quantum field theory exhibits phenomena with no analogue in flat space, such as particle production \cite{Hawking:1975vcx, Unruh:1976db, Grishchuk:1990bj}.

A particularly important problem concerns the behaviour of light scalar fields on superhorizon scales. Once physical wavelengths exceed the Hubble radius, interactions without derivatives continue to evolve, generating secular growth and enhancing infrared effects. As a consequence, perturbation theory breaks down at late times even for weakly coupled theories \cite{Ford:1984hs,Antoniadis:1985pj, Morikawa:1989xz, Tsamis:1993ub, Mukhanov:1996ak,  Brandenberger:2002sk, Tsamis:2005hd,Wu:2006xp,  Enqvist:2008kt, Burgess:2009bs, Burgess:2010dd, Giddings:2011zd, Burgess:2015ajz, Moss:2016uix, Tokuda:2017fdh, Brandenberger:2018fdd,  Tokuda:2018eqs, Baumgart:2019clc, Gorbenko:2019rza,  Brahma:2020zpk, Cohen:2020php, Green:2020txs, Cohen:2021fzf, Cespedes:2020xqq, Cespedes:2023aal, Beneke:2023wmt, Burgess:2024eng, Huenupi:2024ksc, Kamenshchik:2024ybm, Launay:2024qsm, Launay:2025lnc, Palma:2025oux, Lopez:2025arw, Cespedes:2025ple, Beneke:2026rtf, Beneke:2026ksj, Christie:2025knc}. This signals that late-time dynamics cannot be captured within a standard perturbative expansion and must be reorganised at the level of the effective description.

A natural reorganisation is provided by stochastic inflation \cite{Starobinsky:1986fx,Starobinsky:1994bd}, in which long-wavelength modes are described by a stochastic process sourced by short modes as they cross the horizon. This framework resums the leading infrared contributions responsible for this breakdown and yields finite late-time observables~\cite{Baumgart:2019clc, Gorbenko:2019rza, Mirbabayi:2019qtx, Mirbabayi:2020vyt, Cohen:2020php, Cohen:2021fzf, Cohen:2021jbo, Cespedes:2023aal}. However, while its practical success is well established, its status as an effective field theory is still under investigation. In particular, the origin of the stochastic terms, such as noise and drift, and their organisation within a controlled expansion are not fully understood from a path integral perspective. In this work we show that the stochastic description is the leading infrared limit of an open effective field theory obtained by coarse-graining the short modes in the Schwinger-Keldysh contour. This identifies stochastic inflation within a Wilsonian framework: the coarse-graining procedure defines a renormalisation-group flow for the reduced density matrix, in which integrating out modes at the scale $\Lambda(t)$ generates the dissipative and diffusive structures of the infrared theory. In the regime where the effective description becomes local, this flow reduces to the Fokker--Planck equation. In this sense, stochastic inflation is not an independent framework, but the infrared limit of a Wilsonian renormalisation-group (RG) evolution in real time.

From the effective field theory (EFT) perspective, the long-wavelength sector should admit a description in terms of a reduced set of degrees of freedom defined below a time-dependent coarse-graining scale $\Lambda(t)$. Tracing out short modes yields a reduced density matrix with intrinsically non-unitary dynamics. This coarse-graining necessarily generates both dissipation and fluctuations, and the resulting theory is an open effective field theory formulated in the Schwinger--Keldysh formalism. \\

These questions are closely connected to recent developments in non-equilibrium quantum field theory. Over the past decade, effective field theory formulations based on the Schwinger--Keldysh formalism have provided a systematic framework to describe dissipative systems and fluctuations \cite{Kamenev:2011, Crossley:2015evo, 2016RPPh...79i6001S, Glorioso:2018mmw, Colas:2023wxa, Akyuz:2023lsm, Firat:2025upx, Colas:2025app}. In this approach, the dynamics is expressed in terms of a doubled set of fields, and symmetry principles constrain the allowed interactions. Applied to cosmology, this framework provides a natural language to describe the effect of short-wavelength modes acting as an environment for the long-wavelength sector, generating both noise and dissipation \cite{Colas:2023wxa, Salcedo:2024smn,Salcedo:2024nex,Salcedo:2025ezu,Colas:2025app,Li:2025azq}. In contrast to approaches that focus directly on the evolution equation for the probability distribution, our starting point is the effective action for the reduced density matrix, from which both the operator structure and the corresponding evolution equations follow.

At the same time, the coarse-graining procedure admits an alternative interpretation as a RG flow. The separation between long- and short-wavelength modes defines a time-dependent cutoff, and the evolution of the theory as this scale is varied can be viewed as a flow for the reduced density matrix. This perspective has been developed in \cite{Burgess:2009bs,Serreau:2013psa,Guilleux:2015pma, Burgess:2015ajz,Prokopec:2017vxx,Cespedes:2023aal,Green:2025hmo, Alexandre:2025ixz}, where it has been used, for instance, to capture effects such as the dynamical generation of an effective mass and aspects of the infrared evolution. In~\cite{Cespedes:2023aal} it was further argued that stochastic inflation corresponds to a semiclassical regime of this flow. 
However, these approaches do not formulate the problem as an open effective field theory, nor do they determine the operator content and hierarchy of the infrared dynamics from a controlled coarse-graining procedure. In particular, the emergence of stochastic noise and drift as part of a local Schwinger--Keldysh EFT is not made explicit. \\

In this work we develop a systematic effective field theory description of this regime. Our starting point is a coarse-grained description of the long-wavelength sector, obtained by integrating out short modes across a time-dependent scale $\Lambda(t)$ within the Schwinger--Keldysh formalism. This procedure defines an open effective field theory for the reduced density matrix, in which dissipative and diffusive interactions are generated dynamically. In addition, we formulate the evolution of this effective theory as a renormalisation-group flow for the density matrix.

Our results show that stochastic inflation captures the infrared
behaviour of the coarse-grained theory. In this regime, the dynamics
becomes local and dominated by diffusion, with stochastic fluctuations
controlling the evolution over deterministic response. This behaviour
arises when the evolution is dominated by modes crossing the
coarse-graining scale in the super-Hubble regime, which organises the
theory into a controlled expansion. Within this framework, the resulting
infrared dynamics admits a diffusive scaling description and reduces to
a Fokker--Planck equation. In this sense, stochastic inflation provides
the appropriate resummation of infrared effects, with corrections
systematically organised by the same expansion.

The construction presented here can be viewed as a Wilsonian coarse-graining procedure performed in real time. The short-wavelength modes act as environmental degrees of freedom, and integrating them out determines the coefficients of the operators in the open effective action in terms of short-distance correlators. Unlike standard vacuum EFTs, this matching is performed at the level of the reduced density matrix, leading to an effective theory that is intrinsically non-unitary. In this way, the operator content and its hierarchy are derived from first principles, rather than postulated phenomenologically.

In the present context, three ingredients govern this expansion. First, the coarse-graining is performed over a narrow shell in momentum space, which controls the degree of non-locality in time. Second, spatial gradients are suppressed relative to the cut-off scale, allowing for a systematic expansion in momenta. Third, the dynamics simplifies in the superhorizon regime, where the ratio of physical momenta to the Hubble scale becomes small. Making this structure explicit, and showing how it organises the effective theory, is the central goal of this work.

The interplay between these perspectives provides a unified picture of the infrared dynamics. In particular, we show that dissipative and diffusive terms originate from the continuous flow of modes across the cut-off, and that their relative importance is controlled parametrically in the superhorizon regime, where diffusion dominates. In this way, stochastic inflation emerges as the leading local, semiclassical limit of a more general open EFT, and its regime of validity and corrections can be determined systematically.

The paper is organised as follows. In Sec.~\ref{sec:coarsegraining} we derive the effective dynamics of the long-wavelength modes by integrating out short modes and identify the resulting operator structure. In Sec.~\ref{sec:RGPolchinski} we derive the RG flow equation for the reduced density matrix and show how it reduces to a Fokker--Planck equation in the appropriate limit. Finally, Sec.~\ref{sec:conclussions} contains our conclusions.

\paragraph{Summary of main results.}

Our main object of study is the reduced density matrix obtained by
coarse-graining a scalar field over a time-dependent scale $\Lambda(t)$.
For concreteness we focus on a $\lambda \phi^4$ theory, although the
derivation relies only on the structure of the Schwinger--Keldysh path
integral and therefore extends straightforwardly to more general
interactions and field content.

We begin by performing an explicit split of the field into long- and short-wavelength modes. Because the cut-off is time dependent, this decomposition induces mixing between the two sectors already at the quadratic level. Integrating out the short modes within the Schwinger--Keldysh path integral yields an influence functional for the long-wavelength degrees of freedom. This procedure plays the role of a matching calculation: it determines the coefficients of the operators in the open effective action in terms of short-wavelength correlators evaluated at the coarse-graining scale. The resulting description is an open effective field theory in which dissipative and diffusive interactions arise from the continuous flow of modes across the shell $k=\Lambda(t)$. In particular, diffusive and dissipative coefficients are determined by short-mode two-point functions evaluated at the cut-off scale.

A central result of this work is that this long-wavelength theory admits a controlled effective field theory expansion. The simplification proceeds in stages. The narrowness of the shell enforces locality in time, while spatial gradients are suppressed by powers of $p/\Lambda$, leading to an approximately local description. In the super-Hubble regime, the dynamics is controlled by the small parameter
\begin{align}
\epsilon = \frac{\Lambda}{aH} \ll 1\,,
\end{align}
which measures the separation between the coarse-graining scale and the Hubble scale $H$, $a$ being the scale factor, and organises the infrared expansion.

More generally, the effective theory is defined within a combined expansion in three small parameters,
\begin{align}
\frac{\Delta\Lambda}{\Lambda}\ll 1, 
\qquad 
\frac{p}{\Lambda}\ll 1, 
\qquad 
\epsilon=\frac{\Lambda}{aH}\ll 1,
\end{align}
which respectively control memory effects, spatial non-locality, and the super-Hubble hierarchy\footnote{Note that, within the separate-universe approach \cite{Wands:2000dp}, spatial locality can be understood as a consequence of the super-Hubble hierarchy. We nevertheless keep the two hierarchies conceptually distinct here to clarify the derivation.}. These parameters are simultaneously taken to be small, and together define the regime of validity of the EFT and the systematic ordering of corrections.

At leading order in this combined expansion, the effective theory becomes local and is dominated by diffusive interactions. The corresponding action derived in Section \ref{subsec:IF} and given by
\begin{align}
S_0 + S_{\mathrm{int}} &= \int \dd^3x \int \dd t\, a^3(t)\left[
-\partial_\mu\varphi_r^L\partial^\mu\varphi_a^L
-\frac{\lambda}{3!} (\varphi_r^L)^3\varphi_a^L
-\frac{\lambda}{4!}\varphi_r^L(\varphi_a^L)^3
\right],
\label{eq:sum}
\end{align}
where $\varphi^L_{r/a}$ are the coarse-grained retarded/advanced field defined in \eqref{eq_+-2ar} and \eqref{eq:window}, together with the influence functional
\begin{align}
S_{\mathrm{IF}}=\frac{i}{2}\int \dd^3x\int \dd t\, a^3(t) \bigg\{&
(\dot{\varphi}_a^{L})^2
+ \frac{\lambda}{3H^2}f(\epsilon)(\varphi_r^L)^2(\dot{\varphi}_a^{L})^2
- \frac{\lambda}{H}(\varphi_r^L)^2 \varphi_a^L \dot{\varphi}_a^{L} 
\nonumber \\
-& i 
\frac{\lambda}{H^2} \varphi_r^L \varphi_a^L (\dot{\varphi}_a^{L})^2
+ \mathcal{O}\!\left[\epsilon,\lambda^2,(\varphi_a^L)^4\right]
\bigg\}\,,
\label{eq:Intro}
\end{align}
with $f(\epsilon)= -4 + 2 \gamma_E + 2\log\epsilon$.  While some operators, such as the field-dependent correction to the noise kernel $(\varphi_r^L)^2(\dot{\varphi}_a^{L})^2$, are already familiar from the literature \cite{Beneke:2012kn, Gorbenko:2019rza, Mirbabayi:2020vyt, Cohen:2021fzf, Cable:2023gdz}, genuinely new operators arise from the conversion of short modes into long modes at order $\lambda$. These operators form the local basis of the effective theory; their relative importance is determined by the super-Hubble expansion, with higher powers of $\varphi_a^L$ parametrically suppressed. The dynamics is dominated by diffusive terms, whereas dissipative effects, gradient corrections, and higher-order noise contributions appear systematically at subleading order in the expansion.

Only after these approximations does the effective theory admit a stochastic interpretation. In this regime, the leading diffusive term $(\dot{\varphi}_a^L)^2$ generates a Langevin-type evolution, while higher powers of $\varphi_a$ encode non-Gaussian noise. The stochastic limit corresponds to a semiclassical approximation of the Schwinger--Keldysh path integral, in which the action is truncated at leading order in $\varphi_a$. In this limit, the evolution of the reduced density matrix reduces to the standard Fokker--Planck equation of stochastic inflation.

We then turn to the evolution of the theory under changes of the cut-off. So far, this construction determines the effective theory at a fixed cut-off. Imposing the invariance of the generating functional, we derive a Polchinski-type RG equation,
\begin{align}
\frac{\partial e^{iS^\Lambda_{\mathrm{eff}}[\varphi_r^L,\varphi_a^L]}}{\partial \log \Lambda}
=
-\frac{i}{2}\,
\frac{\partial G^{\alpha \beta}_\Lambda}{\partial \log \Lambda} \circ
\left[
- \frac{\delta^2 e^{i S^\Lambda_{\mathrm{eff}}}}{\delta \varphi^L_{\beta}\delta \varphi^L_{\alpha}}
+2i\frac{\delta}{\delta\varphi^L_\alpha}
\left(
\frac{\delta S^\Lambda_{0}}{\delta \varphi^L_{\beta}}
\,e^{iS^\Lambda_{\mathrm{eff}}}
\right)
\right] \,,
\label{eq:RGdiffIntro}
\end{align}
where $S_{\mathrm{eff}}^{\Lambda}$ denotes the effective functional obtained after integrating out modes above the cut-off, with $S^\Lambda_{0}$ its free part. The fields $\varphi_{\alpha,\beta}^L$ collectively denote the retarded and advanced components of the coarse-grained field, while $G^{\alpha \beta}_\Lambda$ is the matrix of coarse-grained propagators defined in \eqref{eq:propagwindow}. Finally, $\circ$ denotes a convolution product, following the notation of \cite{Kamenev:2011}.
This equation is exact prior to any gradient or semiclassical approximation.

The RG flow generates dynamically the same operator structure identified in the coarse-grained theory. In the regime where the EFT becomes local, the flow drives the theory towards the effective action \eqref{eq:Intro}, showing that the stochastic description arises as the infrared limit of this evolution.

Finally, the RG formulation leads naturally to an evolution equation for the probability distribution function. In the same semiclassical and superhorizon limits, this reduces to the Fokker--Planck equation. Our construction therefore makes explicit both the origin of stochastic inflation and the structure of the corrections beyond it, providing a controlled framework to incorporate quantum effects systematically. 

\paragraph{Notation and conventions.}
We use bold symbols to denote spatial 3-vectors, together with the
shorthand notation $k\equiv |\bmk|$. The symbol $\circ$ denotes the
convolution of space-time coordinates,
\begin{align}
(A \circ B)(x,z)
\equiv
\int \mathrm{d}y \,
A(x,y)\,B(y,z)\,,
\end{align}
following the conventions of \cite{Kamenev:2011}. Momentum integrals
are written as
\begin{align}
\int_{\bmk}
\equiv
\int \frac{\mathrm{d}^3k}{(2\pi)^3}\,.
\end{align}
Throughout the paper we mostly work in natural units,
$c=\hbar=1$, and use the metric signature $(-,+,+,+)$.

%%%%%%%%%%%%%%%%%%%%%%%%%%%%%%%%%%%%%%%%%%%%%%%%%%%%%%%%%%%%%%%%%%%%%%%%%%%%%%%%%%%%%%%%%%%%%%%%%%%%%%%%%%%%%%%%%%%%%%%%%%%%%%%%%%

\section{Top down: stochastic inflation in the Schwinger--Keldysh contour}
\label{sec:coarsegraining}

We now derive the dynamics of long-wavelength modes in de Sitter, by coarse-graining the Schwinger--Keldysh path integral at a time-dependent scale $\Lambda(\eta)$. Integrating out the short modes produces an effective description for the long sector in which the short modes act as an environment. The resulting theory is therefore naturally formulated as an open effective field theory, written in terms of the doubled Schwinger--Keldysh fields (see \cite{Lombardo:2004fr,Lombardo:2005iz, PerreaultLevasseur:2013kfq, Crossley:2015evo, Liu:2018kfw, 2016RPPh...79i6001S, Colas:2025app, Li:2025azq, Tasinato:2025zqt}).

The coarse-graining procedure described here can be interpreted as a Wilsonian matching performed in real time. Integrating out short-wavelength modes determines the coefficients of the operators in the effective action in terms of short-mode correlators evaluated at the scale $\Lambda(\eta)$. In particular, diffusive and dissipative terms arise from the Keldysh and retarded components of these correlators, respectively. In this sense, the influence functional provides a first-principles determination of the Wilson coefficients of the open effective field theory governing the long-wavelength sector.

The origin of this structure is twofold. First, because the cut-off is time dependent, the separation between long and short modes is not fixed: modes continuously cross the scale $\Lambda(\eta)$ even in the free theory. Second, tracing out the short modes generates an influence functional, so that the long-wavelength dynamics is generically non-local in time and involves both branches of the Schwinger--Keldysh contour.

The derivation of the infrared theory then proceeds in three steps. Integrating out the short modes produces interactions that are non-local in time. These interactions become local due to the thin-shell nature of the coarse graining. Finally, an expansion in the super-Hubble parameter $\epsilon = \frac{\Lambda}{aH} \ll 1$ organises the resulting theory as a controlled effective field theory. At leading order, the theory is local and dominated by diffusive terms, reproducing stochastic inflation. Subleading corrections are systematically suppressed and generate dissipative effects as well as higher-order stochastic contributions. \\

In order to start the discussion, let us consider a scalar field $\phi$ in the Poincar\'e patch of de Sitter spacetime,
\begin{align}
ds^2 = a^2(\eta)(-\dd \eta^2 + d\bm{x}^2),
\end{align}
with conformal time $\eta$ and scale factor $a(\eta)=-(H\eta)^{-1}$.
Our starting point is the density matrix defined on the Schwinger--Keldysh contour. The state evolves forward in time from $\eta_{\mathrm{ini}}=-\infty$ to $\eta_0$, and then backward along a second branch. We take the initial state to be the Bunch--Davies vacuum $\lvert\mathrm{BD}\rangle$, so that
\begin{align}
\langle\mathcal{O}(\eta)\rangle
=
\mathrm{Tr}\!\left[U(\eta)\rho_0 U^\dagger(\eta)\mathcal{O}\right],
\end{align}
with $\rho_0=\vert\mathrm{BD}\rangle\langle\mathrm{BD}\vert$. In the path-integral representation, fields on the two branches are denoted by $\varphi_+$ and $\varphi_-$. The density matrix at the final time $\eta_0$ reads
\begin{align}
\rho_{\mathrm{UV}}[\phi_+,\phi_-]
=
\int_{\mathrm{BD}}^{\varphi_+(\eta_0)=\phi_+}\!\mathcal{D}\varphi_+
\int_{\mathrm{BD}}^{\varphi_-(\eta_0)=\phi_-}\!\mathcal{D}\varphi_-
\,
e^{iS_{\mathrm{UV}}[\varphi_+] - iS_{\mathrm{UV}}[\varphi_-]} \, .
\label{eq:unitrho}
\end{align}
Though the initial UV theory does not mix the two branches of the Schwinger-Keldysh contour, the introduction of a cutoff and the coarse-graining of short-modes in real-time will generally introduce mixing between branches. This entails the emergence dissipation and noise through a non-vanishing Feynman-Vernon influence functional \cite{FEYNMAN1963118} 
\begin{align}
    S_{\mathrm{UV}}[\varphi_+] - S_{\mathrm{UV}}[\varphi_-] \quad \rightarrow \quad S[\varphi_+,\varphi_-] = S_{\mathrm{IR}}[\varphi_+] - S_{\mathrm{IR}} [\varphi_-]  + F[\varphi_+, \varphi_-]\,,
\end{align}
where $F[\varphi_+, \varphi_-]$ is non-separable.

To make the physics transparent, it is convenient to work in the advanced/retarded (Keldysh) basis,
\begin{align}
\phi_r=\frac{1}{2}(\phi_++\phi_-)\ ,\qquad 
\phi_a=\phi_+-\phi_- \, ,
\label{eq_+-2ar}
\end{align}
where $\phi_r$ captures the average configuration and $\phi_a$ encodes fluctuations. 
In this basis, the diagonal element of the density matrix ($\phi_+=\phi_-=\phi$) becomes
\begin{align}
\rho[\phi]
=
\int_{\mathrm{BD}}^{\varphi_r(\eta_0)=\phi}\!\mathcal{D}\varphi_r
\int_{\mathrm{BD}}^{\varphi_a(\eta_0)=0}\!\mathcal{D}\varphi_a
\,
e^{iS[\varphi_r,\varphi_a]} \, .
\label{eq:unitrhora}
\end{align}
The effective action satisfies 
\begin{align}
&S[\phi_r,\phi_a=0]=0\ ,\qquad \mathrm{Im}\, S[\phi_r,\phi_a]\geq 0\ ,\nonumber\\
&\qquad \quad S[\phi_r,\phi_a]=-S^*[\phi_r,-\phi_a]\ ,
\label{constr_ar}
\end{align}
which reflect normalisation, positivity, and hermiticity respectively \cite{Calzetta:2008iqa}. This basis makes the open-system structure explicit. Terms quadratic in $\phi_a$ define the noise (diffusive) sector and encode fluctuations induced by the traced-out modes. Terms linear in $\phi_a$ encode the response of the long modes and split into a time-symmetric part, which renormalises the deterministic evolution, and a time-antisymmetric part, which captures dissipation. The appearance of these terms signals that the effective dynamics is non-unitary once the short sector is integrated out.

\subsection{Coarse-graining in the Keldysh basis}\label{sec:SIFderiv}

To construct the reduced density matrix for the long-wavelength sector, we split the field into long and short components using a smooth window function $\Omega_\Lambda(k)$,
\begin{align}\label{eq:window}
\varphi^L(\bm{k},\eta)=\Omega_{\Lambda}(k)\varphi(\bm{k},\eta),
\qquad
\varphi^S(\bm{k},\eta)=\bar\Omega_{\Lambda}(k)\varphi(\bm{k},\eta),
\end{align}
with $\bar{\Omega}_\Lambda(k)=1-\Omega_\Lambda(k)$. This decomposition is well defined for a general smooth choice of window. Later, when discussing the local long-wavelength EFT, we will further assume that the external long modes vary on scales well below the shell selected by the cut-off.

For a fixed cut-off at tree level, one would expect the long-wavelength sector to be described by an approximately unitary Wilsonian EFT, with the effect of the short modes encoded in local operators. Here, however, the coarse-graining scale $\Lambda$ depends explicitly on time. The separation between long and short modes is therefore itself time dependent, and this already induces couplings between the two sectors at quadratic order. Physically, modes continuously cross the coarse-graining scale as the background evolves, so the long sector is not closed even in the free theory.

The coupling between long and short modes does not rely on interactions: it is already present in the quadratic theory due to the time dependence of the cut-off. This can already be seen by working with the free Schwinger--Keldysh action,
\begin{align}
S_0 = \frac{1}{2} \int \dd \eta \int \dd^3 \bmx\, a^2(\eta)
\left\{
\left[{\varphi'_+}^2-(\partial_i\varphi_+)^2\right]
-
\left[{\varphi'_-}^2-(\partial_i\varphi_-)^2\right]
\right\}.
\end{align}
Passing to the Keldysh $r/a$ basis, this becomes
\begin{align}\label{eq:free}
S_0
=
\int \dd \eta \int \dd^3\bmx\, a^2(\eta)
\left(
\varphi_r' \varphi_a'
-
\partial_i\varphi_r\,\partial_i\varphi_a
\right).
\end{align}
This basis makes the causal structure explicit and, in particular, renders the action linear in $\varphi_a$, which will be useful when implementing the long--short decomposition.

Upon decomposing the fields as $\varphi_{r,a}=\varphi^L_{r,a}+\varphi^S_{r,a}$ using the window functions in Eq.~\eqref{eq:window}, the mixing arises directly from the time dependence of the cut-off: derivatives act on the window functions as well as on the fields, generating long--short couplings already at quadratic order \cite{Li:2025azq, Li:2026XXX}
\begin{align}
S_{\mathrm{cross}}=\int_{\bm{k}}\int \dd \eta\, a^2(\eta)\Omega_\Lambda^{\prime}(k)\bigg[&-\Omega_\Lambda \varphi_r^{\prime L}(\bmk, \eta)\,\varphi_a^S(-\bmk, \eta)
-\Omega_\Lambda \varphi_a^{\prime L }(\bmk, \eta)\,\varphi_r^S(-\bmk, \eta) \nonumber \\
&+\varphi_a^L(\bmk, \eta)\,\bar\Omega_\Lambda \varphi_r^{\prime S }(-\bmk, \eta)
+\varphi_r^L(\bmk, \eta)\,\bar\Omega_\Lambda \varphi_a^{\prime S}(-\bmk, \eta)\bigg],
\label{eq:crossinteractions}
\end{align}
where the prime acts only on the field, while the explicit factor of $\Omega_\Lambda^\prime$ comes from differentiating the time-dependent window.\footnote{Expanding $(\Omega_\Lambda\varphi)'=\Omega_\Lambda\varphi'+\Omega_\Lambda'\varphi$ and $(\bar\Omega_\Lambda\varphi)'=\bar\Omega_\Lambda\varphi'-\Omega_\Lambda'\varphi$, the cross terms proportional to one power of $\Omega_\Lambda'$ are precisely those that couple one long field to one short field. 
}

These terms make the physical origin of the open-system description explicit. Because the cut-off evolves, modes continuously cross from the short sector into the long one, so the two sectors remain coupled even in the free theory. Once the short modes are traced over, the long-wavelength sector behaves as an open system coupled to an environment formed by the short-wavelength fluctuations. This results in an effective action, also called influence functional, for the long sector that is, in general, non-local in time \cite{Colas:2022kfu, Colas:2024ysu, Cruces:2026yvs} and contains terms mixing the two branches of the Schwinger--Keldysh contour.

\paragraph{Propagators.} The basic ingredients for the construction of the influence functional are the propagators of the scalar field theory. It is useful to distinguish between the \emph{bulk} propagators of the free scalar and the \emph{filtered short-mode propagators} that actually enter the coarse-grained effective action. In spatial Fourier space, the quadratic action of the free theory may be written as
\begin{align}
S_0[\varphi_r,\varphi_a]
=
-\frac12 \int_{\bm k}\int \dd\eta\, \dd\tilde\eta\,
\begin{pmatrix}
\varphi_r(\bm k,\eta) & \varphi_a(\bm k,\eta)
\end{pmatrix}
\begin{pmatrix}
0 & \widehat \dd^{A} \\
\widehat \dd^{R} & -2i\,\widehat \dd^{K}
\end{pmatrix}
\begin{pmatrix}
\varphi_r(-\bm k,\tilde\eta) \\
\varphi_a(-\bm k,\tilde\eta)
\end{pmatrix}.
\label{eq:free_short_action}
\end{align}
Here $\widehat D^R$ and $\widehat D^A$ are fixed by the quadratic action and determine the causal propagators. The Keldysh component $\widehat D^K$, instead, is fixed by the choice of initial state on the Schwinger--Keldysh contour \cite{Keldysh:1964ud, Kamenev:2011, 2016RPPh...79i6001S, Kaplanek:2025moq}. For the Bunch--Davies vacuum of a free massless scalar, it is determined by the corresponding mode functions.

The corresponding bulk propagators are defined by
\begin{align}
\widehat D^{R}\, \circ \, G^{R} &= \delta(\eta-\tilde\eta),\\
\widehat D^{A}\,\circ \, G^{A} &= \delta(\eta-\tilde\eta),
\end{align}
with $G^A(\eta,\tilde\eta)=G^R(\tilde\eta,\eta)$, while $G^K$ is the symmetric two-point function of the Gaussian state. The objects that actually appear in the influence functional are their short-sector projections,
\begin{align}
G_S^X(k,\eta,\tilde\eta)
=
\bar\Omega_{\Lambda(\eta)}(k)\,
\bar\Omega_{\Lambda(\tilde\eta)}(k)\,
G^X(k,\eta,\tilde\eta),
\qquad X=R,A,K.
\label{eq:filtered_short_prop}
\end{align}
These filtered propagators are the ones that contract the short legs generated by Eq.~\eqref{eq:crossinteractions}.

In the Schwinger--Keldysh path integral, the bulk propagators have the standard interpretation
\begin{align}
G^{R}(k,\eta,\tilde\eta) &= \langle \varphi_r(\bm k,\eta)\varphi_a(-\bm k,\tilde\eta)\rangle,\\
G^{A}(k,\eta,\tilde\eta) &= \langle \varphi_a(\bm k,\eta)\varphi_r(-\bm k,\tilde\eta)\rangle,\\
G^{K}(k,\eta,\tilde\eta) &= \langle \varphi_r(\bm k,\eta)\varphi_r(-\bm k,\tilde\eta)\rangle.
\end{align}
The retarded and advanced propagators encode causal response, while the Keldysh propagator encodes fluctuations. This is precisely the split that will later give rise to the response sector and the noise sector of the long-wavelength EFT.

For a free massless scalar in de Sitter, the bulk propagators in conformal time are
\begin{align}
G^{K}(k;\eta,\tilde\eta)
&=
i\frac{H^2}{2k^3}
\left[
(1+k^2\eta\tilde\eta)\cos k(\eta-\tilde\eta)
+
k(\eta-\tilde\eta)\sin k(\eta-\tilde\eta)
\right],
\label{eq:masslessGK}
\\
G^{R}(k;\eta,\tilde\eta)
&=
\Theta(\eta-\tilde\eta)\,
\frac{H^2}{k^3}
\left[
(1+k^2\eta\tilde\eta)\sin k(\eta-\tilde\eta)
-
k(\eta-\tilde\eta)\cos k(\eta-\tilde\eta)
\right],
\label{eq:masslessGR}
\\
G^{A}(k;\eta,\tilde\eta)
&=
-\Theta(\tilde\eta-\eta)\,
\frac{H^2}{k^3}
\left[
(1+k^2\eta\tilde\eta)\sin k(\eta-\tilde\eta)
-
k(\eta-\tilde\eta)\cos k(\eta-\tilde\eta)
\right].
\label{eq:masslessGA}
\end{align}

These expressions simplify considerably in the super-Hubble regime, $|k\eta|\ll1$ and $|k\tilde\eta|\ll1$, which is the region relevant for the long-wavelength EFT. Expanding Eqs.~\eqref{eq:masslessGK}--\eqref{eq:masslessGA} at small arguments, one finds
\begin{align}
G^{K}(k;\eta,\tilde\eta)
&=
i\frac{H^2}{2k^3}
\left[
1 + \mathcal{O}\big((k\eta)^2,(k\tilde\eta)^2\big)
\right],
\\
G^{R}(k;\eta,\tilde\eta)
&=
\Theta(\eta-\tilde\eta)\,
\frac{H^2}{3}(\eta^3-\tilde\eta^3)
+\mathcal{O}\big(k^2\eta^3\tilde\eta^3\big),
\\
G^{A}(k;\eta,\tilde\eta)
&=
-\Theta(\tilde\eta-\eta)\,
\frac{H^2}{3}(\tilde\eta^3-\eta^3)
+\mathcal{O}\big(k^2\eta^3\tilde\eta^3\big).
\end{align}
In this regime the Keldysh propagator approaches a leading constant $\sim k^{-3}$, reflecting the freezing of super-Hubble modes. By contrast, the retarded and advanced propagators vanish at equal times and retain only their causal time dependence. As a result, the symmetric correlator is enhanced in the infrared, while the causal propagators contribute only through their time variation.

These are the ingredients needed to construct the influence functional. In the next subsection we use Eq.~\eqref{eq:crossinteractions} together with the filtered propagators \eqref{eq:filtered_short_prop} to derive the bilocal kernels that govern the effective dynamics of the coarse-grained long-wavelength theory.

\subsection{Influence functional}\label{subsec:IF}

The influence functional $S_{\mathrm{IF}}$~\cite{FEYNMAN1963118} is obtained by integrating out the short modes from the density matrix. In the path-integral formulation this is given by
\begin{align}
e^{iS_{\mathrm{IF}}[\varphi_r^L,\varphi_a^L]} =
\int \dd \varphi
\int_{\mathrm{BD}}^\varphi \mathcal{D}\varphi^S_r
\int_{\mathrm{BD}}^0 \mathcal{D}\varphi^S_a
\, e^{iS_0[\varphi_r^S,\varphi_a^S]}
\, e^{iS_{\mathrm{cross}}[\varphi_r^L,\varphi_a^L;\varphi_r^S,\varphi_a^S]} .
\label{eq:inflFunctional}
\end{align}
This expression involves functional integration over the short modes and their boundary configurations.  Since the short-mode action $S_0$ is quadratic, the integrals are Gaussian and fully determined by the short-sector propagators. Expanding the long--short interaction to second order, one finds
\begin{align}
e^{iS_{\mathrm{IF}}}
= 1 + i \langle S_{\mathrm{cross}} \rangle_S
- \frac12 \langle S_{\mathrm{cross}}^2 \rangle_S + \cdots ,
\end{align}
where the subindex $S$ indicates that the expectation value is taken over the short modes through the path integral in~\eqref{eq:inflFunctional}. Since the short-sector dynamics is Gaussian, \textit{i.e.} quadratic in $\varphi^S$, and the tadpole has been removed by an appropriate choice of background or counterterm, one has $\langle S_{\mathrm{cross}} \rangle_S=0$.
The leading contribution to the influence functional is therefore
\begin{align}
S_{\mathrm{IF}} \simeq \frac{i}{2}\langle S_{\mathrm{cross}}^2\rangle_S +\mathcal{O}(S_{\mathrm{cross}}^3) .
\end{align}
At this order, the effective action is obtained by contracting the internal short-mode fields using their propagators. This naturally organizes the result according to the Schwinger--Keldysh structure of the short-sector two-point function. Let us write the influence functional at this order as 
\begin{align}
S_{\mathrm{IF}} = S_{\mathrm{IF}}^{\rm diff} + S_{\mathrm{IF}}^{\rm drift} + S_{\mathrm{IF}}^{\rm diss},
\end{align}
where the contributions have the following interpretation. The contraction with the Keldysh propagator $G^K_S$ generates a contribution $S_{\mathrm{IF}}^{\rm diff}$ defined in Eq. \eqref{eq:Sdiff}, quadratic in $\varphi_a$, which contributes to the imaginary part of the influence functional and gives rise to the diffusion term of the long-wavelength effective dynamics. By contrast, contractions with the retarded and advanced propagators $G^R_S$ and $G^A_S$ generate terms linear in $\varphi_a$, which determine the response of the long modes. These are naturally decomposed into a symmetric part, denoted $S_{\mathrm{IF}}^{\rm drift}$ written down in Eq.\eqref{eq:Sdrift_bilocal}, which encodes the time-symmetric response and renormalises the deterministic evolution and an antisymmetric part, denoted $S_{\mathrm{IF}}^{\rm diss}$ defined in Eq.\eqref{eq:Sdiss_bilocal}, which 
which encodes the time-antisymmetric (dissipative) response~\cite{Colas:2025ind}. 

As previously discussed, before any further approximation, integrating out the short modes produces kernels that are non-local both in time and in space: the influence functional is bilocal in the two time arguments $\eta$ and $\tilde\eta$, and still involves an integral over the internal momentum $k$. The emergence of time locality is tied to the sharpness of the coarse-graining window. Since the short sector is selected by $\bar\Omega_\Lambda\equiv 1-\Omega_\Lambda$, we take $\bar\Omega_\Lambda(k)\simeq \theta(k-\Lambda)$, so that $\bar\Omega_\Lambda'(k)\simeq -\Lambda aH\,\delta(k-\Lambda)$ up to corrections controlled by the finite width of the window.

This is the crucial point. Each insertion of the long--short coupling carries one factor of $\bar\Omega'_\Lambda(k)$, so at second order the bilocal kernel is proportional to $\bar\Omega^\prime_{\Lambda}(k)\bar\Omega^\prime_{\tilde{\Lambda}}(k)\propto \delta(k-\Lambda)\delta(k-\tilde{\Lambda})$. This localisation in time arises entirely from the sharpness of the window function and should be clearly distinguished from the infrared expansion discussed below.
 Because both delta functions constrain the same internal momentum $k$, they have support only when the two shell scales coincide, $\tilde\Lambda=\Lambda$. Since $\Lambda(\eta)$ is monotonic, this in turn forces the two times to coincide. More explicitly,
\begin{align}
\delta(k-\Lambda)\delta(k-\tilde\Lambda)
=
\delta(k-\Lambda)\,
\frac{\delta(\tilde\eta-\eta)}{\left|\partial_\eta \Lambda(\eta)\right|}.
\end{align}
For $\Lambda(\eta)=-\epsilon/\eta$, this becomes
\begin{align}
\delta(k-\Lambda)\delta(k-\tilde\Lambda)
=
-\frac{\eta}{\Lambda}\,\delta(k-\Lambda)\delta(\tilde\eta-\eta),
\end{align}
so one of the two time integrals is removed and the original bilocal kernel in $(\eta,\tilde\eta)$ collapses onto equal times.

The coarse-graining also introduces a small parameter
\begin{align}
\epsilon \equiv \frac{\Lambda}{aH}\ll1,
\end{align}
which, for light fields in de Sitter, corresponds to modes well outside the horizon. This expansion will be used below to organise the relative size of the equal-time operators that arise after the kernel has been localised. The two steps should be kept distinct: the sharpness of the window controls the localisation in time, while the expansion in $\epsilon$ controls the hierarchy among the resulting local terms.\footnote{More generally, a finite-width kernel generates additional
local terms containing higher time derivatives. These are controlled by
the ratio between the infrared relaxation rate and the microscopic
memory scale, $\partial_t/H\sim\mu_{\rm IR}/H\ll1$, and are therefore
subleading in the Markovian super-Hubble regime. This suppression is
dynamical and does not rely on the existence of a conserved charge, as
in hydrodynamic derivative expansions.}

There is a further simplification at coincident times. The undifferentiated retarded and advanced propagators vanish at equal times, $G^R_S(k,\eta,\eta)=G^A_S(k,\eta,\eta)=0$, while the Keldysh propagator remains finite. When derivatives act on the propagators, however, the coincident-time limit is more subtle: time derivatives of $G^R_S$ or $G^A_S$ need not vanish even though the propagators themselves do. This is why derivative couplings can still generate non-trivial local terms after the time integrals collapse, and why the equal-time limit must be taken only after the relevant derivatives have been evaluated.

The resulting locality in time should be understood as an approximate statement, valid at leading order in the sharp-window limit. If the window has a small but finite width, the delta functions are broadened into kernels with finite temporal support, so the influence functional is no longer exactly local in time and the dynamics acquires weak memory effects. These corrections can be computed systematically by introducing a smooth window and expanding in its width, following~\cite{Gorbenko:2019rza}. 

\paragraph{Diffusion.}
As a first application of the localisation discussed above, we consider the diffusive contribution arising from the contraction with the short-mode Keldysh propagator. Before taking the equal-time limit, it reads
\begin{align}
S_{\mathrm{IF}}^{\rm diff}
&=
\int_{\bm{k}}\int \dd \eta \, a^2(\eta)\bar\Omega'_\Lambda(k)
\int \dd \tilde\eta \,\bar\Omega'_{\tilde\Lambda}(k)\, a^2(\tilde\eta)
\bigg\{
\varphi_a^{\prime L}(\bm k,\eta)\varphi_a^{\prime L}(-\bm k,\tilde\eta)\, G^K_S(k,\eta,\tilde\eta)
\nonumber\\
&\qquad
- 2\,\varphi_a^{\prime L}(\bm k,\eta)\varphi_a^L(-\bm k,\tilde\eta)\,
\partial_{\tilde\eta}G^K_S(k,\eta,\tilde\eta)
+ \varphi_a^L(\bm k,\eta)\varphi_a^L(-\bm k,\tilde\eta)\,
\partial_\eta \partial_{\tilde\eta} G^K_S(k,\eta,\tilde\eta)
\bigg\}.
\label{eq:Sdiff}
\end{align}
Using the shell localisation, the kernel collapses to $\tilde\eta=\eta$, and the short propagator and its derivatives can be evaluated at coincident times. One then obtains
\begin{align}
S_{\mathrm{IF}}^{\rm diff}
&=
\int_{\bm{k}}\int \dd\eta \, a^5(\eta)\,
\bigl[-\bar\Omega_\Lambda'(k)\bigr]\,
\frac{iH^3}{2 k^2}
\bigg\{
\left[1 + (-\eta k)^2\right]
\varphi_a^{\prime L}(\bm k,\eta)\varphi_a^{\prime L}(-\bm k,\eta)
\nonumber\\
&\qquad\qquad
+ 2 k (-\eta k)\,
\varphi_a^{\prime L}(\bm k,\eta)\varphi_a^L(-\bm k,\eta)
+ k^2 (-\eta k)^2\,
\varphi_a^L(\bm k,\eta)\varphi_a^L(-\bm k,\eta)
\bigg\}.
\label{eq:diff}
\end{align}

Since the shell enforces $k\sim\Lambda$, the dimensionless combination $-\eta k$ is of order $\epsilon=\Lambda/(aH)\ll1$. It follows directly from \eqref{eq:diff} that the mixed and non-derivative terms are suppressed by one and two powers of $\epsilon$, respectively, while the term proportional to $\varphi_a^{\prime L}\varphi_a^{\prime L}$ is unsuppressed. Hence, in the super-Hubble regime, the dominant contribution is
\begin{align}
S_{\mathrm{IF}}^{\rm diff}
\;\longrightarrow\;
\int_{\bm{k}}\int \dd\eta \, a^5(\eta)\,
\bigl[-\bar\Omega_\Lambda'(k)\bigr]\,
\frac{iH^3}{2 k^2}\,
\varphi_a^{\prime L}(\bm k,\eta)\varphi_a^{\prime L}(-\bm k,\eta),
\end{align}
which is the operator that survives in the stochastic limit for light fields in de Sitter.

At this stage the operator is local in time, but it remains bilocal in space because the shell still selects a finite band of momenta around the coarse-graining scale.
Passing to position space gives
\begin{align}
&\quad S_{\mathrm{IF}}^{\rm diff}
=
\frac{iH^3}{2}\int \dd\eta\, a^5(\eta)\int \dd^3\bm x\,\dd^3\bm y
\int \frac{\dd^3\bm k}{(2\pi)^3}\,
\bigl[-\bar\Omega_\Lambda'(k)\bigr]\,
e^{i\bm k\cdot(\bm x-\bm y)}
\\
&\times
\bigg[
\frac{1+(-\eta k)^2}{k^2}\,
\varphi_a^{\prime L}(\bm x,\eta)\varphi_a^{\prime L}(\bm y,\eta)
+\frac{2(-\eta k)}{k}\,
\varphi_a^{\prime L}(\bm x,\eta)\varphi_a^L(\bm y,\eta) +(-\eta k)^2\,
\varphi_a^L(\bm x,\eta)\varphi_a^L(\bm y,\eta)
\bigg]. \nonumber
\label{eq:diff_position}
\end{align}
This form is useful because it makes the spatial non-locality explicit while still keeping the dependence on the coarse-graining profile completely general. For a smooth window, the kernel is the Fourier transform of the shell profile $-\bar\Omega'_\Lambda(k)$ and is therefore localised over distances of order $(\Delta\Lambda)^{-1}$, where $\Delta\Lambda$ is the width of the shell.

Spatial locality  requires a smooth shell of finite width $\Delta\Lambda$, so that the Fourier kernel is localised over distances of order $(\Delta\Lambda)^{-1}$. For external long modes satisfying $p\ll \Delta\Lambda$, the bilocal interaction can then be expanded systematically in derivatives around the midpoint $\bm X=(\bm x+\bm y)/2$.
Assuming therefore a smooth shell, one may expand in the relative separation $\bm r=\bm x-\bm y$ around the midpoint $\bm X=(\bm x+\bm y)/2$. For example, 
\begin{align}
\varphi_a^{\prime L}\!\left(\bm X+\frac{\bm r}{2},\eta\right)
\varphi_a^{\prime L}\!\left(\bm X-\frac{\bm r}{2},\eta\right)
&=
\left[\varphi_a^{\prime L}(\bm X,\eta)\right]^2
-\frac{r^i r^j}{4}\,
\partial_i\varphi_a^{\prime L}(\bm X,\eta)\,
\partial_j\varphi_a^{\prime L}(\bm X,\eta)
+\cdots .
\end{align}
To implement this procedure, one must project the resulting effective action onto the separate-universe limit, in which spatial gradients are neglected, and specify the corresponding definition of local operators in the coarse-grained theory. We develop this expansion systematically in \App{app:local} and show that, at leading order in the super-Hubble expansion, it reproduces the Gaussian white-noise term.
\begin{align}
S_{\mathrm{IF}}^{\rm diff}
\simeq
\frac{i}{2}\int \dd^3\bm x\int \dd\eta\, a^2(\eta)\,
\left[\varphi_a^{\prime L}(\bm x,\eta)\right]^2 .
\label{eq:diffAc}
\end{align}
Corrections involving spatial gradients are organized as a systematic derivative expansion around the separate-universe limit and are suppressed by powers of $k_L^2/(\Delta\Lambda)^2$, where $\Delta\Lambda$ sets the spatial resolution scale of the shell integration, as detailed in Appendix~\ref{app:local}.

It is important to keep track of the distinct approximations that enter this result. 
Time locality arises from the shell-width expansion controlled by the sharpness of the window; subleading corrections reintroduce a finite temporal support and hence memory effects. 
Spatial locality, by contrast, requires a smooth shell and a gradient expansion, and its corrections are suppressed by powers of $p^2/\Lambda^2$. 
The local stochastic limit in Eq.~\eqref{eq:diffAc} is obtained only after combining these ingredients with the super-Hubble hierarchy $\epsilon\ll1$, which selects the leading equal-time operator.

Usually is useful to keep the non-local version of the diffusion vertex. This can be obtained by considering the sharp-shell limit,
\begin{align}
-\bar\Omega_\Lambda'(k)\to \delta(k-\Lambda),
\end{align}
for which the momentum integral can be performed explicitly. One then finds
\begin{align}
S_{\mathrm{IF}}^{K}
&=
\frac{iH^3}{4\pi^2}\int \dd^3 \bmx \int \dd^3 \bmy \int \dd\eta\, a^5(\eta)\,
j_0(\epsilon aH |\bm{x}-\bm{y}|)
\bigg[
(1 + \epsilon^2)\,\varphi_a^{\prime L}(\bmx, \eta) \varphi_a^{\prime L}(\bmy, \eta)
\nonumber\\
&\qquad\qquad\qquad\qquad
+ 2 \epsilon^2 a H\,\varphi_a^{\prime L}(\bmx, \eta) \varphi_a^L(\bmy, \eta)
+ \epsilon^4 a^2 H^2\,\varphi_a^L(\bmx, \eta) \varphi_a^L(\bmy, \eta)
\bigg],
\label{eq:diffzerothcoor}
\end{align}
where $j_0(z)=\sin z/z$. This expression makes the oscillatory structure of the kernel fully explicit, recovering well established results \cite{Starobinsky:1994bd}. However, it also illustrates an important subtlety: the sharp-shell limit is convenient for displaying the kernel, but it is not the appropriate starting point for a controlled local derivative expansion.\footnote{
In the strictly sharp limit $-\bar\Omega'_\Lambda(k)\to\delta(k-\Lambda)$ the kernel becomes oscillatory, $K(r)\sim \Lambda^2 j_0(\Lambda r)$, and is not localised around $r=0$. As a result, the integral over $\bm r$ is not dominated by short separations, and a derivative expansion in external momenta is not controlled. Equivalently, taking the sharp-shell limit and performing the local expansion do not commute.
}
A local spatial EFT does not follow from the sharp-shell expression itself. In the strictly sharp limit the kernel is oscillatory and non-local in position space, so a derivative expansion is not controlled. In the case when $\epsilon\ll 1$ we recover the known expression for the non-local noise~\cite{Starobinsky:1994bd} 

\paragraph{Response functions.}
We now analyse the response sector, which follows the same localisation and expansion as in the diffusion case, but arises from contractions with the retarded and advanced propagators $G^R_S$ and $G^A_S$ \cite{Colas:2022hlq, Burgess:2024heo, Colas:2025ind}. These contributions are linear in $\varphi_a$ and therefore govern the response of the long modes. It is useful to separate them according to their symmetry under interchange of the time arguments. The symmetric combination defines the drift term, $S_{\mathrm{IF}}^{\rm drift}$, which renormalises the deterministic evolution and can be viewed as a correction to the effective Hamiltonian of the long-wavelength sector. The antisymmetric combination defines the dissipative term, $S_{\mathrm{IF}}^{\rm diss}$, which mixes the two branches of the Schwinger--Keldysh contour and therefore captures the genuinely non-unitary part of the dynamics. Both arise from the same response sector; the distinction lies in the symmetry of the kernel.

Before any approximation, these terms are bilocal in time and space. The dissipative contribution is \footnote{The retarded and advanced propagators are not interchangeable in these expressions: each appears with a fixed assignment inherited from the Schwinger--Keldysh structure, and exchanging $G^R_S$ and $G^A_S$ would modify the bilocal kernel.}
\begin{align}
S_{\mathrm{IF}}^{\rm diss}
&=
\int_{\bm{k}}\int \dd \eta\, a^2(\eta)\bar\Omega^\prime_\Lambda(k)
\int \dd \tilde\eta\, \bar\Omega^\prime_{\tilde\Lambda}(k)\, a^2(\tilde\eta)
\bigg\{
\varphi_a^{\prime L}(\eta)\varphi_r^{\prime L}(\tilde\eta)\,G^A_S(k,\eta,\tilde\eta)
\nonumber\\
&\qquad
-
\left[
\varphi_r^{\prime L}(\eta)\varphi_a^L(\tilde\eta)
-
\varphi_r^L(\eta)\varphi_a^{\prime L}(\tilde\eta)
\right]\partial_\eta G^A_S(k,\eta,\tilde\eta)
+
\varphi_a^L(\eta)\varphi_r^L(\tilde\eta)\,
\partial_\eta\partial_{\tilde\eta}G^A_S(k,\eta,\tilde\eta)
\bigg\},
\label{eq:Sdiss_bilocal}
\end{align}
whereas the drift contribution is
\begin{align}
S_{\mathrm{IF}}^{\rm drift}
&=
\int_{\bm{k}}\int \dd \eta\, a^2(\eta)\bar\Omega^\prime_\Lambda(k)
\int \dd \tilde\eta\, \bar\Omega^\prime_{\tilde\Lambda}(k)\, a^2(\tilde\eta)
\bigg\{
\varphi_a^{\prime L}(\eta)\varphi_r^{\prime L}(\tilde\eta)\,G^R_S(k,\eta,\tilde\eta)
\nonumber\\
&\qquad
-
\left[
\varphi_r^{\prime L}(\eta)\varphi_a^L(\tilde\eta)
+
\varphi_r^L(\eta)\varphi_a^{\prime L}(\tilde\eta)
\right]\partial_\eta G^R_S(k,\eta,\tilde\eta)
-
\varphi_a^L(\eta)\varphi_r^L(\tilde\eta)\,
\partial_\eta\partial_{\tilde\eta}G^R_S(k,\eta,\tilde\eta)
\bigg\}.
\label{eq:Sdrift_bilocal}
\end{align}
As in the diffusion sector, the product $\bar\Omega^\prime_\Lambda(k)\bar\Omega^\prime_{\tilde\Lambda}(k)$ localises the bilocal kernel around $\tilde\eta=\eta$. For a smooth window this localisation has a finite width, but at leading order one may evaluate the kernels at equal times. The important difference with diffusion is that
\begin{align}
G^R_S(k,\eta,\eta) &= 0, \\
G^A_S(k,\eta,\eta) &= 0,
\end{align}
so the undifferentiated propagators drop out in the coincident-time limit. The surviving local contributions arise from derivatives acting on the retarded and advanced propagators, whose equal-time limits remain finite.

Using these equal-time limits, one finds
\begin{align}
S_{\mathrm{IF}}^{\rm drift}
&=
2\int_{\bm{k}}\int \dd\eta\, a^3(\eta)\,\bigl[-\bar\Omega_\Lambda'(k)\bigr]\,
Hk
\left[
\varphi_r^{\prime L}(\bm k,\eta)\varphi_a^L(-\bm k,\eta)
+
\varphi_r^L(\bm k,\eta)\varphi_a^{\prime L}(-\bm k,\eta)
\right],
\label{eq:drift_smooth}
\\
S_{\mathrm{IF}}^{\rm diss}
&=
-
\int_{\bm{k}}\int \dd\eta\, a^3(\eta)\,\bigl[-\bar\Omega_\Lambda'(k)\bigr]\,
Hk
\left[
\varphi_r^{\prime L}(\bm k,\eta)\varphi_a^L(-\bm k,\eta)
-
\varphi_r^L(\bm k,\eta)\varphi_a^{\prime L}(-\bm k,\eta)
\right].
\label{eq:diss_smooth}
\end{align}
These are the response-sector analogues of the local-in-time diffusion operator: the non-locality in time has been reduced to corrections controlled by the width of the shell, while the operators remain non-local in space because the momentum integral is supported over a finite band around $\Lambda$.

The physical distinction between the two structures is worth making explicit. The drift term contains the symmetric combination
\begin{align}
\varphi_r^{\prime L}\varphi_a^L+\varphi_r^L\varphi_a^{\prime L},
\end{align}
which modifies the deterministic evolution. This becomes more transparent in the $+/-$ basis, where the drift term can be written as
\begin{align}
S_{\mathrm{IF}}^{\rm drift}
&=
\int_{\bm{k}}\int \dd\eta\, a^3(\eta)\,\bigl[-\bar\Omega_\Lambda'(k)\bigr]\,
Hk
\left\{
\frac{\dd}{\dd\eta}\bigl[\varphi_+^L(\eta)\bigr]^2
-
\frac{\dd}{\dd\eta}\bigl[\varphi_-^L(\eta)\bigr]^2
\right\},
\label{eq:drift_pm}
\end{align}
which shows that it acts independently on each branch of the contour. In flat spacetime this would reduce to a total derivative, but in an FLRW background the time-dependent prefactor prevents it from being discarded, so it contributes to the effective Hamiltonian.
By contrast, the dissipative term does not factorise. In the $+/-$ basis it retains the branch-mixing structure
\begin{align}
-\varphi_+^{\prime L}\varphi_-^L+\varphi_-^{\prime L}\varphi_+^L,
\end{align}
which is precisely what encodes non-unitary evolution.

The sharp-shell limit, $-\bar\Omega_\Lambda'(k)\to\delta(k-\Lambda)$, makes the scaling explicit. In this limit the drift term becomes
\begin{align}
S_{\mathrm{IF}}^{\rm drift}
&=
2\int_{\bm{k}}\int \dd\eta\, a^3(\eta)\,\delta(k-\Lambda)\,H\Lambda
\left[
\varphi_r^{\prime L}(\bm k,\eta)\varphi_a^L(-\bm k,\eta)
+
\varphi_r^L(\bm k,\eta)\varphi_a^{\prime L}(-\bm k,\eta)
\right],
\label{eq:drift_sharp}
\end{align}
and similarly for the dissipative term.
It is then clear that both response contributions vanish as the cut-off is lowered. After the momentum integral, the shell contributes a factor $\Lambda^2$, which combines with the explicit factor of $\Lambda$ to give an overall scaling $\sim \Lambda^3$. Hence $S_{\mathrm{IF}}^{\rm drift}$ and $S_{\mathrm{IF}}^{\rm diss}$ disappear as $\Lambda\to0$, provided the long fields remain regular.

As in the diffusion sector, the remaining non-locality is spatial. Fourier transforming \eqref{eq:drift_smooth} and \eqref{eq:diss_smooth} gives bilocal kernels in $\bm x-\bm y$, showing that the response sector is local in time (up to shell-width corrections) but spread over spatial separations set by the inverse width of the momentum shell. A local spatial description requires a smooth window function, so that the kernel has finite spatial support, together with external momenta satisfying $p\ll\Lambda$. In that regime one performs a gradient expansion around the midpoint $\bm X=(\bm x+\bm y)/2$, producing a local derivative series in even powers of $\bm r=\bm x-\bm y$. Following \App{app:local}, For a narrow shell centred, the first gradient corrections are suppressed by $p^2/\Lambda^2$.

The logic is therefore the same as for diffusion. The shell-width expansion controls the departure from exact time locality, producing memory corrections when subleading terms are kept, while the gradient expansion, which is well defined for a smooth window, controls the departure from spatial locality, producing higher-derivative corrections suppressed by $p^2/\Lambda^2$. The only real difference is physical: diffusion is controlled by the Keldysh sector and survives as the dominant stochastic effect in the infrared, whereas the response sector splits into a Hamiltonian correction and a genuinely dissipative term, both of which are parametrically suppressed as the cut-off is lowered.

%%%%%%%%%%%%%%%%%%%%%%%%%%%%%%%%%%%%%%%%%%%%%%%%%%

\paragraph{Stochastic regime.}

The previous analysis shows that lowering the coarse-graining scale into the superhorizon regime reorganises the effective theory. When
\begin{align}
\epsilon\equiv \frac{\Lambda}{aH}\ll1,
\end{align}
the response sector becomes parametrically suppressed with respect to the diffusive one. Comparing the leading local operators obtained above, one finds
\begin{align}
\frac{\text{dissipation}}{\text{diffusion}}
\sim \epsilon^3.
\label{eq:FDT}
\end{align}
As a result, the long-wavelength dynamics is dominated by diffusion, while drift and dissipation provide only subleading corrections.

This hierarchy may be viewed in three equivalent ways. At the level of local operators, the leading response terms are suppressed relative to diffusion by $\epsilon^3$ \cite{Launay:2024trh}. At the level of propagators, diffusion is controlled by the equal-time Keldysh correlator, which remains infrared enhanced as the shell is lowered, whereas response terms arise from differentiated retarded and advanced propagators and therefore carry additional powers of $k\sim\Lambda$. Equivalently, in infrared power counting, the Keldysh sector remains leading while operators with additional response legs are suppressed. These statements all express the same physical conclusion: on super-Hubble scales the influence functional is dominated by diffusion, with response effects entering only as subleading corrections.

Physically, diffusion describes the stochastic sourcing of long modes by short fluctuations continuously crossing the coarse-graining scale. The response sector instead encodes how the long-wavelength dynamics reacts to these modes. While both effects are comparable deep inside the horizon, once $\epsilon\ll1$ the stochastic sourcing persists whereas the response rapidly switches off.

The same result can be understood as a change in power counting. On subhorizon scales, $k\gg aH$, all propagators scale similarly and no particular Schwinger--Keldysh structure is singled out. On superhorizon scales, however, the equal-time Keldysh propagator scales as $G^K_S\sim k^{-3}$, whereas the response sector is less infrared-enhanced. Equivalently, the long-wavelength theory is organised by the infrared dimensions $\Delta_r=0$ and $\Delta_a=3$, so that operators controlled by the Keldysh kernel remain leading, while those arising from differentiated response kernels are suppressed by powers of $\epsilon$.

This shows that, on superhorizon scales, the influence functional reduces at leading order to its diffusive sector, with response effects systematically suppressed in the infrared.

\paragraph{EFT at first order in $\lambda$.}
We now turn to the quartic interaction,
\begin{align}\label{eq:NL}
    S_{\mathrm{int}} = -  \frac{\lambda}{4!}\int \dd^4 x\sqrt{-g} \left( \varphi^4_+  -\varphi^4_- \right)
    = - \frac{\lambda}{3!} \int \dd^4 x\sqrt{-g} \left( \varphi^3_r \varphi_a  + \frac{1}{4} \varphi_r \varphi_a^3\right).
\end{align}
Our goal is to analyse the effective functional obtained after integrating out the short modes to first order in $\lambda$.

At this stage we keep the full Schwinger--Keldysh operator content of the coarse-grained theory. In particular, both $(\varphi_r^{L})^3\varphi_a^L$ and $\varphi_r^{L} (\varphi_a^{L})^3$ are part of the effective action. The semiclassical (stochastic) limit will only be imposed later, when interpreting the effective theory, and should not be assumed in the derivation of the operator content.

At this order, three qualitatively different classes of contributions arise. First, the long-wavelength theory still contains the quartic operators already present in the microscopic unitary theory, namely $(\varphi_r^{L})^3\varphi_a^L$ and $\varphi_r^{L} (\varphi_a^{L})^3$. The former encodes the classical non-linearities entering the effective equation of motion, whereas the latter is the purely quantum vertex already present in the Schwinger--Keldysh action. In a stochastic language this operator is associated with non-Gaussian noise, but it is important to stress that it is not generated by coarse graining: it is already part of the original unitary dynamics.

Second, integrating out the short modes dresses the Gaussian influence functional itself. This happens when two short fields in the long--short decomposition of the quartic interaction are contracted against one another. Expanding $S_{\mathrm{int}}$ to quadratic order in the short fields, one finds
\begin{align}
S^{(2)}_{\mathrm{int}} &=  - \frac{\lambda}{3!} \int \dd^4 x\sqrt{-g} \bigg[
3 (\varphi^L_r)^2 \varphi^S_r \varphi_a^S
+ 3 \varphi^L_r \varphi^L_a (\varphi^S_r)^2
+ \frac{3}{4} (\varphi^L_a)^2 \varphi^S_a \varphi^S_r
+ \frac{3}{4} \varphi^L_a \varphi^L_r (\varphi^S_a)^2
\bigg].
\label{eq:effectiveaction1}
\end{align}
In what follows we restrict to the same local approximation used in the Gaussian sector. Namely, we keep the leading shell-crossing contribution, evaluate the induced bilocal kernels at equal times, and retain only the leading term in the long-wavelength gradient expansion. The operators displayed below should therefore be understood as the local limit of the corresponding bilocal influence functional, with memory effects controlled by the shell width and higher-derivative corrections suppressed by powers of $p^2/\Lambda^2$. 

Within this approximation, the first term in \eqref{eq:effectiveaction1} is particularly important because, from the viewpoint of the short modes, it acts as an effective mass insertion \cite{Cable:2023gdz},
\begin{align}
m^2_{\mathrm{eff}}=\lambda (\varphi_r^L)^2 .
\end{align}
Since the long mode evolves slowly compared with the short shell modes, it may be treated as approximately frozen when computing the short-mode propagator. This resums part of the interaction into a background-dependent shift of the short-sector two-point functions. For a light effective mass, $m^2_{\mathrm{eff}}\ll H^2$, the super-Hubble Keldysh propagator becomes
\begin{align}
    G_S^K(k;\eta,\teta)=\frac{iH^2}{2k^3}\left\{1+\frac{m_{\mathrm{eff}}^2}{3H^2}\left[-4+2\gamma_E+\log(4 k^2\eta\teta)\right]\right\}.
\end{align}
Substituting this dressed propagator into the local diffusion term obtained previously, and evaluating it on the shell, one finds
\begin{align}\label{eq:correction1}
 S^K_{\mathrm{IF}} = \frac{i}{2} \int \dd^3 \bmx\int \dd\eta\, a^2(\eta)
 \left\{ 1 + \frac{\lambda}{3H^2} f(\epsilon)\,
 \left[\varphi_r^{L}(\bm{x},\eta)\right]^2\right\}
 \left[\varphi_a^{\prime L}(\bm{x},\eta)\right]^2,
\end{align}
where
\begin{align}
f(\epsilon)= -4+2\gamma_E+2\log\epsilon.
\end{align}
This shows that the tadpole-type dressing of the short propagator generates a quartic diffusive operator of the form
\begin{align}
\lambda\,(\varphi_r^L)^2(\varphi_a^{\prime L})^2.
\end{align}
Physically, this means that the diffusion coefficient becomes dependent on the slowly varying background value of the coarse-grained field. Equivalently, the amplitude of the stochastic noise is field dependent.

The same effective mass also modifies the principal-value and spectral propagators of the short modes, and therefore induces quartic corrections to drift and dissipation. However, in the super-Hubble regime these response-sector corrections remain suppressed relative to the quartic diffusive term, by the same power counting that made diffusion dominate over dissipation already at Gaussian order. For that reason, the field-dependent correction to the noise kernel is the leading new effect in the local infrared theory.

Third, there are genuinely new operators that do not arise from simply dressing the Gaussian kernels. These originate from the conversion of short modes into long modes through insertions of the long--short mixing term $S_{\mathrm{cross}}$. Concretely, one first decomposes the interaction into terms with a fixed number of short fields and then contracts one of these short fields with a short leg from $S_{\mathrm{cross}}$. This mechanism converts short response fields into local long-wavelength operators.

The structure of these contributions is controlled by two ingredients: the hierarchy of short propagators and the conversion rule for short response legs. In the super-Hubble regime, contractions involving short advanced fields are suppressed, while the leading contributions arise from Keldysh contractions of short response fields. As a result, only contractions involving $\varphi_r^S$ are unsuppressed at leading order. The conversion of a short response leg then produces local structures involving $\varphi_a^{\prime L}$ and $\varphi_a^L$, organised in powers of $\epsilon=\Lambda/(aH)$

Using the local approximation and shell localisation described above, this procedure generates a tower of diffusive operators. Their explicit derivation, including the hierarchy of contributions and the resulting operator basis, is presented in \App{app:ope}. At leading order, the dominant operators include
\begin{align}
(\varphi_r^L)^2 \varphi_a^L \varphi_a^{\prime L}, 
\qquad  (\varphi_a^L)^3 \varphi_a^{\prime L}, \qquad
\varphi_r^L \varphi_a^L (\varphi_a^{\prime L})^2,
\qquad
\varphi_a^L (\varphi_a^{\prime L})^3,
\end{align}
with subleading corrections suppressed by powers of $\epsilon^2$ (see Table~\ref{tab:ope}).

These induced operators are conceptually distinct from the two effects discussed above. They are not the original non-linearities of the microscopic action, and they are not simply the field-dependent dressing of the Gaussian kernels. Rather, they are new composite operators produced by integrating out shell modes in the interacting theory. In particular, operators involving higher powers of $\varphi_a^L$ are generated. According to the super-Hubble power counting discussed earlier, such terms are increasingly irrelevant in the infrared. In the stochastic (semiclassical) regime, where $\varphi_r^L\gg \varphi_a^L$, operators with multiple powers of $\varphi_a^L$ correspond to higher-order corrections to the Fokker--Planck description and therefore give only subleading contributions to the dynamics. For this reason, in the main text we focus on the operators that are at most cubic in $\varphi_a^L$.

\begin{tcolorbox}[%
			enhanced, 
			breakable,
			skin first=enhanced,
			skin middle=enhanced,
			skin last=enhanced,
			before upper={\parindent15pt},
			]{}

\vspace{0.05in}

\paragraph{Summary.}
Assuming a sharp window function, and working in the local super-Hubble approximation described above, the long-wavelength dynamics at leading order in $\epsilon$ and first order in $\lambda$ is given by $S=S_0+S_{\mathrm{int}}+S_{\mathrm{IF}}$, where
\begin{align}
S_0 + S_{\mathrm{int}} &=\int \dd^3x \int \dd t\, a^3(t)\left[-\partial_\mu\varphi_r^L\partial^\mu\varphi_a^L-\frac{\lambda}{3!} (\varphi_r^L)^3\varphi_a^L-\frac{\lambda}{4!}\varphi_r^L(\varphi_a^L)^3\right] ,
\label{eq:fin1}
\end{align}
and
\begin{align}
S_{\mathrm{IF}}=\frac{i}{2}\int \dd^3x\int \dd t\, a^3(t) \bigg\{&
(\dot{\varphi}_a^{L})^2
+ \frac{\lambda}{3H^2}f(\epsilon)(\varphi_r^L)^2(\dot{\varphi}_a^{L})^2
- \frac{\lambda}{H}(\varphi_r^L)^2 \varphi_a^L \dot{\varphi}_a^{L} 
\nonumber \\
-& i 
\frac{\lambda}{H^2} \varphi_r^L \varphi_a^L (\dot{\varphi}_a^{L})^2
+ \mathcal{O}\!\left[\epsilon,\lambda^2,(\varphi_a^L)^4\right]
\bigg\}\,,
\label{eq:fin2}
\end{align}
with $f(\epsilon)= -4+2\gamma_E+2\log\epsilon$. Here we have rewritten the functional in cosmic time, $\dd t=a\,\dd\eta$ and $\dot\varphi_a^L=\varphi_a^{\prime L}/a$, and separated the unitary terms already present in the microscopic theory, \eqref{eq:fin1}, from the diffusive terms induced by integrating out the short modes, \eqref{eq:fin2}. Even at tree level, the coarse-grained long-wavelength theory is therefore not purely Hamiltonian: integrating out the short sector generates non-unitary stochastic contributions, and at first order in $\lambda$ the leading new effect is a field-dependent correction to the diffusion kernel. 
\end{tcolorbox}

%%%%%%%%%%%%%%%%%%%%%%%%%%%%%%%%%%%%%%%%%%%%%%%%%%%%%%%%%%%%%%%%%%%%%%%%%%%%%%%%%%%%%%%%%%%%%%%%%%%%%%%%%%%%%%%%%%%%%%%%%%%%%%%%%%%%%%%%%%%%%

\begin{figure}[t]
\centering
\begin{tikzpicture}[scale=0.74, line cap=round, line join=round]

\definecolor{looporange}{RGB}{230,125,45}

\tikzset{
  ext/.style={draw=black, line width=1.4pt},
  extd/.style={draw=black, line width=1.4pt, dash pattern=on 3.5pt off 3pt},
  loop/.style={draw=looporange, line width=1.8pt},
  loopd/.style={draw=looporange, line width=1.8pt, dash pattern=on 4pt off 3pt},
  vtx/.style={circle, fill=black, inner sep=0pt, minimum size=3.6pt}
}

% Quadrants
\newcommand{\UL}[1]{\draw[#1] (180:1.15) arc[start angle=180,end angle=90,x radius=1.15,y radius=1.15];}
\newcommand{\UR}[1]{\draw[#1] ( 90:1.15) arc[start angle= 90,end angle=  0,x radius=1.15,y radius=1.15];}
\newcommand{\LR}[1]{\draw[#1] (  0:1.15) arc[start angle=  0,end angle=-90,x radius=1.15,y radius=1.15];}
\newcommand{\LL}[1]{\draw[#1] (-90:1.15) arc[start angle=-90,end angle=-180,x radius=1.15,y radius=1.15];}

\node at (0,3.5) {Unitary};
\node at (7.3,3.5) {Stochastic};
\node at (14.6,3.5) {Cancelling};

\draw (3.65,-4.1) -- (3.65,2.9);
\draw (10.95,-4.1) -- (10.95,2.9);

% -------- LEFT TOP --------
\begin{scope}[shift={(0,1.5)}]
  \coordinate (L) at (-1.15,0);
  \coordinate (R) at ( 1.15,0);

  \draw[ext]  (L) -- ++(-1.3, 1.1);
  \draw[ext]  (L) -- ++(-1.4,-1.1);
  \draw[ext]  (R) -- ++( 1.3, 1.1);
  \draw[extd] (R) -- ++( 1.4,-1.1);

  \UL{loopd}\UR{loop}\LR{loop}\LL{loop}

  \node[vtx] at (L) {};
  \node[vtx] at (R) {};
  \node at (0,-1.9) {$(\varphi_r^L)^3\varphi_a^L$};
\end{scope}

% -------- LEFT BOTTOM --------
\begin{scope}[shift={(0,-2.2)}]
  \coordinate (L) at (-1.15,0);
  \coordinate (R) at ( 1.15,0);

  \draw[extd] (L) -- ++(-1.3, 1.1);
  \draw[extd] (L) -- ++(-1.4,-1.1);
  \draw[extd] (R) -- ++( 1.3, 1.1);
  \draw[ext]  (R) -- ++( 1.4,-1.1);

  \UL{loop}\UR{loop}\LR{loopd}\LL{loop}

  \node[vtx] at (L) {};
  \node[vtx] at (R) {};
  \node at (0,-1.9) {$\varphi_r^L(\varphi_a^L)^3$};
\end{scope}

% -------- MIDDLE TOP --------
\begin{scope}[shift={(7.3,1.4)}]
  \coordinate (L) at (-1.15,0);
  \coordinate (R) at ( 1.15,0);

  \draw[ext]  (L) -- ++(-1.3, 1.1);
  \draw[extd] (L) -- ++(-1.4,-1.1);
  \draw[ext]  (R) -- ++( 1.3, 1.1);
  \draw[extd] (R) -- ++( 1.4,-1.1);

  \UL{loop}\UR{loop}\LR{loop}\LL{loop}

  \node[vtx] at (L) {};
  \node[vtx] at (R) {};
  \node at (0,-1.9) {$(\varphi_r^L)^2(\varphi_a^L)^2$};
\end{scope}

% -------- MIDDLE BOTTOM --------
\begin{scope}[shift={(7.3,-2.2)}]
  \coordinate (L) at (-1.15,0);
  \coordinate (R) at ( 1.15,0);

  \draw[ext]  (L) -- ++(-1.3, 1.1);
  \draw[extd] (L) -- ++(-1.4,-1.1);
  \draw[ext]  (R) -- ++( 1.3, 1.1);
  \draw[extd] (R) -- ++( 1.4,-1.1);

  \UL{loopd}\UR{loop}\LR{loop}\LL{loopd}

  \node[vtx] at (L) {};
  \node[vtx] at (R) {};
  \node at (0,-1.9) {$(\varphi_r^L)^2(\varphi_a^L)^2$};
\end{scope}

% -------- RIGHT TOP --------
\begin{scope}[shift={(14.6,1.5)}]
  \coordinate (L) at (-1.15,0);
  \coordinate (R) at ( 1.15,0);

  \draw[ext] (L) -- ++(-1.3, 1.1);
  \draw[ext] (L) -- ++(-1.4,-1.1);
  \draw[ext] (R) -- ++( 1.3, 1.1);
  \draw[ext] (R) -- ++( 1.4,-1.1);

  \UL{loopd}\UR{loop}\LR{loopd}\LL{loop}

  \node[vtx] at (L) {};
  \node[vtx] at (R) {};
  \node at (0,-1.9) {$(\varphi_r^L)^4$};
\end{scope}

% -------- RIGHT BOTTOM --------
\begin{scope}[shift={(14.6,-2.2)}]
  \coordinate (L) at (-1.15,0);
  \coordinate (R) at ( 1.15,0);

  \draw[extd] (L) -- ++(-1.3, 1.1);
  \draw[extd] (L) -- ++(-1.4,-1.1);
  \draw[extd] (R) -- ++( 1.3, 1.1);
  \draw[extd] (R) -- ++( 1.4,-1.1);

  \UL{loopd}\UR{loop}\LR{loopd}\LL{loop}

  \node[vtx] at (L) {};
  \node[vtx] at (R) {};
  \node at (0,-1.9) {$(\varphi_a^L)^4$};
 
\end{scope}

\end{tikzpicture}
 \caption{One-loop diagrams at order $\lambda^2$. The $\varphi^S$ propagators are shown in orange and the $\varphi^L$ fields in black. Solid orange lines denote $G^K_S$ propagators and dashed orange lines denote $G^R_S$ propagators. \textit{Right:} one-loop correction to $\lambda$ already present in the unitary theory; \textit{Middle:} emergence of the stochastic operator in the non-equilibrium theory; \textit{Right:} Forbidden operators from unitarity and causality.}
    \label{fig:diagram}
    \end{figure}
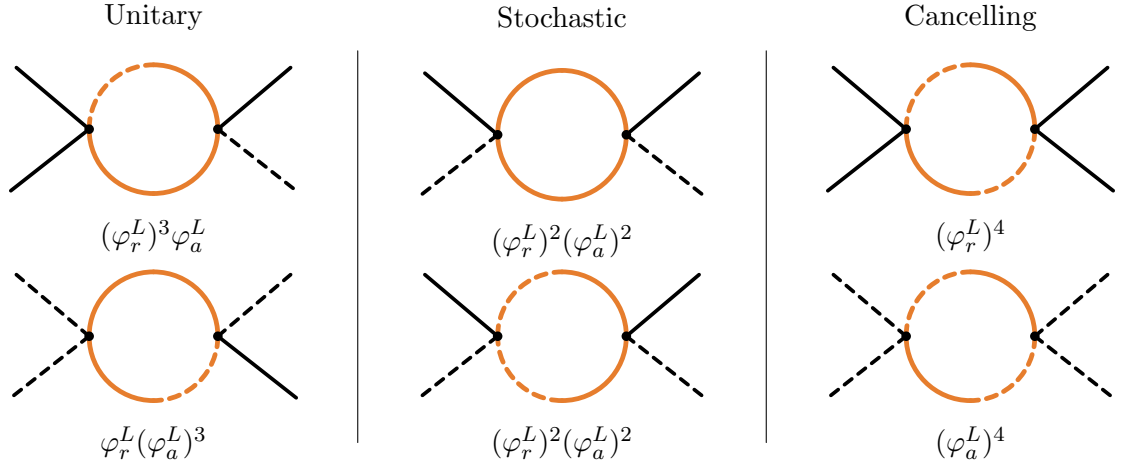

\paragraph{Higher orders in $\lambda$.} 
Higher orders in $\lambda$ can be analyzed in the same way. On the Schwinger-Keldysh contour, dissipative and stochastic operators are generated alongside the familiar unitary operator of the standard $\lambda \varphi^4$ theory \cite{Frangi:2025xss, Panda:2025tpu}. For example, at order $\lambda^2$, the unitary operator $(\varphi_r^L)^3 \varphi_a^L + \varphi_r^L(\varphi_a^L)^3/4$ receives a contribution from a loop involving $G_S^R$ and $G_S^K$, corresponding to the usual renormalization of $\lambda$ in a unitary theory. In addition, a stochastic operator $(\varphi_r^L)^2 (\varphi_a^L)^2$, which mixes the two branches of the path integral, is generated by the non-vanishing loop with two $G_S^K$ or two $G_S^R$ propagators. This mechanism is illustrated in \Fig{fig:diagram}.

As discussed in \cite{Cespedes:2025ple} and illustrated in the last collum of \Fig{fig:diagram}, the causality structure of the theory, encapsulated in the theta functions appearing in the retarded and advanced Green's function $G^R_S$ and $G^A_S$ ensure the avoidance of operators that would violate unitarity of the UV theory such as $(\varphi^L_r)^4$. The symmetry between $r\leftrightarrow a$ in the vertex structure makes that  $(\varphi^L_a)^4$ is neither generated at order $\lambda^2$. It does not mean that this operator is generally forbidden yet illustrate the utility of unitarity and causality of constrain the structure of the EFT, in particular through the largest-time equation (LTE) \cite{Gao:2018bxz}.

Finally, some operators can be generated by the conversion of a short mode into a long mode through $S_{\mathrm{cross}}$, as discussed above and shown explicitly in \App{app:ope}. It turns out that at leading order in $\epsilon$ these operators were already appearing at order $\mathcal{O}(\lambda)$. Hence, the $\mathcal{O}(\lambda^2)$ contributions simply add up to the $\mathcal{O}(\lambda)$ contributions in the running of these coefficients.

%%%%%%%%%%%%%%%%%%%%%%%%%%%%%%%%%%%%%%%%%%%%%%%%%%%%%%%%%%%%%%%%%%%%%%%%%%%%%%%%%%%%%%%%%%%%%%%%%%%%%%%%%%%%%%%%%%%%%%

\subsection{Stochastic inflation from the open EFT}\label{subsec:recover}

So far, we have derived the local effective theory obtained by integrating out the short modes and organising the result in the combined narrow-shell, long-wavelength, and super-Hubble limits. We now show how this open Schwinger--Keldysh EFT reproduces the tradition Langevin equation from stochastic inflation and how its corrections arise in a controlled way.

The stochastic description follows from a semiclassical expansion of the
Schwinger--Keldysh effective action. This expansion is naturally
organised in powers of the response field $\varphi_a^L$, which controls
deviations from the classical trajectory. The physical evolution is
defined by the saddle of the path integral,
\begin{align}
\frac{\delta S_{\rm eff}^{\rm IR}}{\delta \varphi_a^L}=0,
\end{align}
which gives the equation of motion for $\varphi_r^L$. In this sense,
$\varphi_a^L$ enforces the dynamics of the long-wavelength field at the
level of the saddle. Terms linear in $\varphi_a^L$ therefore determine
the classical evolution, while higher powers encode fluctuations around
this trajectory.

The structure of the effective action separates deterministic dynamics
from fluctuations. The real part, linear in $\varphi_a^L$, fixes the
classical equation of motion, while the imaginary part begins at
quadratic order and controls the statistical weight of different
histories. In particular, the leading imaginary term is quadratic in
$\varphi_a^L$ and defines a Gaussian distribution of fluctuations.
Retaining these terms corresponds to describing the dynamics as
classical evolution supplemented by Gaussian noise, which is the semiclassical limit of the theory.
The consistency of this truncation follows from the infrared scaling of
the coefficients multiplying these operators. The linear term is
controlled by the retarded kernel and is set by the Hubble scale,
schematically $\sim H^2$. The quadratic term, instead, is proportional
to the shell-localised Keldysh correlator,
\begin{align}
\partial_{\log\Lambda} G^K_\Lambda(t,t)
\sim
\frac{H^3}{4\pi^2},
\end{align}
which is fixed by modes crossing the coarse-graining scale. This
enhancement compensates the additional power of $\varphi_a^L$, so the
quadratic term contributes at the same parametric order as the linear
one. By contrast, operators with higher powers of $\varphi_a^L$ are not
associated with such enhanced coefficients and are suppressed in the
combined expansion. This is the Martin--Siggia--Rose limit\footnote{also known as the Martin--Siggia--Rose--Janssen--de Dominicis (MSRJD) formalism.} of the Schwinger-Keldysh formalism \cite{Kamenev:2009jj, 2016RPPh...79i6001S}.

The stochastic limit is therefore obtained by truncating the action to
terms linear and quadratic in $\varphi_a^L$, while higher-order terms
generate subleading non-Gaussian corrections.
At leading order, the local infrared action in cosmic time takes the
form
\begin{align}
S_{\rm eff}^{\rm IR}
&=
\int \dd t\,\dd^3\bmx\, a^3(t)
\left[
\dot\varphi_r^L \dot\varphi_a^L
-\frac{1}{a^2}
\partial_i\varphi_r^L\,
\partial_i\varphi_a^L
-\frac{\lambda}{3!}
(\varphi_r^L)^3
\varphi_a^L
\right]
\nonumber\\
&\qquad
+
\frac{i}{2}
\int \dd t\,\dd^3\bmx\,a^3(t)
\left[
(\dot\varphi_a^L)^2
\right]_\Omega
+\cdots .
\label{eq:local_IR_action}
\end{align}
The ellipsis denotes corrections organised by the expansion parameters
\begin{align}
\frac{\Delta\Lambda}{\Lambda},
\qquad
\frac{p}{\Delta\Lambda},
\qquad
\epsilon=\frac{\Lambda}{aH},
\end{align}
which respectively control residual memory effects, higher-derivative
corrections to the local expansion, and subleading contributions in the
super-Hubble expansion. The stochastic description corresponds to the
leading term in this combined expansion.

\paragraph{Recovering stochastic inflation.}
 Since this noise operator depends only
on the response field $\varphi_a$,  it lies entirely in the Keldysh sector and can be treated using
a Hubbard–Stratonovich (HS) transformation.
Keeping only the leading term in the super-Hubble expansion of
Eq.~\eqref{eq:diffzerothcoor}, the quadratic Keldysh contribution reads in conformal time
\begin{align}
S_{\mathrm{IF}}^{K}
&=
\frac{iH^3}{4\pi^2}
\int \dd\eta\,a^5(\eta)
\int \dd^3\bmx\,\dd^3\bmy\,
j_0\!\left(\epsilon aH|\bmx-\bmy|\right)
\varphi_a^{\prime L}(\bmx,\eta)
\varphi_a^{\prime L}(\bmy,\eta).
\label{eq:leading_nonlocal_Keldysh}
\end{align}
Since this term is quadratic in the response field, it may be
linearised through a Hubbard--Stratonovich transformation. Introducing
the kernel
\begin{align}
{\cal N}_\epsilon(\eta;\bmx,\bmy)
=
\frac{H^3}{2\pi^2}
a^5(\eta)\,
j_0\!\left(\epsilon aH|\bmx-\bmy|\right),
\end{align}
one may write
\begin{align}
&
\exp\left[
-\frac12
\int \dd\eta
\int \dd^3\bmx\,\dd^3\bmy\,
\varphi_a^{\prime L}(\bmx,\eta)\,
{\cal N}_\epsilon(\eta;\bmx,\bmy)\,
\varphi_a^{\prime L}(\bmy,\eta)
\right]
\nonumber\\
&\qquad=
{\cal N}_0
\int {\cal D}\eta_\xi\,
\exp\left[
-\frac12
\int \dd\eta
\int \dd^3\bmx\,\dd^3\bmy\,
\xi(\bmx,\eta)\,
{\cal N}^{-1}_\epsilon(\eta;\bmx,\bmy)\,
\xi(\bmy,\eta)
\right.
\nonumber\\
&\hspace{5.7cm}\left.
-i\int \dd\eta\,\dd^3\bmx\,
\xi(\bmx,\eta)\,
\varphi_a^{\prime L}(\bmx,\eta)
\right].
\label{eq:HS_nonlocal}
\end{align}
The auxiliary field is therefore Gaussian, with vanishing mean and
spatially non-local two-point function
\begin{align}
\left\langle
\xi(\bmx,\eta)
\xi(\bmy,\tilde\eta)
\right\rangle
&=
\delta(\eta-\tilde\eta)\,
{\cal N}_\epsilon(\eta;\bmx,\bmy)
\nonumber\\
&=
\frac{H^3}{2\pi^2}
a^5(\eta)\,
j_0\!\left(\epsilon aH|\bmx-\bmy|\right)
\delta(\eta-\tilde\eta).
\label{eq:nonlocal_noise_correlator}
\end{align}
Thus, at this stage the stochastic source is local in time but remains
spatially correlated over the coarse-graining scale.

After integrating the noise term by parts, one finds
\begin{align}
S_{\rm HS}
=
\int
\dd t\,\dd^3\bmx\,a^3(t)
\left[
\dot\varphi_r^L
\dot\varphi_a^L
-\frac{1}{a^2}
\partial_i\varphi_r^L
\partial_i\varphi_a^L
-\frac{\lambda}{3!}
(\varphi_r^L)^3
\varphi_a^L
+
(\partial_t+3H)\xi\,
\varphi_a^L
\right].
\label{eq:HS_action_local}
\end{align}
This form makes explicit how the stochastic dynamics arises. The
response field $\varphi_a^L$ appears linearly and therefore enforces the
dynamics of $\varphi_r^L$ in the presence of the stochastic source
$\xi$. At the same time, the kinetic structure remains second order in
time derivatives, reflecting the underlying infrared EFT prior to any
further reduction. In particular, the action still contains both a field
and its time derivative as independent components of the dynamics.

The usual stochastic inflation description is instead first order. This is
not imposed, but follows from the hierarchy of scales in the
super-Hubble regime. The second-order structure found above originates from the
kinetic term $\dot\varphi_r^L \dot\varphi_a^L$, which introduces a
velocity sector in addition to the field itself. This sector contains a
fast component whose dynamics is controlled by Hubble friction and
therefore relaxes on a time scale of order $H^{-1}$, so it does not
represent an independent degree of freedom at long times.
In this overdamped regime, $\omega \ll H$, this fast component can be eliminated, leaving an effective description
in terms of the slow field $\varphi_r^L$ alone. Spatial gradients are
parametrically suppressed for $p/(aH)\ll1$, but are retained as
subleading contributions in the effective theory.

Crucially, this reduction must be performed at the level of the path
integral, since the statistical properties of the noise are defined by
the quadratic Keldysh term in the action. A controlled implementation is
obtained by rewriting the theory in phase-space form and integrating out
the fast velocity mode. The details of this procedure are given in
Appendix~\ref{sec:nonlocal_langevin}.

After this reduction, the action becomes first order in time derivatives. The response field in this formulation, denoted $\chi_a^L$, is related to the original response field by the phase-space constraint, $\chi_a^L=-\dot\varphi_a^L$, see \App{sec:nonlocal_langevin}. In the first-order theory it appears linearly and imposes the stochastic equation of motion for $\varphi_r^L$ at the level of the path integral. The resulting action is of Martin--Siggia--Rose type,
\begin{align}
S_{\rm red}
=
\int \dd t \dd^3\bmx\, a^3(t)\,
\chi_a^L
\left[
\eta
-\dot\varphi_r^L
-\frac{1}{3H}V_{,\varphi}(\varphi_r^L)
+\frac{1}{3H}\frac{\partial^2}{a^2}\varphi_r^L
\right]
+\cdots .
\label{eq:first_order_action_main}
\end{align}
The noise variable $\eta$ appearing in the first-order theory is obtained
from the Hubbard--Stratonovich field $\xi$ after the overdamped
reduction. Its correlator is fixed by the quadratic Keldysh term and
remains local in time.

Since $\chi_a^L$ appears linearly, the path integral enforces the
stochastic equation\footnote{
An equivalent formulation can be given in phase space, in which both
$\varphi_r^L$ and its velocity are retained as dynamical variables. In
the overdamped super-Hubble regime, integrating out the velocity
reproduces the configuration-space Langevin equation shown here.}
\begin{align}
\dot\varphi_r^L
=
-\frac{V_{,\varphi}(\varphi_r^L)}{3H}
+
\frac{1}{3H}\frac{\partial^2}{a^2}\varphi_r^L
+
\eta
+\cdots ,
\end{align}
with Gaussian white noise
\begin{align}
\langle \eta(t,\bmx)\eta(t',\bmy)\rangle
=
\frac{H^3}{4\pi^2}
\delta(t-t')\delta^3(\bmx-\bmy).
\end{align}
This is the standard stochastic inflation equation, obtained here as the overdamped limit of the Schwinger--Keldysh infrared EFT. 
The fluctuation-dissipation theorem, which relates dissipation and noise, and its implementation in the Schwinger-Keldysh path integral through the dynamical Kubo–Martin–Schwinger (dKMS) symmetry \cite{Crossley:2015evo, Liu:2018kfw}, are discussed in \App{app:dKMS}.
Corrections are systematically organised by the expansion in
$\omega/H$, $p/(aH)$ and $\epsilon$. 

\medskip

More generally, the role of different operators follows from the
structure of the Schwinger--Keldysh expansion. Terms quadratic in the response field define the noise kernel and modify the statistical distribution of $\eta$, while terms linear in the response field determine the deterministic drift. Higher powers of the response field generate non-Gaussian corrections, which are parametrically suppressed in the super-Hubble expansion and will be discussed below.

\paragraph{Field-dependent Gaussian noise.}

At order $\lambda$, the leading correction to the stochastic dynamics
arises from the dressing of the short-wavelength Keldysh propagator by
the long-wavelength background. As discussed above, this effect can be
understood as a background-dependent effective mass,
\begin{align}
m_{\mathrm{eff}}^2 = \lambda (\varphi_r^L)^2 ,
\end{align}
which modifies the short-mode fluctuations at horizon crossing.

Substituting the dressed propagator into the local Keldysh kernel and projecting onto the coarse-grained operator basis leads to the leading field-dependent correction to the local diffusion operator, \eqref{eq:correction1},
\begin{align}
S^K_{\mathrm{IF}}
=
\frac{i}{2}
\int
\dd^3\bmx
\int
\dd t\,
a^3(t)
\left[
1+
\frac{\lambda}{3H^2}
f(\epsilon)
(\varphi_r^L)^2
\right]
\left[
(\dot\varphi_a^L)^2
\right]_\Omega .
\end{align}

The key point is that this operator remains quadratic in the response
field. It therefore does not introduce non-Gaussian noise, but instead
modifies the Gaussian diffusion kernel itself. In other words, the
variance of the stochastic force becomes dependent on the background
value of the long-wavelength field.

This interpretation follows directly from performing the same
Hubbard--Stratonovich transformation as at leading order. The effect of the interaction is to promote the local Keldysh kernel to
a field-dependent one. Equivalently, after the
Hubbard--Stratonovich transformation the variance of the Gaussian noise
becomes dependent on the long-wavelength background. After the overdamped reduction, this leads to a
multiplicative correction to the stochastic source in the first-order
theory.

The resulting Langevin equation takes the form
\begin{align}
\dot\varphi_r^L
=
-\frac{V_{,\varphi}(\varphi_r^L)}{3H}
+
\left[
1 + \frac{\lambda}{6H^2} f(\epsilon)
(\varphi_r^L)^2
\right]\eta
+\cdots ,
\label{eq:Langevi_multiplicative_noise}
\end{align}
with noise correlator
\begin{align}
\langle \eta(t,\bmx)\eta(t',\bmy)\rangle
=
\frac{H^3}{4\pi^2}\,
\delta(t-t')\,\delta^3(\bmx-\bmy).
\end{align}
The noise remains local, but the amplitude with which it enters the
equation of motion is now field dependent. This is the precise sense in
which the EFT generates multiplicative Gaussian noise.
Two comments are in order. First, this correction is not optional: it is
required by consistency of the EFT expansion. The same interaction that
generates the drift term at order $\lambda$ also generates a correction
to the diffusion kernel at the same order. Both effects arise from the
same microscopic vertex and are therefore tied together by the
influence-functional derivation.

Second, the appearance of multiplicative noise raises the question of
stochastic interpretation (It\^o vs.\ Stratonovich). In a purely
phenomenological Langevin equation this choice must be specified
separately, since different discretisations correspond to inequivalent
stochastic processes. In the present framework, however, the dynamics is
defined by the underlying Schwinger--Keldysh path integral, and no such
ambiguity arises \cite{Pinol:2020cdp}.
More concretely, the stochastic theory is obtained from a local
Schwinger--Keldysh action in which the response field appears linearly.
The corresponding Martin--Siggia--Rose functional defines the stochastic
process together with its causal time ordering, inherited from the
original real-time contour. This fixes the discretisation prescription
unambiguously and leads to the It\^o form of the associated Langevin
equation.\footnote{
The It\^o prescription follows from the causal, retarded structure of
the Schwinger--Keldysh contour: response fields are evaluated at earlier
times and equal-time contractions vanish, corresponding to a pre-point
discretisation of the stochastic process. Equivalent Stratonovich
representations may be obtained after the standard noise-induced drift
redefinition. See e.g.\ \cite{Kamenev:2009jj,Tauber:2014zca,Pinol:2020cdp} for the relation between Schwinger--Keldysh/MSR formalisms and stochastic calculus.}
Equivalently, the MSR action \eqref{eq:first_order_action_main} provides
the fundamental definition of the stochastic dynamics, and the Langevin
equation \eqref{eq:Langevi_multiplicative_noise} should be understood as
a shorthand for this path-integral formulation.

This correction does not alter the Gaussian structure of the noise: the
stochastic force remains fixed by its two-point function, but its
amplitude becomes field dependent. By contrast, operators with higher
powers of the response field modify the noise distribution itself and
therefore generate genuinely non-Gaussian noise, to which we now turn.

\paragraph{Non-Gaussian noise.}

Higher powers of $\varphi_a^L$ generate genuinely non-Gaussian
corrections to the stochastic dynamics. The explicit derivation of these
contributions, including their hierarchy and kernel structure, is
presented in Appendix~\ref{app:ope}. Here we summarise their role in the
reduced infrared theory.

The operator $\varphi_r^L(\varphi_a^L)^3$, already present in
\Eq{eq:fin1}, does not contribute within the Gaussian
truncation discussed above. Its effect becomes visible only after the
reduction to the first-order theory, where it projects onto the response
sector of the reduced action, as described in
Appendix~\ref{sec:nonlocal_langevin}.

In this formulation, the response field appears linearly in the action
and enforces the stochastic dynamics at the level of the path integral.
Insertions of the response field then act as functional derivatives with
respect to the physical field in the evolution equation for the
probability functional.
This structure is captured by the Martin--Siggia--Rose functional,
\begin{align}
e^{\,iS_{\rm noise}[\phi^L,\chi_a^L]}
=
\Big\langle
\exp\!\Big[
-i\int \dd t\,\dd^3\bmx\,
a^3(t)\,\chi_a^L(\bmx,t)\,\eta(\bmx,t)
\Big]
\Big\rangle_\eta ,
\end{align}
whose logarithm generates the connected correlators of the stochastic
force. A cubic response vertex produces a term proportional to
$(\chi_a^L)^3$, corresponding to a non-vanishing three-point cumulant of
the noise. More generally, operators of the form $(\varphi_a^L)^n$
generate the $n$-th cumulant of $\eta$ after reduction.

The role of $\varphi_r^L(\varphi_a^L)^3$ is to generate the leading
non-Gaussian correction to the stochastic dynamics, rather than to
modify the deterministic drift. At the level of the probability
functional, this appears as a higher-order functional derivative in the
Kramers--Moyal expansion.

The corresponding coefficients are not fixed by the reduced EFT alone
and must be obtained by matching to the underlying Schwinger--Keldysh
theory. This matching, together with the associated scaling in the
super-Hubble expansion, is carried out in
Appendix~\ref{sec:nonlocal_langevin}, where the relevant kernel is obtained
from diagrams involving one Keldysh and multiple retarded propagators.

Non-Gaussian noise is thus encoded in higher-response interaction vertices
of the coarse-grained Schwinger--Keldysh effective action. These terms
generate higher-order functional derivatives in the evolution of the
probability functional, as discussed for instance
in~\cite{Riotto:2011sf,Cohen:2021fzf,Achucarro:2021pdh}.

%%%%%%%%%%%%%%%%%%%%%%%%%%%%%%%%%%%%%%%%%%%%%%%%%%%%%%%%%%%%%%%%%%%%%%%%%%%%%%%%%%%%%%%%%%%%%%%%%%%%%%%%%%%%%%%

\section{Bottom up: from Polchinski to Fokker-Planck equation}
\label{sec:RGPolchinski}
    
%%%%%%%%%%%%%%%%%%%%%%%%%%%%%%%%%%%%%%%%%%%%%%%%%%%%%%%%%%%%%%%%%%%%%%%%%%%%%%%%%%%%%%%%%%%%%%%%%%%%%%%%%%%%%%%%%%%%%%

In this section, we reformulate the coarse-graining procedure as a Wilsonian RG flow on the Schwinger--Keldysh contour and derive the corresponding evolution equation for the effective action. This provides a bottom-up description of the infrared dynamics, in which the stochastic limit and its corrections are organised directly by the RG flow.
This can be implemented by adapting the Wilsonian RG framework of~\cite{Polchinski:1983gv} to real time \cite{Berges:2000ew,  Berges:2012ty}. This provides a systematic evolution equation for $S_{\mathrm{eff}}^\Lambda$ and allows one to track how diffusive and dissipative structures are generated along the flow. Related constructions were presented in~\cite{Cotler:2022fze, Cespedes:2023aal, Goldman:2024cvx, Green:2024cmx, Green:2025hmo, Bowen:2025kyo}.\footnote{It will be interesting to understanad the precise relation between the RG flow and the Lindbladian evolution (see \cite{Goldman:2024cvx} for some recent discussion) . In this framework, Markovian completely positive and trace-preserving dynamical maps emerge naturally from the RG flow, without requiring the imposition of a dynamical Kubo-Martin-Schwinger symmetry in the Schwinger-Keldysh path integral. This, in turn, opens the door to exploring regimes far from thermal equilibrium.}

As in the top-down construction, the effective action $S_{\mathrm{eff}}^\Lambda$ describes an open system obtained after integrating out short modes, and therefore contains non-unitary Schwinger--Keldysh structures such as noise and dissipation. Because the flow is formulated in real time, the RG equation generates kernels that are bilocal in the time arguments, reflecting causal propagation along the contour.

The RG evolution proceeds by integrating out a thin shell of modes crossing the time-dependent cut-off, $k\simeq \Lambda(\eta)$. This shell structure determines the support of the kernels and, in particular, leads to localisation in time once the equal-time limit is taken. Combined with the super-Hubble expansion $\epsilon=\Lambda/(aH)\ll1$, this organises the hierarchy among the resulting operators. At leading order this yields the same local stochastic EFT dominated by diffusion, with response effects parametrically suppressed.

%%%%%%%%%%%%%%%%%%%%%%%%%%%%%%%%%%%%%%%%%%%%%%%%%%%%%%%%%%%%%%%%%%%%%%%%%%%%%%

\subsection{Non-equilibrium Polchinski equation}

In this part, we review generic features of coarse-grained theories in finite time QFT and derive a real-time Wilsonian RG equation on the Schwinger--Keldysh contour. While the construction parallels the Euclidean Polchinski equation, its implementation and physical interpretation differ in important ways. In particular, the flow is intrinsically non-unitary, bilocal in time, and directly generates the noise and response structures characteristic of open systems.

We want to study how the theory evolves as the cut-off scale $\Lambda$ is varied. This amounts to tracking how the effective action changes as the short-wavelength sector is integrated out. The flow is controlled by the shell of modes crossing the cut-off, while physical observables remain invariant under this separation of scales. The requirement that observables be independent of $\Lambda$ can be expressed as a condition on the coarse-grained Schwinger--Keldysh generating functional,
\begin{align}
    Z_\Lambda[J_r,J_a] &= \int \dd \phi^L_\bmk \int^{\phi^L_\bmk} \mathcal{D} \varphi_r^L \int^{0} \mathcal{D} \varphi_a^L \exp\left\{i S_{\mathrm{eff}}^\Lambda[\varphi_r^L, \varphi_a^L]\right\} \nonumber \\
    &\quad\times \exp\left\{i \int \dd \eta\int \frac{\dd^3 \bmk}{(2\pi)^3} \sqrt{-g} \left[\varphi^L_r(\bmk, \eta) J_a(-\bmk, \eta) + \varphi^L_a(\bmk, \eta) J_r(-\bmk, \eta)\right]\right\}\, ,
    \label{eq:smearedPF}
\end{align}
where $S_{\mathrm{eff}}^{\Lambda}$ denotes the effective functional obtained after integrating out modes above the cut-off. 
Unlike in the standard in–out Wilsonian procedure, the quadratic functional is allowed to be non-local in time, reflecting the causal structure of the Schwinger--Keldysh contour.
Demanding that observables computed from this generating functional do not depend on the particular choice of $\Lambda$ leads to the consistency condition
\begin{align} 
\label{eq:RGE}
\frac{d}{d\log \Lambda}Z_{\Lambda}[J_r,J_a]=0\, .
\end{align}
This statement determines how the effective action must vary as the cut-off is shifted. As $\Lambda$ is lowered, the change in $S_{\mathrm{eff}}^\Lambda$ compensates the change in the integration domain, so that correlation functions remain invariant. This is the real-time, or in–in, analogue of the usual Wilsonian flow encapsulated by Polchinski’s equation~\cite{Polchinski:1983gv}, although its implementation differs due to the causal structure of the Schwinger--Keldysh contour.
In particular, the flow governs an open effective action and therefore generates deterministic, dissipative, and stochastic contributions simultaneously.

To make this condition explicit, we decompose the effective action into a quadratic and an interacting piece,
\begin{align}
    S_{\mathrm{eff}}^\Lambda[\varphi^L_r, \varphi^L_a] = S_0^\Lambda[\varphi^L_r, \varphi^L_a]+S_{\mathrm{int}}^\Lambda[\varphi^L_r, \varphi^L_a]\, ,
\end{align}
where $S_0^\Lambda$ is quadratic in the fields and $S_{\mathrm{int}}^\Lambda$ contains cubic and higher interactions.\footnote{In general, the differential operators appearing in $S_0^\Lambda$ are not simply the bare ones; they can already include $\Lambda$-dependent parameters, reflecting the running of couplings or wavefunction renormalization. These coefficients are fixed by the initial conditions of the flow.}
In contrast with the standard Polchinski construction, the quadratic sector already encodes the Schwinger--Keldysh structure, including statistical correlations and causal propagation, and therefore participates non-trivially in the RG flow.
Explicitly, the quadratic functional takes the form,
\begin{align}
S_0^\Lambda[\varphi^L_r, \varphi^L_a]= - \frac{1}{2}  \int_{\bm{k}}\int \dd \eta\, \int \dd \teta\, 
\left(\pr^L(\bm{k},\eta),\pa^L(\bm{k},\eta)\right)  
\begin{pmatrix}
0 & \widehat{D}^{A}_\Lambda \\ 
\widehat{D}^{R}_\Lambda & -2i\widehat{D}^{K}_\Lambda
\end{pmatrix}
\begin{pmatrix}
\pr^L(-\bm{k},\teta) \\ 
\pa^L(-\bm{k},\teta)
\end{pmatrix} \, ,
\label{eq:free_part}
\end{align}
where the operators $\widehat \dd^{R,A,K}_\Lambda$ encode the retarded, advanced, and Keldysh components of the dynamics. It is important to make the convention in \Eq{eq:free_part} explicit. The bilocal kernels $\widehat D_\Lambda^{R,A,K}$ are defined with respect to the flat measure $\dd\eta\,\dd\tilde\eta$, and therefore already include the appropriate factors of the FLRW measure and the time-dependent coefficients of the quadratic operator. In particular, if one starts from the local scalar action in an FLRW background, the factors of $a(\eta)$ generated by $\sqrt{-g}$ and by integrations by parts are absorbed into the definition of $\widehat D_\Lambda^{R,A,K}$. With this convention, the inverse relation \Eq{eq:propagwindow} holds as written, without extra measure factors. This convention fixes the exact form of the RG flow. When we later isolate the leading shell-crossing contribution, the relevant factors $a^2(\eta)\bar\Omega'_\Lambda(k)$ and $a^2(\tilde\eta)\bar\Omega'_{\tilde\Lambda}(k)$ will be extracted explicitly from the action of the quadratic operators on the filtered propagators. They are related to the propagators by
\begin{align}
\label{eq:propagwindow}
\widehat{D}_{\Lambda\,\alpha\gamma} \,\circ\, G_\Lambda^{\gamma\beta}
=
\delta_\alpha^{~\beta}\,\delta(\eta-\tilde\eta)\, .
\end{align}

The coarse-graining is implemented by attaching the cut-off window to both time arguments,
\begin{align}
G^{\alpha\beta}_\Lambda(k, \eta, \teta)
=
\Omega_{\Lambda(\eta)}(k)\,
\Omega_{\Lambda(\teta)}(k)\,
G^{\alpha\beta}(k,\eta,\teta)\, .
\end{align}
equivalently as done in \eqref{eq:filtered_short_prop} for the short modes propagator. 
This implementation of the cut-off defines the coarse-grained theory at a fixed scale $\Lambda$. We now study how this description evolves as the cut-off is varied, requiring that physical observables remain invariant under this change. 

\paragraph{Renormalization Group Equation.}

Let us now unpack Eq.~\eqref{eq:RGE}. Taking the derivative of the generating functional with respect to $\Lambda$, one finds
\begin{align}\label{eq:begin}
    &\Lambda \frac{\dd}{ \dd \Lambda}  Z_{\Lambda}[J_r,J_a] = \int \dd \phi^L_\bmk \int^{\phi^L_\bmk} \mathcal{D} \varphi_r^L \int^{0} \mathcal{D} \varphi_a^L \ee^{i S^\Lambda_{\mathrm{eff}} [\pr^L,\pa^L]} \ee^{i \int_{\bm{k}}\int \dd \eta \sqrt{-g} \left(\pa^L J_r + \pr^L J_a \right)} \\
    &\times \Bigg[\Lambda \frac{\partial}{\partial \Lambda} S^\Lambda_{\mathrm{int}} [\pr^L,\pa^L] 
    - \frac{i}{2} \int_{\bm{k}} \int \dd \eta  \int \dd \teta  
    \left(\pr^L,\pa^L\right) 
    \Lambda \frac{\partial}{\partial \Lambda} 
    \begin{pmatrix}
    0 & \widehat{D}^{A}_\Lambda \\ 
    \widehat{D}^{R}_\Lambda & -2i\widehat{D}^{K}_\Lambda
    \end{pmatrix}
    \begin{pmatrix}
    \pr^L \\ 
    \pa^L
    \end{pmatrix} 
    \Bigg]\, .\nonumber
\end{align}
Requiring the right-hand side to vanish such that \eqref{eq:RGE} holds leads to
\begin{tcolorbox}[colframe=white,arc=5pt]
\begin{align}
\label{eq:PolchinskiEq}
\Lambda \frac{\partial}{\partial \Lambda} \ee^{i S^\Lambda_{\mathrm{int}}[\varphi_r^L,\varphi_a^L]}
=
\frac{i}{2}
\int_{\bm{k}}\int \dd \eta\,\dd \teta \,
\frac{\partial G^{\alpha \beta}_\Lambda(k;\eta,\teta)}{\partial \log \Lambda}
\frac{\delta^2 \ee^{i S^\Lambda_{\mathrm{int}}[\varphi_r^L,\varphi_a^L]}}
{\delta \varphi^L_{\beta}(\bm{k},\teta)\,
 \delta \varphi^L_{\alpha}(-\bm{k},\eta)}\, .
\end{align}
\end{tcolorbox}

This relation follows by rewriting the variation of the quadratic action in terms of functional derivatives and integrating by parts in field space. The term quadratic in $(\widehat D_\Lambda\circ \varphi)$ cancels the variation of the quadratic kernel, while the remaining contributions are either field-independent or reduce to contact terms. Boundary terms vanish for a regulated Gaussian measure. A detailed derivation is given in \App{app:derivation_polchinski}.

Equation~\eqref{eq:PolchinskiEq} determines the flow of the effective action and generates the full Schwinger–Keldysh structure of the theory. In particular, the second functional derivative produces the different SK sectors: the $aa$ sector associated with noise, and the mixed $ar/ra$ sectors associated with causal response. A key feature of the RG equation is that it is bilocal in time: the kernel $\partial_{\log\Lambda}G_\Lambda^{\alpha\beta}(\eta,\tilde\eta)$ couples fields evaluated at different times. This bilocality reflects causal propagation along the Schwinger--Keldysh contour and leads to memory effects along the flow.
The RG flow is controlled by the variation $\partial_{\log\Lambda}G_\Lambda$, which isolates the shell of modes crossing the cut-off. This makes explicit that the flow is driven by the continuous integration of modes at the scale $\Lambda$.

It is sometimes useful to rewrite the flow in terms of the short-wavelength propagator,
\begin{align}
     \bar G^{\alpha\beta}_\Lambda(k, \eta, \teta)
    =
    \bar\Omega_{\Lambda(\eta)}(k)\,
    \bar\Omega_{\Lambda(\teta)}(k)\,
    G^{\alpha\beta}(k,\eta,\teta)\, ,
\end{align}
with $\bar\Omega_\Lambda=1-\Omega_\Lambda$. Since $G_\Lambda+\bar G_\Lambda=G$, one has
\begin{align}
\frac{\partial G_\Lambda^{\alpha\beta}}{\partial\log\Lambda}
=
-\frac{\partial \bar G_\Lambda^{\alpha\beta}}{\partial\log\Lambda}\, .
\end{align}
This relation makes explicit that lowering the cut-off removes modes from the long sector and integrates them out. Accordingly, the RG flow can be written either in terms of the variation of the long-wavelength propagator $G_\Lambda$, or equivalently in terms of the modes being integrated out, encoded in $\partial_{\log\Lambda}\bar G_\Lambda$. The relative minus sign reflects this change of perspective, so that the RG equation can be written equivalently as
\begin{align}
\Lambda \frac{\partial}{\partial \Lambda} \ee^{i S^\Lambda_{\mathrm{int}}}
=
-\frac{i}{2}
\int_{\bm{k}}\int \dd \eta\,\int \dd \teta \,
\frac{\partial \bar G^{\alpha \beta}_\Lambda}{\partial \log \Lambda}
\frac{\delta^2 \ee^{i S^\Lambda_{\mathrm{int}}}}
{\delta \varphi^L_{\beta}(\bm{k},\teta)\,
 \delta \varphi^L_{\alpha}(-\bm{k},\eta)}\, .
\end{align}
In this form the RG flow is explicitly driven by the shell of modes being integrated out, as encoded in $\partial_{\log\Lambda}\bar G_\Lambda$.

%%%%%%%%%%%%%%%%%%%%%%%%%%%%%%%%%%%%%%%%%%%%%%%%%%%%%%%%%%%%%%%%%%%%%%%%%%%%%%%%%%%

\paragraph{de Sitter limit and local flow.}

We now specialise the RG equation \eqref{eq:PolchinskiEq} to light scalar
fields in de Sitter and show how the stochastic regime emerges directly
from the structure of the flow. The key point is that the bilocal kernel
generated by the RG equation becomes effectively local in time once one
focuses on the shell of modes crossing the cut-off, and that this
localisation, together with the super-Hubble expansion, organises the
hierarchy among the resulting operators.

The statement we establish is the following. In the limit
\begin{align}
\epsilon \equiv \frac{\Lambda}{aH} \to 0,
\end{align}
the local part of the RG flow is dominated by the $aa$ sector, while the
mixed $ar$ and $ra$ components are parametrically suppressed. As a
consequence, the leading local contribution is controlled by a single
diffusive structure. In this sense, the RG flow becomes local in time
after shell localisation, local in space after the finite-width
reduction, and diffusion-dominated at leading order in the infrared.

We begin from the bilocal equation
\begin{align}
\Lambda \frac{\partial}{\partial \Lambda} e^{i S^\Lambda_{\mathrm{int}}}
=
-\frac{i}{2}
\int_{\bm{k}}\int \dd \eta\, d \tilde\eta \,
\frac{\partial \bar G^{\alpha \beta}_\Lambda}{\partial \log \Lambda}
\frac{\delta^2 e^{i S^\Lambda_{\mathrm{int}}}}
{\delta \varphi^L_{\beta}(\bm{k},\tilde\eta)\,
 \delta \varphi^L_{\alpha}(-\bm{k},\eta)}\, .
\end{align}
For a narrow filter, the $\Lambda$-derivative isolates a thin shell of
modes crossing the cut-off,
\begin{align}
\partial_{\log\Lambda}\bar\Omega_{\Lambda(\eta)}(k)
\simeq
\Lambda(\eta)\,\delta\big(k-\Lambda(\eta)\big),
\end{align}
so the flow is controlled by modes with $k\simeq\Lambda(\eta)$. The
product of the two shell factors,
\begin{align}
\delta\big(k-\Lambda(\eta)\big)\,
\delta\big(k-\Lambda(\tilde\eta)\big),
\end{align}
is sharply peaked when both times correspond to the same crossing scale.
Since $\Lambda(\eta)$ is monotonic, this localises the bilocal kernel near $\tilde\eta = \eta$.
This is again the origin of the Markovian limit: locality in time follows
directly from the shell structure of the short propagator.

After this step, the flow becomes local in time but remains
non-local in space. Spatial locality does not follow from this step and must be established separately, as we discuss below. Performing the angular integral in momentum space,
\begin{align}
\int \frac{\dd^3\bm k}{(2\pi)^3}\,
\delta(k-\Lambda)\,e^{i\bm k\cdot(\bm x-\bm y)}
=
\frac{\Lambda^2}{2\pi^2}\,
j_0\!\big(\Lambda|\bm x-\bm y|\big),
\end{align}
one obtains a local RG equation of the form
\begin{align}
\Lambda \frac{\partial}{\partial \Lambda} e^{i S^\Lambda_{\mathrm{int}}}
=
\int \dd \eta \int \dd^3 x \int \dd^3y\;
j_0 \big(\Lambda|\bm x-\bm y|\big)\,
\mathcal C^{\alpha\beta}(\eta)\,
\frac{\delta^2 e^{i S^\Lambda_{\mathrm{int}}}}
{\delta \varphi^L_{\beta}(\bm y,\eta)\,
 \delta \varphi^L_{\alpha}(\bm x,\eta)}
+\cdots ,
\end{align}
The coefficients $\mathcal C^{\alpha\beta}(\eta)$ are obtained by
performing the time integral of the shell-localised kernel
$\partial_{\log\Lambda}\bar G^{\alpha\beta}_\Lambda$ after the
bilocal RG equation has been reduced to its local-in-time form.
They therefore encode the same Keldysh and retarded/advanced
components of the RG flow that appear in Eq.~\eqref{eq:RGdiff},
but evaluated on the shell $k\simeq \Lambda(\eta)$.

Evaluating these contributions with the free de Sitter propagators and
expanding for $\epsilon \ll 1$, one finds
\begin{align}
\mathcal C^{\alpha\beta}
=
\frac{H}{2\pi^2}
\begin{pmatrix}
1+\mathcal O(\epsilon^2) & \frac{1}{4}\,\epsilon^3+\mathcal O(\epsilon^5) \\
\frac{1}{4}\,\epsilon^3+\mathcal O(\epsilon^5) & 0
\end{pmatrix},
\end{align}
where the rows and columns are ordered as $(r,a)$. The mixed components
are parametrically suppressed relative to the $rr$ component,
\begin{align}
\frac{\mathcal C^{ra}}{\mathcal C^{rr}},
\frac{\mathcal C^{ar}}{\mathcal C^{rr}}
\sim \epsilon^3 \to 0.
\end{align}
This hierarchy shows that, in the infrared, the RG flow becomes diffusion
dominated, and the leading evolution is governed by a single diffusive
operator acting on the probability functional.

To complete the identification of the infrared structure, we now address
spatial locality. In our framework it is not obtained simply by expanding
$j_0(\Lambda|\bm x-\bm y|)$ for small $\Lambda$, but requires the same
finite-width filter analysis used in the derivation of the local EFT in
the previous section and reviewed in Appendix~\ref{app:local}.
Once that analysis is invoked, the bilocal kernel reduces to the same
local operator basis as in the top-down construction.

Under these assumptions, the leading infrared RG flow takes the local
form
\begin{align}
\Lambda \frac{\partial}{\partial \Lambda} e^{i S^\Lambda_{\mathrm{int}}}
=
\frac{H}{2\pi^2}
\int \dd \eta \int \dd^3 x\;
\frac{\delta^2 e^{i S^\Lambda_{\mathrm{int}}}}
{\delta \varphi^L_{r}(\bm x,\eta)^2}
+\mathcal O(\epsilon^3),
\end{align}
up to the subleading corrections generated by the response sector and by
higher-order operators. Under these assumptions, the infrared RG flow
reduces at leading order to a local Gaussian diffusive structure. This is
precisely the RG counterpart of the leading stochastic EFT obtained in
the top-down construction.

%%%%%%%%%%%%%%%%%%%%%%%%%%%%%%%%%%%%%%%%%%%%%%%%%%%%%%%%%%%%%%%%%%%%%%%%%%%%%%%%%%%%%%%%%%%%%%%%%%%%%%%%%%%%%%%%%%%%%%%%%%%%%%%%%%%%%%%%%%%

\subsection{Effective action from the RG flow}

We now show that the real-time RG flow reconstructs the same non-local
Schwinger--Keldysh effective action obtained in the influence-functional
derivation presented in Sec.~\ref{sec:coarsegraining}. More precisely, we will demonstrate that the RG equation~\eqref{eq:PolchinskiEq}
generates exactly the same operator kernels in the $aa$ and
$ar/ra$ sectors, before any equal-time or super-Hubble reduction is
performed.

This result is non-trivial: the RG flow is formulated abstractly in terms
of the variation of the propagator, and does not explicitly refer to a
partial trace over short modes. Nevertheless, once rewritten as an
equation for $e^{iS^\Lambda_{\mathrm{eff}}}$, it contains enough
information to reconstruct the same open-system operator content.

The reason is that the RG equation already contains, in differential
form, the same ingredients that appear in the coarse-graining
construction. The derivative $\partial_{\log\Lambda}G_\Lambda$
restricts the propagator to the infinitesimal shell of modes at
$k\sim\Lambda$, while the functional derivatives acting on
$e^{iS^\Lambda_{\mathrm{eff}}}$ generate the contractions with those
modes. Each RG step therefore integrates out the shell and inserts its
connected correlators into the effective action for the long sector.
From this point of view, the RG flow is not merely a formal evolution
equation for couplings, but the differential version of the same
coarse-graining operation that in Sec.~\ref{sec:coarsegraining} was
implemented by tracing out the short modes. The comparison carried out
below shows that this equivalence holds already at the level of the
bilocal Schwinger--Keldysh kernels, before any infrared approximation is
taken.
The emergence of the local stochastic EFT then follows from the same
shell-crossing and super-Hubble approximations discussed in the previous
subsection. We therefore focus here entirely on the non-local kernel
structure, postponing any local reduction.

Using the exact identity derived in \App{app:derivation_polchinski}, the RG eq \eqref{eq:PolchinskiEq} can be rewritten directly as an equation for the full exponential $e^{iS_{\mathrm{eff}}^\Lambda}$.
\begin{tcolorbox}[colframe=white,arc=0pt]
\begin{align}
\frac{\partial e^{iS^\Lambda_{\mathrm{eff}}[\varphi_r^L,\varphi_a^L]}}{\partial \log \Lambda}
=
-\frac{i}{2}\,
\frac{\partial G^{\alpha \beta}_\Lambda}{\partial \log \Lambda} \circ
\left[
- \frac{\delta^2 e^{i S^\Lambda_{\mathrm{eff}}[\varphi_r^L,\varphi_a^L]}}{\delta \varphi^L_{\beta}\delta \varphi^L_{\alpha}}
+2i\frac{\delta}{\delta\varphi^L_\alpha}
\left(
\frac{\delta S^\Lambda_{0}[\varphi_r^L,\varphi_a^L]}{\delta \varphi^L_{\beta}}
\,e^{iS^\Lambda_{\mathrm{eff}}[\varphi_r^L,\varphi_a^L]}
\right)
\right] \,.
\label{eq:RGdiff}
\end{align}
\end{tcolorbox}
\noindent This form is equivalent to the Polchinski equation for $e^{iS_{\mathrm{int}}^\Lambda}$, but is better suited for identifying the Schwinger--Keldysh tensor structures generated by the running. The overall coefficient in \Eq{eq:RGdiff} is fixed by the exact derivation in App.~\ref{app:derivation_polchinski} and follows from the normalisation of the quadratic action \Eq{eq:free_part}.

Equation~\eqref{eq:RGdiff} is exact and non-local in time. The second
functional derivative generates the different Schwinger--Keldysh sectors. The term involving $\delta S_0^\Lambda / \delta \varphi_\beta^L$ inserts the linear equations of motion from the quadratic action, so that the RG kernel is composed with the quadratic operators $\widehat \dd^{R,A,K}_\Lambda$, rather than acting directly on the fields. As a
result, the RG flow directly produces operator kernels with the same
causal and statistical structure as in the influence-functional
approach. To make this structure explicit at the level of the fields, it is useful to evaluate the functional derivatives of the quadratic action \Eq{eq:free_part}. Because the kernels $\widehat D_\Lambda^{R,A,K}$ are defined as measure-dressed two-time kernels, these derivatives take the form
\begin{align}
\frac{\delta S_0^\Lambda}{\delta \varphi_r^L(\bm k,\eta)}
&=
- \int \dd \tilde\eta\;
\widehat D^A_\Lambda(k;\eta,\tilde\eta)\,\varphi_a^L(-\bm k,\tilde\eta),
\\
\frac{\delta S_0^\Lambda}{\delta \varphi_a^L(\bm k,\eta)}
&=
- \int \dd \tilde\eta\;
\widehat D^R_\Lambda(k;\eta,\tilde\eta)\,\varphi_r^L(-\bm k,\tilde\eta)
+2i\int \dd \tilde\eta\;
\widehat D^K_\Lambda(k;\eta,\tilde\eta)\,\varphi_a^L(-\bm k,\tilde\eta).
\end{align}
This is the exact convention inherited from the previous subsection.
When the leading shell-crossing contribution is isolated, the factors
$a^2(\eta)\bar\Omega'_\Lambda(k)$ and
$a^2(\tilde\eta)\bar\Omega'_{\tilde\Lambda}(k)$ arise explicitly from the
action of the external quadratic operators on the filtered propagators.
This is the RG analogue of the shell-localised kernels that appeared in
the influence-functional derivation.
Substituting these expressions into \Eq{eq:RGdiff}, one finds that the RG
flow reproduces exactly the same structure as in the
influence-functional derivation, with the same assignment of derivatives
and the same dependence on the short Keldysh propagator.

\paragraph{Gaussian theory: noise sector.}
We first consider the $aa$ sector. At quadratic order, the corresponding contribution to the flow is not simply proportional to $\partial_{\log\Lambda}G^K_\Lambda$. Rather, the Keldysh propagator is dressed by the differential operators acting on the two external legs, so one finds
\begin{align}
\frac{\partial S_{\mathrm{eff}}^\Lambda}{\partial\log\Lambda}\Big|_{aa}
&=
-\varphi_a^L\circ \widehat D_\Lambda^A \circ
\frac{\partial G^K_\Lambda}{\partial\log\Lambda}
\circ \widehat D_\Lambda^R \circ \varphi_a^L
+\mathrm{B.T.}\, ,
\label{eq:RGaaKernel}
\end{align}
where B.T.\ denotes boundary terms generated by integrations by parts. This expression is the RG counterpart of the non-local noise kernel that appears in the influence-functional description.

Equation~\eqref{eq:RGaaKernel} is exact at quadratic order. In order to make contact with the influence-functional derivation, it is convenient to rewrite the kernel in terms of the complementary short-sector propagator,
\begin{align}
G_S^X(k,\eta,\tilde\eta)
=
\bar\Omega_{\Lambda(\eta)}(k)\,
\bar\Omega_{\Lambda(\tilde\eta)}(k)\,
G^X(k,\eta,\tilde\eta),
\qquad X=R,A,K.
\end{align}
The next step is to make explicit how the differential operators act on this kernel. After integrating by parts, the time derivatives are redistributed so that they act partly on the long fields and partly on the kernel. Focusing on the contribution in which one time derivative is effectively associated with each external leg, one obtains
\begin{align}\label{eq:RGaadetail}
&\frac{\partial S_{\mathrm{eff}}^\Lambda}{\partial\log\Lambda}\Big|_{aa}
=
\int_{\mathbf{k}}\int \dd\eta\,\dd\tilde\eta\;
a^2(\eta)\bar\Omega'_\Lambda(k)\,
a^2(\tilde\eta)\bar\Omega'_{\tilde\Lambda}(k)\,
\bigg[
\varphi_a^{\prime L}(\mathbf{k},\eta)\,
G_S^K(k,\eta,\tilde\eta)\,
\varphi_a^{\prime L}(-\mathbf{k},\tilde\eta)
\\
&\quad
-2\,\varphi_a^{\prime L}(\mathbf{k},\eta)\,
\partial_{\tilde\eta}G_S^K(k,\eta,\tilde\eta)\,
\varphi_a^L(-\mathbf{k},\tilde\eta)
+\varphi_a^L(\mathbf{k},\eta)\,
\partial_\eta\partial_{\tilde\eta}
G_S^K(k,\eta,\tilde\eta)\,
\varphi_a^L(-\mathbf{k},\tilde\eta)
\bigg]
+\mathrm{B.T.}\, . \nonumber
\end{align}
In this expression, the factors $a^2(\eta)\bar\Omega'_\Lambda(k)$ and $a^2(\tilde\eta)\bar\Omega'_{\tilde\Lambda}(k)$ arise from the action of the differential operators on the window functions, while the derivatives acting on $G_S^K$ reflect the remaining action of the operators after the integrations by parts.

Equation~\eqref{eq:RGaadetail} makes explicit how the noise sector is
generated along the flow. Because the kernel is entirely controlled by
the short-sector Keldysh propagator and is quadratic in
$\varphi_a^L$, the operator produced at this step is a diffusive
Schwinger--Keldysh kernel. In other words, the RG flow identifies the
fluctuations of the shell with the source of noise in the long-distance
effective theory. The importance of Eq.~\eqref{eq:RGaadetail} is
therefore not only that it matches the influence-functional result, but
that it shows how that result is built up shell by shell as the cut-off
is lowered.

The three terms in \Eq{eq:RGaadetail} reproduce exactly the non-local
diffusive operator obtained in the influence-functional computation,
Eq.~\eqref{eq:Sdiff}. In particular, the arrangement of derivatives
acting on the long fields and on the short Keldysh propagator matches
term by term, and the same shell factors multiply the kernel. The RG
flow therefore reconstructs the same noise operator before any infrared
reduction is performed.

It should be emphasised that Eq.~\eqref{eq:RGaadetail} isolates the part of the exact kernel that
is dominated by shell-crossing and therefore directly comparable to the
influence-functional result. Additional contributions are present in
\eqref{eq:RGaaKernel} and are discussed in App.~\ref{app:derivation_polchinski}, but they either correspond to contact terms or modify the coefficients without changing the operator structure. These arise from the action of the differential operators on the smooth part of the propagator and from lower-derivative pieces of $\widehat D_\Lambda^{R,A}$.

For a massless scalar in de Sitter, these additional contributions are parametrically suppressed in the shell-crossing regime. In that regime, the derivatives of the window functions are sharply localised on the moving shell and dominate over derivatives acting on the smooth super-Hubble propagator. As a result, Eq.~\eqref{eq:RGaadetail} captures the dominant shell-localised part of the kernel and reproduces the corresponding contribution to the influence-functional result \eqref{eq:Sdiff}.
\paragraph{Gaussian theory: response sector.}
The same RG equation determines the mixed $ar/ra$ sector. As in the noise
sector, the flow generates bilocal kernels that are identical to those
obtained in the influence-functional derivation, now involving the
retarded and advanced propagators.
\begin{align}
\frac{\partial S_{\mathrm{eff}}^\Lambda}{\partial\log\Lambda}\Big|_{ar}
=
-\int \dd^4x \int \dd^4y\;
\varphi_a^L(x)\,
\frac{\partial \widehat G^R_\Lambda(x,y)}{\partial\log\Lambda}\,
\varphi_r^L(y)
+\mathrm{B.T.}\, ,
\label{eq:RGar}
\end{align}
with the advanced counterpart obtained by \(R\leftrightarrow A\). Using the inverse relation \(\widehat D_\Lambda\circ \, G_\Lambda=1\), the running of the mixed kernel may be rewritten in terms of the short propagators \(G_S^{R,A}\). As in the diffusion sector, the comparison with the influence functional requires isolating explicitly the leading shell-crossing piece. The same shell-crossing approximation used in the $aa$ sector produces the factors $a^2(\eta)\bar\Omega'_\Lambda(k)$ and $a^2(\tilde\eta)\bar\Omega'_{\tilde\Lambda}(k)$, so that after the same integrations by parts used above one obtains bilocal kernels of the same form as in the influence-functional derivation:
\begin{align}
\frac{\partial S_{\mathrm{eff}}^\Lambda}{\partial\log\Lambda}\Big|_{ar}
&=
\int_{\mathbf{k}}\int \dd\eta\,\dd\tilde\eta\;
a^2(\eta)\bar\Omega'_\Lambda(k)\,
a^2(\tilde\eta)\bar\Omega'_{\tilde\Lambda}(k)\,
\bigg\{
\varphi_a^{\prime L}(\eta)\varphi_r^{\prime L}(\tilde\eta)\,
G^A_S(k,\eta,\tilde\eta)
\nonumber\\
&\qquad
-
\left[
\varphi_r^{\prime L}(\eta)\varphi_a^L(\tilde\eta)
-
\varphi_r^L(\eta)\varphi_a^{\prime L}(\tilde\eta)
\right]
\partial_\eta
G^A_S(k,\eta,\tilde\eta)
\nonumber\\
&\qquad
+
\varphi_a^L(\eta)\varphi_r^L(\tilde\eta)\,
\partial_\eta\partial_{\tilde\eta}
G^A_S(k,\eta,\tilde\eta)
\bigg\}
+\mathrm{B.T.},
\label{eq:RGdiss_bilocal}
\\
\frac{\partial S_{\mathrm{eff}}^\Lambda}{\partial\log\Lambda}\Big|_{ra}
&=
\int_{\mathbf{k}}\int \dd\eta\,\dd\tilde\eta\;
a^2(\eta)\bar\Omega'_\Lambda(k)\,
a^2(\tilde\eta)\bar\Omega'_{\tilde\Lambda}(k)\,
\bigg\{
\varphi_a^{\prime L}(\eta)\varphi_r^{\prime L}(\tilde\eta)\,
G^R_S(k,\eta,\tilde\eta)
\nonumber\\
&\qquad
-
\left[
\varphi_r^{\prime L}(\eta)\varphi_a^L(\tilde\eta)
+
\varphi_r^L(\eta)\varphi_a^{\prime L}(\tilde\eta)
\right]
\partial_\eta
G^R_S(k,\eta,\tilde\eta)
\nonumber\\
&\qquad
-
\varphi_a^L(\eta)\varphi_r^L(\tilde\eta)\,
\partial_\eta\partial_{\tilde\eta}
G^R_S(k,\eta,\tilde\eta)
\bigg\}
+\mathrm{B.T.}.
\label{eq:RGdrift_bilocal}
\end{align}
These expressions coincide with the bilocal operators obtained in
Eqs.~\eqref{eq:Sdiss_bilocal} and \eqref{eq:Sdrift_bilocal}. At this
stage it is better to think of the mixed sector as a single bilocal
response kernel rather than as already split into drift and dissipation.
That split becomes transparent only after decomposing the kernel into
parts symmetric and antisymmetric under interchange of the two time
arguments. The symmetric combination gives
\begin{align}
\varphi_r^{\prime L}(\eta)\varphi_a^L(\eta)
+
\varphi_r^L(\eta)\varphi_a^{\prime L}(\eta),
\end{align}
which corresponds to drift, namely a renormalisation of the effective
Hamiltonian evolution of the long modes. By contrast, the antisymmetric
combination gives
\begin{align}
\varphi_r^{\prime L}(\eta)\varphi_a^L(\eta)
-
\varphi_r^L(\eta)\varphi_a^{\prime L}(\eta),
\end{align}
which corresponds to dissipation. Thus the RG flow reproduces not only
the existence of a response sector, but also its decomposition into
Hamiltonian renormalisation and dissipative branch-mixing terms.

\paragraph{First order in $\lambda$.}
The same logic extends to the interacting theory. The RG flow generates
bilocal kernels with one insertion of the interaction, and these again
match exactly the corresponding operators obtained in the
influence-functional derivation. Writing
\begin{align}
S_{\mathrm{eff}}^\Lambda
=
S_{\mathrm{eff},0}^\Lambda
+
\lambda\,S_{\mathrm{eff},1}^\Lambda
+\cdots ,
\end{align}
the flow generates corrections both in the $aa$ sector, associated with
noise, and in the mixed $ar/ra$ sector, associated with causal response.
In the $aa$ sector, the first correction is obtained by replacing one of
the Gaussian insertions by one insertion of the interaction. One finds
\begin{align}
\frac{\partial S_{\mathrm{eff}}^\Lambda}{\partial\log\Lambda}\Big|_{aa}^{(1)}
&=
-
\frac{\delta S_{\mathrm{int}}^\Lambda}{\delta\varphi_r}
\circ
\frac{\partial G_S^K}{\partial\log\Lambda}
\circ
\widehat D_\Lambda^R\circ \varphi_a
-
\varphi_a\circ \widehat D_\Lambda^A
\circ
\frac{\partial G_S^K}{\partial\log\Lambda}
\circ
\frac{\delta S_{\mathrm{int}}^\Lambda}{\delta\varphi_r}
+\cdots ,
\label{eq:RGaa1}
\end{align}
This has the same Schwinger--Keldysh structure as the Gaussian noise
kernel, but now dressed by one interaction insertion. Since
\begin{align}
\frac{\delta S_{\mathrm{int}}^\Lambda}{\delta\varphi_r}
=
-\sqrt{-g}\,\lambda
\left[
\frac12 \varphi_r^2\varphi_a
+\frac1{24}\varphi_a^3
\right],
\end{align}
the bilocal kernel generated by \Eq{eq:RGaa1} coincides with the
interacting correction to the noise sector derived in the
influence-functional approach. After the same shell-crossing and local
super-Hubble approximation used above, its leading contribution reduces
to a field-dependent correction to the diffusion term,
\begin{align}
\lambda\,(\varphi_r^L)^2(\varphi_a^{\prime L})^2,
\end{align}
which is precisely the quartic diffusive operator obtained in the
influence-functional derivation. Terms containing additional powers of
$\varphi_a^L$ are also generated, but according to the super-Hubble
power counting they are subleading and may be collected separately in
Appendix~\ref{app:ope}.

The mixed sector is corrected in an analogous way. Replacing one Gaussian insertion by one interaction insertion gives
\begin{align}
\frac{\partial S_{\mathrm{eff}}^\Lambda}{\partial\log\Lambda}\Big|_{ar}^{(1)}
&=
-
\frac{\delta S_{\mathrm{int}}^\Lambda}{\delta\varphi_r}
\circ
\frac{\partial G_S^R}{\partial\log\Lambda}
\circ
\widehat D_\Lambda^R\circ \varphi_r
-
\varphi_a\circ \widehat D_\Lambda^A
\circ
\frac{\partial G_S^A}{\partial\log\Lambda}
\circ
\frac{\delta S_{\mathrm{int}}^\Lambda}{\delta\varphi_a}
+\cdots ,
\label{eq:RGar1}
\end{align}
again up to boundary terms and terms that are subleading in the local super-Hubble limit. As in the Gaussian theory, Eq.~\eqref{eq:RGar1} should first be understood as defining a bilocal mixed kernel. After decomposing that kernel into parts symmetric and antisymmetric under interchange of the time arguments, one finds interacting corrections both to drift and to dissipation.

However, these interacting response terms remain subleading once the cut-off is lowered outside the horizon. The reason is the same as at quadratic order. The mixed sector is controlled by retarded and advanced kernels and their derivatives, and therefore carries additional powers of the shell momentum $k\sim \Lambda$, whereas the $aa$ sector is controlled by the infrared-enhanced Keldysh propagator. After the same local shell reduction used above, the interacting response operators are therefore suppressed by powers of $\epsilon=\Lambda/(aH)$ relative to the leading quartic correction to diffusion.\\

To summarise, the main result is that the RG flow reconstructs the same bilocal Schwinger--Keldysh kernels as the influence-functional derivation, in both the noise and response sectors, already before any equal-time or super-Hubble approximation is taken. In this sense, the RG
equation provides the differential implementation of the same coarse-graining procedure: at each step, the derivative
$\partial_{\log\Lambda} \bar{G}_\Lambda$ isolates the two-point correlator of the modes in the thin shell at $k\sim\Lambda$ being integrated out, while the functional derivatives insert this contribution into the long-wavelength effective action with the appropriate
Schwinger--Keldysh tensor structure.

The local stochastic EFT then follows as a further infrared reduction of these non-local kernels. After the shell-crossing and super-Hubble expansion, the $aa$ sector gives the dominant diffusive terms, while the $ar/ra$ sector yields subleading response contributions, including drift and dissipation. The RG flow therefore reproduces not only the existence of the open EFT, but also the hierarchy of operators that characterises its stochastic limit.

%%%%%%%%%%%%%%%%%%%%%%%%%%%%%%%%%%%%%%%%%%%%%%%%%%%%%%%%%%%%%%%%%%%%%%%%%%%%%%%%%%%

\paragraph{Diffusive infrared regime of the RG flow.}

The RG analysis above shows that, once the cut-off is lowered into the
super-Hubble regime, the flow becomes strongly hierarchical: the $aa$
sector, controlled by the equal-time Keldysh kernel, remains
unsuppressed, while the mixed $ar/ra$ sectors are suppressed by powers
of $\epsilon=\Lambda/(aH)$. As a result, each RG step predominantly
injects stochastic fluctuations into the long sector, while the causal
response is subleading. This provides the RG interpretation of the
stochastic regime derived in the coarse-graining approach.

To connect this result with the effective description used in the main
text, we recall the reduced first-order action derived in
Sec.~\ref{subsec:recover} and analysed in
Appendix~\ref{sec:nonlocal_langevin},
\begin{align}
S_{\rm red}
=
\int \dd t \int \dd^3\bmx\, a^3(t)\,
\chi_a^L
\left[
\eta
-\dot\varphi_r^L
+\frac{1}{3H}\frac{\lambda}{3!}(\varphi_r^L)^3
-\frac{1}{3H}\frac{\partial^2}{a^2}\varphi_r^L
\right]
+\cdots .
\label{eq:first_order_action_main_repeat}
\end{align}
The stochastic source $\eta$ arises from the $aa$ sector, while the other  terms in the square brackets originate from the microscopic action and determine the response ($ar/ra$) sector. 
The RG flow does not determine the existence of the latest, but it does determine its infrared hierarchy. 
As we will now see, the RG hierarchy therefore explains why stochastic fluctuations dominate over deterministic response at late times. Nevertheless, if one considers the Hubble friction as a manifestation of dissipation, some form of the fluctuation-dissipation theorem can be obtained \cite{Rigopoulos:2016oko}. It follows from the emergence of the dynamical KMS symmetry \cite{Crossley:2015evo, Liu:2018kfw} discussed in \App{app:dKMS}.

The infrared scaling is fixed by the quadratic part of
\eqref{eq:first_order_action_main_repeat},
\begin{align}
S_{\rm red}^{(2)}
=
\int \dd t \int \dd^3\bmx\, a^3(t)\,
\chi_a^L
\left[
\eta
-\dot\varphi_r^L
-\frac{1}{3H}\frac{\partial^2}{a^2}\varphi_r^L
\right].
\label{eq:RGFixed}
\end{align}
After integrating out the Gaussian noise, this gives a quadratic
Schwinger--Keldysh theory in which the Keldysh sector generates the
noise term $(\chi_a^L)^2$, while the retarded sector fixes the leading
operators $\chi_a^L \partial_t \varphi_r^L$ and
$\chi_a^L \nabla_{\rm phys}^2 \varphi_r^L$. Since the quadratic theory
contains one time derivative but two spatial derivatives, it selects a
diffusive scaling in which time scales as two powers of space. Retaining
the leading spatial term, the anisotropic scaling is\footnote{
The scaling dimensions above refer to the local diffusive
Schwinger--Keldysh theory prior to the final overdamped stochastic
reduction. In the strict super-Hubble limit the physical field
$\varphi_r^L$ behaves as an effectively dimension-zero stochastic
variable, consistent with the equilibrium Fokker--Planck distribution.
This does not invalidate the present power counting, whose role is to
organise the hierarchy of operators in the open-system effective action.
In particular, the relevant hierarchy is controlled by response fields,
time derivatives, and powers of the coarse-graining parameter. Spatial
gradients are exponentially suppressed after horizon crossing and become
subleading after a few e-folds. The distinction between the scaling
dimensions of $\varphi_r^L$ and $\chi_a^L$ nevertheless remains
important, as it controls the hierarchy between semiclassical and subleading vertices in the Schwinger--Keldysh effective theory.}
\begin{align}
\bmx \to b\,\bmx,
\qquad
t\to b^z t,
\end{align}
with
\begin{align}
z=2,
\qquad
[\varphi_r^L]=\frac{d-2}{2},
\qquad
[\chi_a^L]=\frac{d+2}{2}.
\end{align}
The RG flow therefore drives the reduced density matrix into a Gaussian
diffusive infrared regime, with the scaling discussed in
Sec.~\ref{sec:SIFderiv}. This corresponds to the stochastic
inflationary regime. While the microscopic theory is relativistic, the open-system dynamics
reorganises its infrared behaviour into a diffusive one.

This clarifies the relation between the $\epsilon$-expansion and the
infrared relevance. The former determines which operators are generated
at the matching scale, while the latter determines which of them govern
the infrared dynamics. As shown in
Appendix~\ref{app:sh-ir-scaling}, these orderings do not coincide. In
particular, the operator $(\varphi_r^L)^3\varphi_a^L$ becomes relevant
in $d=4$, whereas operators with additional response fields are
increasingly irrelevant. The RG flow therefore not only explains the emergence of a stochastic EFT from a conservative microscopic theory, but also explains
the hierarchy of operators that controls its infrared dynamics. This is
closely analogous to state-dependent non-equilibrium RG analyses, where
the quadratic structure of the state modifies the scaling of
interactions relative to vacuum power counting
\cite{Rosenhaus:2025bgy, Nagy:2025pqs}.

Another interesting feature of the infrared effective action
\eqref{eq:RGFixed} is that, after introducing the canonically
normalized response field $\chi_a$, it exhibits an emergent dynamical
KMS (dKMS) symmetry \cite{Crossley:2015evo}. The dKMS symmetry is the real-time counterpart of
thermal equilibrium in the Schwinger--Keldysh formalism: it combines
time reversal with a shift of the response field proportional to the
equation of motion, leaving the effective action invariant. This
symmetry enforces the fluctuation--dissipation relation between the
dissipative and stochastic sectors of the theory and therefore implies
the existence of an effective temperature. As shown in
App.~\ref{app:dKMS}, the corresponding temperature is of order
$T_{\rm eff}\sim H$, or equivalently $\beta_{\rm eff}\sim H^{-1}$, in
agreement with the equilibrium distribution obtained from the
Fokker--Planck equation.
It is important to stress that the present analysis only establishes
the dKMS symmetry within the infrared approximation considered in this
work. The transformation is exact in the classical field
$\varphi_r$, while only the leading terms in the response field
$\chi_a$ have been retained. Higher-order terms in $\chi_a$ correspond
to non-Gaussian noise cumulants and are expected to modify the dKMS
transformation. Similarly, subleading derivative operators and
non-Markovian corrections inherited from the underlying bilocal
influence functional may also break the symmetry. The appearance of
dKMS should therefore be viewed as an emergent property of the
infrared dynamics rather than an exact symmetry of the microscopic
theory \cite{Bucca:2026xmh}. Understanding how this symmetry is corrected beyond the
leading infrared approximation, and whether it can be systematically
extended to the full effective theory, is left for future work.

This diffusive infrared regime is reached by the RG flow
of the reduced density matrix derived from the Schwinger--Keldysh
coarse-graining, and provides the infrared limit of the open effective
field theory constructed in this work. 
The crucial insight from the RG is that interactions which are perturbative in the microscopic relativistic theory become relevant with respect to the infrared diffusive scaling. In particular, if an operator has positive scaling dimension with respect to this infrared scaling, its coupling grows under coarse-graining, so that its contribution to the effective action is enhanced relative to the quadratic terms. Conversely, operators that are initially marginal in the microscopic theory such as $\varphi_r^L(\varphi_a^L)^3$ may become increasingly irrelevant nearby the diffusive IR fixed point.
In this sense, the long-wavelength theory flows toward the Gaussian diffusive fixed point, as interactions that are relevant under the diffusive scaling grow along the RG flow. In particular, the steady state obtained recovers the equilibrium solution derived from the Fokker-Planck equation in \cite{Starobinsky:1994bd}. The corresponding resummation of this infrared dynamics \cite{Tsamis:1993ub, Boyanovsky:1998aa, Burgess:2009bs, Serreau:2013psa, Garbrecht:2013coa, Garbrecht:2014dca, Burgess:2015ajz, Green:2020txs, Andersen:2021lii, Colas:2024xjy, Burgess:2024eng, Miao:2024shs, Bhattacharya:2025jtp, Nath:2025vhg, Ye:2026saa}, which reduces to the stochastic/Fokker--Planck description, will be derived in the next subsection.

%%%%%%%%%%%%%%%%%%%%%%%%%%%%%%%%%%%%%%%%%%%%%%%%%%%%%%%%%%%%%%%%%%%%%%%%%%%%%%%%%%%%%%%%%%%%%%%%%%%%%%%%%%%%%%%

\subsection{Flow equation for the probability distribution function}

We now show that the real-time RG equation \eqref{eq:RGdiff} induces a
closed evolution equation for the probability distribution functional
$P_\Lambda[\phi^L]$. In this way, the stochastic dynamics can be derived
directly from the path integral: the Starobinsky--Yokoyama equation
emerges as the infrared limit of the Wilsonian RG flow.

This construction generalises previous derivations of stochastic
inflation based on infrared resummation
\cite{Gorbenko:2019rza,Cespedes:2023aal}. In particular, we show that the
stochastic description arises as the infrared limit of the RG flow of the
reduced density matrix, rather than being postulated as an effective
dynamical equation.

The connection to observables follows from the Schwinger--Keldysh path
integral, which projects onto the diagonal elements of the reduced
density matrix. These define the probability distribution functional
$P_\Lambda[\phi^L]$ from which equal-time scalar correlators are
computed. We now make this relation explicit and derive the corresponding
evolution equation.

\paragraph{Probability distribution.}

To extract an evolution equation for physical observables, we consider
the reduced density matrix of the long-wavelength sector obtained after
integrating out the short modes. Observables accessible in cosmology are
functions of the field configuration at a fixed time, and their
expectation values can be written as
\begin{align}
\langle \mathcal O^\ell(\eta_0) \rangle
=
\mathrm{Tr}\!\left[\widehat{\mathcal O}^\ell \widehat{\rho}_\Lambda(\eta_0)\right]
=
\int \mathcal D\phi^L\;
\mathcal O^\ell[\phi^L]\,
P_\Lambda[\phi^L]\,,
\end{align}
where
\begin{align}
P_\Lambda[\phi^L]
\equiv
\langle \phi^L | \widehat{\rho}_\Lambda(\eta_0) | \phi^L \rangle\ ,
\end{align}
is the diagonal element of the density matrix in the field basis.

This definition remains valid for mixed states, which arise naturally in
the present context due to the non-unitary evolution induced by
coarse-graining over short modes. In the Schwinger--Keldysh formulation,
$P_\Lambda[\phi^L]$ admits the exact path-integral representation
\begin{align}\label{eq:PDFsimpl_clean}
P_\Lambda[\phi^L]
=
\int \mathcal{D}\varphi_r^L \,\mathcal{D}\varphi_a^L \;
e^{i S_{\mathrm{eff}}^\Lambda[\varphi_r^L, \varphi_a^L]}\,
\delta\big[\varphi_r^L(t_0)-\phi^L\big]\,
\delta\big[\varphi_a^L(t_0)\big] ,
\end{align}
which corresponds to the projection of the Schwinger--Keldysh path
integral onto the diagonal of the density matrix at the final time. The
delta functionals implement the boundary conditions
\begin{align}
\varphi^L_r(\mathbf{k}, t_0)=\phi^L(\mathbf{k}),
\qquad
\varphi^L_a(\mathbf{k}, t_0)=0\ .
\end{align}
This condition enforces $\varphi_a^L(t_0)=0$ and selects the configuration
relevant for diagonal matrix elements.

We now insert the RG flow equation \eqref{eq:RGdiff} into the path
integral. Functional derivatives acting on $e^{iS_{\mathrm{eff}}^\Lambda}$
are integrated by parts in field space and transferred to the boundary
projector. Since the explicit dependence on $\phi^L$ enters only through
the final-time constraint $\varphi_r^L(t_0)=\phi^L$, this converts
functional derivatives with respect to $\varphi_r^L$ into derivatives
with respect to $\phi^L$, localised at the final time:\footnote{The replacement in Eq.~\eqref{eq:reprule_clean} is
dimensionally consistent. A functional derivative has the dimension
required to compensate the integration measure and the field variation
in $\delta{\cal F}=\int d^Dx\,(\delta{\cal F}/\delta\phi)\delta\phi$.
Thus, for $[\varphi_r^L]=1$, the four-dimensional functional derivative
has dimension $4-[\varphi_r^L]=3$, whereas the functional derivative on
the final-time hypersurface has dimension
$3-[\phi^L]=2$. The missing unit of mass dimension is supplied by
$[\delta(t-t_0)]=1$, so that both sides of
Eq.~\eqref{eq:reprule_clean} have dimension three.}
\begin{align}\label{eq:reprule_clean}
\frac{\delta}{\delta\varphi^L_r(t,\mathbf{x})}
\;\rightarrow\;
\delta(t-t_0)\frac{\delta}{\delta\phi^L(\mathbf{x})}.
\end{align}

By contrast, functional derivatives with respect to $\varphi_a^L$ do not
generate physical derivatives after projection, since the constraint
$\varphi_a^L(t_0)=0$ removes any dependence on its final value. As a
result, only components with $\alpha=r$ contribute after the projection.

\paragraph{Diffusion term.}

The first term in \Eq{eq:RGdiff} contains two functional derivatives.
After integrating by parts, both derivatives act on the boundary
projector and collapse the time integrals to the final time
$t=t'=t_0$. One finds
\begin{align}
\frac{\partial P_\Lambda[\phi^L]}{\partial\log\Lambda}\Big|_{\rm diff}
&=
i
\int \dd^3x\,\dd^3y\;
\frac{\partial G^{K}_\Lambda(\vert\bm{x}-\bm{y}\vert,t_0,t_0)}{\partial\log\Lambda}\,
\frac{\delta^2P_\Lambda[\phi^L]}
{\delta\phi^L(\bm{x})\,\delta\phi^L(\bm{y})},
\label{eq:FP-diffusion-short-clean}
\end{align}
where we used $G^{rr}_\Lambda=2G^K_\Lambda$. This corresponds to a
diffusion term localised at the final time slice, whose magnitude is
controlled by the amplitude of fluctuations at the cut-off.

\paragraph{Drift term.}

For the second term in \Eq{eq:RGdiff}, only $\alpha=r$ contributes after
projection. Moreover, at $\varphi_a^L=0$ only $\beta=a$ survives, since
\begin{align}
\left.
\frac{\delta S^\Lambda_0}{\delta\varphi^L_r}
\right|_{\varphi_a^L=0}
=0.
\end{align}
The remaining contribution defines the effective equation-of-motion
kernel
\begin{align}
E^\Lambda[\phi^L]
=
\left.
\frac{\delta S^\Lambda_{0}[\varphi^L_r,\varphi^L_a]}
{\delta\varphi^L_a}
\right|_{\varphi^L_r=\phi^L,\;\varphi^L_a=0}.
\label{eq:ELambda-short-clean}
\end{align}
In this case, only the outer functional derivative collapses to the
boundary, while the inner variation remains evaluated at an intermediate
time $t'$. This yields
\begin{align}
\frac{\partial P_\Lambda[\phi^L]}{\partial\log\Lambda}\Big|_{\rm drift}
&=
\int \dd^3x\,\dd^3y\;\int^{t_0}\dd t'\;
\frac{\partial G^{R}_\Lambda(\vert\bm{x}-\bm{y}\vert;t_0,t')}{\partial\log\Lambda}\,
\frac{\delta}{\delta\phi^L(\bm{x})}
\Big[
E^\Lambda(t',\bm{y};[\phi^L])\,
P_\Lambda[\phi^L]
\Big].
\label{eq:FP-drift-short-clean}
\end{align}

Combining \Eqs{eq:FP-diffusion-short-clean}, we
obtain the exact RG-induced flow equation for the probability
distribution,
\begin{tcolorbox}[colframe=white,arc=0pt]
\begin{align}\label{eq:RGGFP_clean}
\frac{\partial P_\Lambda[\phi^L]}{\partial\log\Lambda}
&=
i
\int \dd^3x\, \dd^3y\;
\frac{\partial G^{K}_\Lambda(\vert\bm{x}-\bm{y}\vert;t_0,t_0)}{\partial\log\Lambda}\,
\frac{\delta^2 P_\Lambda[\phi^L]}
{\delta\phi^L(\bm{x})\,\delta\phi^L(\bm{y})}
\nonumber\\
&\quad
+
\int \dd^3x\, \dd^3y\;\int^{t_0} \dd t'\,
\frac{\partial G^{R}_\Lambda(\vert\bm{x}-\bm{y}\vert;t_0,t')}{\partial\log\Lambda}\,
\frac{\delta}{\delta\phi^L(\bm{x})}
\Big[
E^\Lambda(t',\bm{y};[\phi^L])\,
P_\Lambda[\phi^L]
\Big].
\end{align}
\end{tcolorbox}
Equations~\eqref{eq:PDFsimpl_clean}--\eqref{eq:RGGFP_clean} are exact
consequences of the RG flow and of the projection onto the diagonal
density matrix. No semiclassical or super-Hubble approximation has been
used so far.
In particular, Eq.~\eqref{eq:RGGFP_clean} is the probability-space counterpart of the Polchinski
flow. It is bilocal in space and retains memory of intermediate times
through the retarded kernel. The equal-time Keldysh propagator controls
diffusion, while the retarded propagator controls causal response. In
this sense, \Eq{eq:RGGFP_clean} is a generalised non-local
Fokker--Planck equation derived directly from the RG flow, without
introducing a Langevin description. This equation is analogous to the one found in \cite{Hu:1991di, Hu:1993qa, Collins:2017haz} for generalised Brownian motion. Its appearance here shows that is a consequence of the Wilsonian RG flow generating noise and dissipation.

The exact equation \eqref{eq:RGGFP_clean} already captures the
Gaussian stochastic regime. To identify the first correction beyond it, one must now evaluate the Schwinger--Keldysh weight itself in the same long-wavelength regime used to derive the local EFT. The same
super-Hubble hierarchy that organises the shell-localised RG kernel
therefore also controls the first non-Gaussian correction to the
probability flow.

\paragraph{Super-Hubble limit of the RG flow.}

Note that equations of the form \eqref{eq:RGGFP_clean} are not specific to stochastic inflation and more generally arise in exact renormalisation-group descriptions of reduced density matrices or probability distributions, including in flat space. What is specific to stochastic inflation is the additional coarse-graining procedure associated with super-Hubble dynamics, whereby fluctuations crossing the coarse-graining scale are continuously transferred into the long-wavelength sector. In particular, the stochastic interpretation emerges only once the long-wavelength fields are understood as spatially coarse-grained variables.

To recover the stochastic-inflation equation, we now evaluate
\Eq{eq:RGGFP_clean} in the same infrared regime in which the local EFT
was derived. The point of this step is not to introduce a new
approximation, but to apply to the probability-flow equation the same
shell-crossing and super-Hubble expansion that were used above to derive
the local effective action itself. In particular, the hierarchy between
diffusion and response in the probability flow must follow from the same
hierarchy already established for the shell-localised RG kernel.

For clarity, we first work at tree level. The exact flow equation
\eqref{eq:RGGFP_clean} contains two qualitatively different structures.
The first is the Keldysh contribution,
\begin{align}\label{eq:contrib1}
\frac{\partial P_\Lambda}{\partial\log\Lambda}\Big|_{\rm diff}
&=
i
\int \dd^3x\, \dd^3y\;
\frac{\partial G^{K}_\Lambda(|\bm{x}-\bm{y}|;t_0,t_0)}{\partial\log\Lambda}\,
\frac{\delta^2 P_\Lambda}{\delta\phi^L(\bm{x})\,\delta\phi^L(\bm{y})},
\end{align}
which is already local in time because both functional derivatives act
on the final-time projector. The second is the retarded contribution,
\begin{align}\label{eq:contrib2}
\frac{\partial P_\Lambda}{\partial\log\Lambda}\Big|_{\rm drift}
&=
\int \dd^3x\, \dd^3y\;\int^{t_0} \dd t'\,
\frac{\partial G^{R}_\Lambda(|\bm{x}-\bm{y}|;t_0,t')}{\partial\log\Lambda}\,
\frac{\delta}{\delta\phi^L(\bm{x})}
\Big[
E^\Lambda(t',\bm{y};[\phi^L])\,
P_\Lambda
\Big],
\end{align}
which is non-local in time because the retarded kernel keeps memory of
the intermediate time $t'$. Such equations are often referred to as generalised Fokker--Planck equations and have appeared previously, for example in the context of quantum Brownian motion \cite{Hu:1991di,Hu:1993qa, Collins:2017haz, Kamenshchik:2024ybm, Kamenshchik:2025ses, Tasinato:2025zqt, Nassiri-Rad:2025dsa, Ahmadi:2025oon, Cho:2026prp}.
It is important to distinguish this
RG-induced contribution from the classical drift already present in the
first-order effective action: the latter arises from the deterministic
equations of motion of the long-wavelength modes, whereas the term above
is generated dynamically by integrating out short modes along the RG
flow.

The relative importance of these two terms in \Eqs{eq:contrib1} and \eqref{eq:contrib2} is controlled by the same
shell-localised coefficients that appeared in the previous subsection. As shown there in the shell-reduced expansion of the
bilocal RG kernel, the corresponding Schwinger--Keldysh coefficients satisfy
\begin{align}
\mathcal C^{aa}\sim \epsilon^0,
\qquad
\mathcal C^{ar},\mathcal C^{ra}\sim \epsilon^3,
\qquad
\epsilon\equiv \frac{\Lambda}{aH}\ll1.
\end{align}
These coefficients were introduced precisely to encode the local
Keldysh and retarded/advanced parts of the shell-reduced RG kernel,
namely the local limit of $\partial_{\log\Lambda}G_\Lambda^{\alpha\beta}$
evaluated on the shell $k\simeq\Lambda$. Their scaling therefore
directly determines the relative importance of the diffusion and
retarded contributions in the probability flow. The statement above
translates into a hierarchy for Eq.~\eqref{eq:RGGFP_clean}: the Keldysh
contribution enters at order $\epsilon^0$, while the retarded
contribution is suppressed by $\epsilon^3$ once the same
shell-localisation is imposed.

It is useful to note the following. 
On super-Hubble scales, the equal-time
Keldysh propagator remains finite, so the diffusion term survives at
leading order. Its first subleading correction comes from the
super-Hubble expansion of the equal-time correlator and is therefore of
order $\epsilon^2$. By contrast, the retarded propagator vanishes at
equal times. The retarded part of \eqref{eq:RGGFP_clean} contributes
only because of the integral over earlier times $t'$, and therefore does
not yet have the form of a local operator. After performing the same
shell-localisation that leads to the local EFT, this integral collapses
onto the final time and produces a local contribution whose coefficient
inherits the $\epsilon^3$ suppression of the retarded sector.

The exact probability-flow equation therefore admits the super-Hubble
expansion
\begin{align}
\frac{\partial P_\Lambda[\phi^L]}{\partial\log\Lambda}
&=
\mathcal \dd^{(0)}_\Lambda P_\Lambda
+
\epsilon^2\,\mathcal \dd^{(2)}_\Lambda P_\Lambda
+
\epsilon^3\,\mathcal R^{(3)}_\Lambda P_\Lambda
+\cdots ,
\label{eq:RGGFP_hierarchy}
\end{align}
where $\mathcal \dd^{(0)}_\Lambda$ is the leading Keldysh diffusion
operator, $\mathcal \dd^{(2)}_\Lambda$ its first super-Hubble correction,
and $\mathcal R^{(3)}_\Lambda$ the leading local contribution inherited
from the retarded sector. In particular, diffusion dominates the
infrared evolution, while the RG-induced drift-like correction is
parametrically subleading.

Keeping only the leading term, one recovers
\begin{align}
\frac{\partial P_\Lambda[\phi^L]}{\partial\log\Lambda}
=
\frac{H^2}{8\pi^2}
\int \dd^3x \int \dd^3y\,
j_0(\Lambda|\bm{x}-\bm{y}|)
\frac{\delta^2 P_\Lambda[\phi^L]}
{\delta\phi^L(\bm{x})\,\delta\phi^L(\bm{y})}
+\mathcal O(\epsilon^2,\epsilon^3),
\label{eq:treeleveDiff_clean}
\end{align}
which reproduces the standard diffusion term of stochastic inflation.
At this stage the flow is local in time but still non-local in space.
As above, spatial locality is not obtained by a naive expansion of
$j_0(\Lambda|\bm x-\bm y|)$, but requires a separate step, namely the
finite-width filter analysis used in the derivation of the local EFT
(see Appendix~\ref{app:local}). Once that reduction is invoked,
the probability flow reduces to the same local operator basis as the
effective action, with the same hierarchy displayed in
\eqref{eq:RGGFP_hierarchy}.
%%%%%%%%%%%%%%%%%%%%%%%%%%%%%%%%%%%%%%%%%%%%%%%%%%%%%%%%%%%%%%%%%%%%%%%%%%%%%%%%%%%%%%%%%%%%%%%%%%%%%%%%%%%%%%%%%%%

%%%%%%%%%%%%%%%%%%%%%%%%%%%%%%%%%%%%%%%%%%%%%%%%%%%%%%%%%%%%%%%%%%%%%%%%%%%%%%%%%%%%%%%%%%%%%%%%%%%%%%%%%%%%%%%%%%%%%%%%%%%%%%%%%%%%%%%%%%
\paragraph{Quantum corrections to the flow equation.}

The exact flow equation \eqref{eq:RGGFP_clean} was obtained without
approximation. To compute corrections to the Gaussian stochastic regime,
however, one must evaluate it in a controlled expansion. The natural
expansion parameter is provided by the same super-Hubble hierarchy that
underlies the local EFT. On long wavelengths, the linear solutions imply
\begin{align}
\frac{\varphi_a^L}{\varphi_r^L}
\sim
\left(\frac{k}{aH}\right)^3
\equiv \epsilon^3
\ll 1,
\end{align}
while the projection onto the diagonal density matrix
already singles out the configuration $\varphi_a^L=0$ as the relevant
Schwinger--Keldysh saddle. The super-Hubble hierarchy does not create
this saddle, but makes the expansion around it parametrically
controlled by suppressing fluctuations with additional powers of
$\varphi_a^L$ \cite{Firat:2025upx}.
Accordingly, the effective action may be decomposed as
\begin{align}
S_{\mathrm{eff}}^\Lambda[\varphi_r^L,\varphi_a^L]
=
S_{\mathrm{MSR}}^\Lambda[\varphi_r^L,\varphi_a^L]
+
S^{(3)}[\varphi_r^L,\varphi_a^L]
+
\mathcal O\!\left((\varphi_a^L)^4\right),
\end{align}
where $S_{\mathrm{MSR}}^\Lambda$ contains the terms linear and quadratic
in $\varphi_a^L$, while $S^{(3)}$ is the leading correction beyond the
Gaussian sector.

For the quartic interaction $\lambda\phi^4$, this first correction is
\begin{align}
S^{(3)}[\varphi_r^L,\varphi_a^L]
=
-\frac{\lambda}{4!}
\int \dd^3\bmx \int \dd t\, a^3(t)\,
\varphi_r^L(\bmx,t)\,[\varphi_a^L(\bmx,t)]^3 .
\end{align}
Relative to the drift vertex $(\varphi_r^L)^3\varphi_a^L$, it is
suppressed by
\begin{align}
\frac{\varphi_r^L(\varphi_a^L)^3}{(\varphi_r^L)^3\varphi_a^L}
\sim
\left(\frac{\varphi_a^L}{\varphi_r^L}\right)^2
\sim
\epsilon^6 .
\end{align}
Thus it is the first genuinely non-Gaussian operator in the
super-Hubble expansion.

To determine the contribution of this vertex to the probability flow, one
inserts $iS^{(3)}$ into the exact RG equation and keeps the term linear
in $S^{(3)}$. At this order, the flow is generated by a single action of
the RG kernel on the cubic interaction, so there is only one differentiated
propagator $\dot G^{\alpha\beta}_\Lambda$. Since the vertex
$S^{(3)}\propto \phi_r^L(\phi_a^L)^3$ contains three $a$-legs, the only
contribution relevant for the projected diagonal density matrix is the one
in which these three response insertions are converted into functional
derivatives with respect to $\phi_r^L$. This is done using the
Schwinger--Keldysh identity
\begin{align}
\varphi_a^L(x)\,e^{iS_{\mathrm{MSR}}^\Lambda}
=
i\int \dd^4y\,
G_R(x,y)\,
\frac{\delta}{\delta\varphi_r^L(y)}
e^{iS_{\mathrm{MSR}}^\Lambda}.
\end{align}
Applying this relation to each of the three $a$-legs converts the cubic
vertex into a structure with three retarded propagators. After this step,
the only component of the RG kernel that contributes to the projected flow
is the Keldysh component, so the resulting non-Gaussian term is governed by
one $\dot G^K_\Lambda$ and three retarded propagators.

Integrating by parts in field space and then projecting onto the final-time
diagonal density matrix gives a third-order Kramers--Moyal contribution,
\begin{align}
\frac{\partial P_\Lambda[\phi^L]}{\partial\log\Lambda}
\supset
\lambda
\int \dd^3\bmx\,
\dd^3\bmx_1\,
\dd^3\bmx_2\,
\dd^3\bmx_3\,
C_\Lambda(\bmx,\bmx_1,\bmx_2,\bmx_3;t_0)\,
\phi^L(\bmx)\,
\frac{\delta^3 P_\Lambda[\phi^L]}
{\delta\phi^L(\bmx_1)\delta\phi^L(\bmx_2)\delta\phi^L(\bmx_3)}.
\end{align}
If the projection onto long modes is written explicitly, the coefficient is
\begin{align}
&C_\Lambda(\bmx,\bmx_1,\bmx_2,\bmx_3;t_0)
\nonumber\\
&\qquad=
\int_{\bm{k},\bm{k}_1,\bm{k}_2,\bm{k}_3}
e^{i\bm{k}\cdot\bmx
+i\sum_{i=1}^{3}\bm{k}_i\cdot\bmx_i}\,
\dot{\bar\Omega}_\Lambda(k,t_0)
\prod_{i=1}^3 \bar\Omega_\Lambda(k_i,t_0)\,
\langle \phi_{\bm{k}}
\phi_{\bm{k}_1}\phi_{\bm{k}_2}\phi_{\bm{k}_3}\rangle_{raaa},
\end{align}
where $\bar\Omega_\Lambda$ denotes the long-wavelength window function.
The shell localisation is instead carried by the differentiated propagator
inside the four-point kernel,
\begin{align}
\langle \phi_{\bm{k}}
\phi_{\bm{k}_1}\phi_{\bm{k}_2}\phi_{\bm{k}_3}\rangle_{raaa}
\equiv
(2\pi)^3
\delta^{(3)}(\bm{k}+\bm{k}_1+\bm{k}_2+\bm{k}_3)
\int^{t_0}\dd t\,a^3(t)\,
 G_{K}(k;t,t_0)
\prod_{i=1}^3 G_R(k_i;t,t_0).
\end{align}
This makes explicit that the kernel is generated by one shell contraction
through $\dot G_{K,\Lambda}$ together with the three retarded propagators
produced by the $(\phi_a^L)^3$ structure of $S^{(3)}$.

A comment is in order. The correlator entering $C_\Lambda$ is only one of
the two tree-level contributions to the full $\lambda\phi^4$ four-point
function, the other coming from the vertex $\phi_r^3\phi_a$. In the
vacuum both structures contribute at the same parametric order, and their
sum is needed to recover the correct analytic structure of the full
four-point function, in particular the absence of folded singularities, as
discussed in \cite{Cespedes:2025ple}. After horizon exit, however, the two
vertices scale differently. The $\phi_r^3\phi_a$ vertex gives the leading
super-Hubble contribution and is the one that carries the secular growth,
whereas the $\phi_r(\phi_a)^3$ vertex is subleading. In the reduced open
EFT this distinction is reflected directly in the probability flow: the
former contributes to the deterministic sector, i.e. to the drift, whereas
the latter generates a genuinely non-Gaussian correction, encoded as a
third-order Kramers--Moyal term.

The RG flow localises the differentiated propagator to the shell
$k\sim\Lambda(t)$, and in the stochastic regime $\Lambda\ll aH$ this
lies entirely in the super-Hubble domain. There the Keldysh propagator is
unsuppressed, while each retarded propagator starts only at order
$\epsilon^3$. Since the microscopic kernel contains three retarded
propagators, it is suppressed as $\epsilon^9$. After matching onto the
local operator basis this becomes the $\epsilon^6$ suppression of the
corresponding local Kramers--Moyal operator quoted above. Thus the first
correction beyond the MSR sector is a third-order derivative term in the
probability flow, and its coefficient is suppressed by the same
super-Hubble hierarchy that organises the local open EFT. \footnote{
The derivation here organises the coarse-graining differently from
Appendix~B of~\cite{Gorbenko:2019rza}. In our case, the leading diffusive
term arises directly from the shell localisation of the non-local
Schwinger--Keldysh kernel, without expanding the projection in the shell
width. By contrast, Gorbenko \emph{et al.} keep a smooth window function
explicit and perform a finite-width expansion in $\delta$, which generates
a Kramers--Moyal tower of higher-derivative terms. The two approaches agree
on the leading local diffusion operator when both are applicable, but they
organise subleading corrections differently.
}

The RG flow localises the differentiated propagator to the shell
$k\sim\Lambda(t)$, and in the stochastic regime $\Lambda\ll aH$ this
lies entirely in the super-Hubble domain. There the Keldysh propagator is
unsuppressed, while each retarded propagator starts only at order
$\epsilon^3$. Since the microscopic kernel contains three retarded
propagators, it is suppressed as $\epsilon^9$. After matching onto the
local operator basis this becomes the $\epsilon^6$ suppression of the
corresponding local Kramers--Moyal operator quoted above. Thus the first
correction beyond the MSR sector is a third-order derivative term in the
probability flow, and its coefficient is suppressed by the same
super-Hubble hierarchy that organises the local open EFT.
%%%%%%%%%%%%%%%%%%%%%%%%%%%%%%%%%%%%%%%%%%%%%%%%%%%%%%%%%%%%%%%%%%%%%%%%%%%%%%%%%%%%%%%%%%%%%%%%%%%%%%%%%%%%%%%%%%%%%%%%%%%%%%%%%%%%%%%%%%%%%%%%%%%%

\paragraph{Recovering the Fokker--Planck equation of stochastic inflation.}

The role of Eq.~\eqref{eq:RGGFP_clean} is to provide the exact
renormalisation-group evolution of the probability functional
$P_\Lambda[\phi^L]$ associated with the diagonal of the reduced density
matrix. By itself, however, this is an equation for the flow with
respect to the coarse-graining scale. To make contact with the standard
stochastic-inflation literature, one must further relate this RG flow to
the physical time evolution of the long-wavelength probability
distribution.

The relevant time dependence arises in two ways. First, the probability
functional evolves explicitly at fixed cut-off. Second, the cut-off
itself is time dependent, since the long-wavelength sector is defined by
a moving coarse-graining scale $\Lambda(t)$ that continuously absorbs
modes as they cross the horizon. The total time derivative is therefore
\begin{align}
\frac{\dd P_\Lambda}{\dd t_0}
=
\left.\frac{\partial P_\Lambda}{\partial t_0}\right|_{\Lambda}
+
\frac{\partial \Lambda}{\partial t_0}
\frac{\partial P_\Lambda}{\partial \Lambda}\Bigg|_{t_0}.
\label{eq:Pevol_clean}
\end{align}

The first term of Eq.~\eqref{eq:Pevol_clean} describes the explicit time evolution at fixed
coarse-graining scale. Since $P_\Lambda[\phi^L]$ is defined as the
diagonal element of the reduced density matrix at the final time $t_0$,
a variation with respect to $t_0$ acts on the boundary fields of the
Schwinger--Keldysh path integral. 
In the local infrared regime, the effective action becomes first order
in time derivatives and linear in the response field, so that
\begin{align}
S_{\rm eff}^{\rm IR}
\supset
\int \dd t\,\dd^3x\,
\phi_a^L
\left[
\dot\phi_r^L
+\frac{1}{3H}V^{\rm eff}_{,\phi}(\phi_r^L)
-\frac{1}{3H}\frac{\partial^2}{a^2}\phi_r^L
+\cdots
\right].
\end{align}
In this regime, in a saddle-point expansion in $\phi_a^L$
truncated at quadratic order, the dynamics reduces to a first-order
deterministic evolution for the long modes. Keeping only terms up to
quadratic order in $\phi_a^L$ (Gaussian truncation), one obtains
\begin{align}
\partial_t \phi^L(\bmx,t)
=
-\frac{1}{3H}V^{\rm eff}_{,\phi}(\phi^L(\bmx,t))
+\frac{1}{3H}\frac{\partial^2}{a^2}\phi^L(\bmx,t)
+\cdots .
\label{eq:det_flow_phi}
\end{align}
This defines the deterministic velocity field
\begin{align}
b[\phi^L](\bmx)
\equiv
\partial_t \phi^L(\bmx,t).
\end{align}
At fixed $\Lambda$, the evolution of the probability functional is
determined by probability conservation under the deterministic flow
$\partial_t\phi^L=b[\phi^L]$, which implies a continuity equation in
configuration space, \footnote{
Equivalently, one may derive \eqref{eq:drift_general} by considering the
time evolution of $\langle \mathcal F[\phi]\rangle$ for arbitrary
functionals $\mathcal F$ and integrating by parts in field space.
}
\begin{align}
\left.\frac{\partial P_\Lambda}{\partial t_0}\right|_{\Lambda}
=
-
\int \dd^3x\,
\frac{\delta}{\delta\phi^L(\bmx)}
\Big[
b[\phi^L](\bmx)\,P_\Lambda[\phi^L]
\Big].
\label{eq:drift_general}
\end{align}
At this level, Eq.~\eqref{eq:drift_general} represents the leading
deterministic contribution to the Kramers--Moyal expansion. Higher
powers of $\phi_a^L$ generate higher-order functional derivatives,
which are neglected in the Gaussian truncation.
Neglecting gradients and subleading derivative terms, this reduces to
\begin{align}
\left.\frac{\partial P_\Lambda}{\partial t_0}\right|_{\Lambda}
=
\int \dd^3x\,
\frac{\delta}{\delta\phi^L(\bmx)}
\left[
\frac{V^{\rm eff}_{,\phi}(\phi^L(\bmx))}{3H}
\,P_\Lambda[\phi^L]
\right].
\label{eq:drift_clean}
\end{align}

The second term in Eq.~\eqref{eq:Pevol_clean} arises from the motion of
the cut-off itself. At leading order, this is determined by the Keldysh
component of the RG flow,
\begin{align}
\frac{\partial \Lambda}{\partial t_0}
\frac{\partial P_\Lambda}{\partial \Lambda}\Bigg|_{t_0}
=
\frac{H^3}{8\pi^2}
\int \dd^3x \dd^3y\;
j_0(\Lambda|\bmx-\bmy|)
\frac{\delta^2 P_\Lambda[\phi^L]}
{\delta\phi^L(\bmx)\,\delta\phi^L(\bmy)}.
\label{eq:diff_clean}
\end{align}

Combining these contributions, one recovers the Fokker--Planck equation
of stochastic inflation as the leading local and Gaussian truncation of
the full evolution,
\begin{align}
\frac{\dd P_\Lambda}{\dd t_0}
&=
\int \dd^3x\,
\frac{\delta}{\delta\phi^L(\bmx)}
\left[
\frac{V^{\rm eff}_{,\phi}(\phi^L(\bmx))}{3H}
\,P_\Lambda[\phi^L]
\right]
\nonumber\\
&\quad
+
\frac{H^3}{8\pi^2}
\int \dd^3x \dd^3y\;
j_0(\Lambda|\bmx-\bmy|)
\frac{\delta^2 P_\Lambda[\phi^L]}
{\delta\phi^L(\bmx)\,\delta\phi^L(\bmy)}.
\label{eq:FP_final_clean}
\end{align}
This shows that the Fokker--Planck equation arises as the leading
infrared truncation of the RG evolution once the dynamics is reduced to a
local, first-order form and truncated at Gaussian order in the response
field. In this regime, the deterministic drift originates from the
saddle-point dynamics at fixed cut-off, while the diffusion term is
generated by the Keldysh component of the RG flow associated with the
continuous crossing of modes through the coarse-graining scale.

%%%%%%%%%%%%%%%%%%%%%%%%%%%%%%%%%%%%%%%%%%%%%%%%%%%%%%%%%%%%%%%%%%%%%%%%%%%%%%%%%%%%%%%%%%%%%%%%%%%%%%%%%%%%%%%%%%%%%%%%%%%%%%%%%%%%%%%%%%%%%%%%%%%%

\section{Conclusion}
\label{sec:conclussions}

In this work we have developed a systematic effective field theory
description of the long-wavelength dynamics of light scalar fields in
de Sitter space. Our starting point was to treat the separation between
long and short modes as a coarse-graining procedure, leading naturally
to a reduced density matrix for the infrared sector. This makes explicit
that the relevant effective description is that of an open quantum
system, rather than a closed, unitary field theory.

Our main result is that stochastic inflation arises as the infrared
regime of the corresponding renormalisation-group flow for the reduced
density matrix. In this regime, the flow becomes dominated by the
diffusive sector and admits an approximate description in terms of a
local stochastic effective theory.
The dynamics of the long-wavelength modes admits a controlled EFT
expansion organised by the coarse-graining parameters. The thin-shell
structure of the cut-off enforces locality in time, spatial gradients
are suppressed in the infrared, and the super-Hubble expansion provides
a hierarchy among operators. This allows the operator content of the
effective theory to be derived explicitly and its scaling to be
determined systematically. \\

We have shown that the effective action contains both dissipative and
diffusive interactions, which arise from the same physical mechanism:
the continuous flow of modes across the coarse-graining scale. Their
relative importance is controlled parametrically, and in the
super-Hubble regime diffusion dominates over dissipation. This provides a precise EFT understanding of why stochastic behaviour emerges at long wavelengths.

Stochastic inflation corresponds to the regime in which the effective
theory becomes local in time and space and is dominated by the leading
diffusive terms. In this limit, the evolution of the reduced density
matrix reduces to a Fokker--Planck equation for the probability
distribution of long-wavelength modes. Hence, the stochastic
description is neither fundamental nor heuristic, but rather the leading term in a systematically improvable expansion.

Beyond this leading limit, the EFT contains well-defined corrections.
These include subleading contributions to the drift and diffusion terms,
as well as higher-order operators associated with non-Gaussian noise.
Their structure follows directly from the Schwinger--Keldysh effective
action and is organised by the same super-Hubble hierarchy, providing a
controlled framework to quantify deviations from the Gaussian
stochastic regime. 

A complementary perspective is provided by the renormalisation-group
approach developed in this work. By studying the flow of the reduced
density matrix under changes of the coarse-graining scale, we have shown
that the same operator structure is generated dynamically. This provides
a Polchinski-type RG equation for an open quantum system formulated on
the Schwinger--Keldysh contour. In this picture, stochastic inflation
emerges as the infrared limit of a more general RG evolution,
establishing a direct connection between stochastic dynamics and
Wilsonian RG ideas.

An important feature of our results is that the Fokker--Planck equation
is obtained from the RG flow of the Schwinger--Keldysh path integral
after taking the local and semiclassical limits. This shows that the
stochastic description can be understood from coarse-graining the Schwinger-Keldysh action, with coefficients controlled by $\beta$-functions determined by the short-distance
dynamics. It also clarifies the regime of validity of the
Starobinsky--Yokoyama equation. In particular, the super-Hubble
hierarchy suppresses higher-order terms in the advanced field, ensuring
that corrections to the stochastic description are parametrically
controlled. 

Taken together, these results provide a unified and controlled EFT
framework for the infrared dynamics in de Sitter space. They clarify the
origin of stochastic behaviour, identify the relevant expansion
parameters, and make explicit the structure of the corrections beyond
the standard stochastic description. Moreover, the infrared effective theory exhibits an emergent dynamical KMS symmetry, implying an effective temperature of order $H$ consistent with the Fokker--Planck description, although the symmetry is presently established only within the leading infrared and Gaussian-noise
approximation.
An important open question is whether this diffusive regime can be
promoted to a genuine infrared fixed point of the RG flow \cite{Berges:2025ccd}. While our
results show that the dynamics is driven towards a regime dominated by
diffusion, establishing a true fixed point requires showing that this
structure is preserved under successive coarse-graining steps.
In practice, this means demonstrating that the effective theory remains
local and of the same form as the cut-off is lowered, with its
coefficients approaching a scale-independent behaviour. It is also
necessary to ensure that corrections, such as non-local effects or
non-Gaussian contributions, do not build up in a way that invalidates
the stochastic description.
Clarifying these points would determine whether the stochastic regime is
simply a good approximation over a range of scales, or the endpoint of
the RG flow in a precise sense.

A lesson of our construction is that the superhorizon effective
theory is intrinsically formulated in terms of a reduced density matrix,
and therefore describes a genuinely mixed state. The Wilsonian
coarse-graining that underlies the EFT amounts to a continuous tracing
over modes as they cross the horizon, so that the long-wavelength sector
is unavoidably an open system. In particular, the limit
$\epsilon \equiv \Lambda/(aH) \to 0$ does not restore a unitary
description within the EFT: instead, it drives the dynamics towards a
regime in which the Keldysh sector dominates and the evolution becomes
effectively diffusive and irreversible. As a consequence, there is no
closed, autonomous pure-state formulation of the infrared theory in
terms of its own degrees of freedom. This has an important conceptual
implication: the particle interpretation of the long-wavelength modes
loses its meaning in this regime. The relevant degrees of freedom are not
propagating quanta associated with well-defined dispersion relations,
but rather collective stochastic variables governed by relaxation and
diffusion. In this sense, stochastic inflation should be understood not
as an approximation to a unitary field theory, but as the natural
infrared fixed point of its open-system renormalisation group flow.

        %%%%%%%%%%%%%%%%%%%%%%%%%%%%%%%%%%%%%%%%%%%%%%%%%%%%%%%%%%%%%%%%%%%%%%%%%%%%%%%%%%%%%%%%%%%%%%%%%%%%%%%%%%%%%%%%%%%%%%%

        \subsection*{Acknowledgements:} We thank Nishant Agarwal, Santiago Ag\"u\'i Salcedo, Senarath de Alwis, Philippe Brax, Clifford P. Burgess, Bei-Lok Hu, Greg Kaplanek, Yoann Lonnay, Yue-Zhou Li, Julien Grain, Daniel Green, Richard Holman, Andrea F. Sanfilippo, Enrico Pajer, Fernando Quevedo, Gerasimos Rigopoulos, Guanhao Sun, Gianmassimo Tasinato, Andrew Tolley, Vincent Vennin and Dong-Gang Wang for  insightful discussions. T.C. acknowledges the Julian Schwinger Foundation for financial support to attend the 2026 Peyresq Spacetime Meeting, where valuable discussions helped shape this work. This work has been supported by STFC consolidated grant ST/X001113/1, ST/T000694/1, ST/X000664/1 and EP/V048422/1. SC is supported in part by the STFC Consolidated Grants ST/T000791/1 and ST/X000575/1 and by a Simons Investigator award 690508.

        \appendix

%%%%%%%%%%%%%%%%%%%%%%%%%%%%%%%%%%%%%%%%%%%%%%%%%%%%%%%%%%%%%%%%%%%%%%%%%%%%%%%%%%%%%%%%%%%%%%%%%%%%%%%%%%%%%%%%%%%%%%%%%%%%%%%%%%%%%%%%%%%%%%%%%%%%%%%%%%%%%%%%%%%%%%%%

\section{Gradient expansion and locality}\label{app:local}

This appendix derives the local influence functional, in particular the local form of the diffusion vertex appearing after integrating out the crossing modes. We show that spatial locality follows from a systematic gradient expansion, valid for long-wavelength fields varying on scales much larger than the Hubble radius. We then estimate the leading corrections.

The time-local diffusion term obtained after integrating out the
crossing modes is
\begin{align}
S_{\mathrm{IF}}^{\rm diff}
&=
i
\int \dd \eta\,a^4(\eta)\Lambda'(\eta)
\int_{\bf k}
\sigma_\Lambda(k,\eta)\,
G_S^K(k,\eta,\eta)\,
\varphi_a^{\prime L}({\bf k},\eta)
\varphi_a^{\prime L}(-{\bf k},\eta).
\label{eq:diffusion_time_local}
\end{align}
Here $\sigma_\Lambda(k,\eta)$ is localized around the stochastic
coarse-graining scale
\begin{equation}
\Lambda(\eta)=\epsilon aH,
\end{equation}
with $\epsilon\ll 1$. This scale determines which modes are transferred
from the short sector to the long sector.

The emergence of a local description is a consequence of the
separate-universe approximation. For sufficiently long wavelengths,
\begin{equation}
k_L \ll aH,
\end{equation}
spatial gradients become subleading compared to the local time
evolution. In this regime the equations of motion admit a systematic
in powers of $\frac{k_L^2}{(aH)^2}$,
To leading order in this expansion, the evolution at a given spatial point depends only on the local values of the fields and not on the configuration of neighbouring regions. Each region therefore evolves approximately as an independent FRW universe, with interactions between different regions appearing only through the gradient corrections.

This observation is independent of the stochastic coarse-graining
procedure. The scale $\Lambda=\epsilon aH$ determines which modes are transferred from the short sector to the
long sector, but it does not by itself imply locality. Locality
follows from the fact that the long fields vary only on scales much larger than the Hubble radius, so that a bilocal interaction generated by integrating out short modes cannot resolve their spatial variation.

In practice this implies that  the bilocal composite operator generated by the RG step may be represented by a local derivative expansion,
\begin{equation}
{\cal B}(X,r;\eta)
=
{\cal B}(X,0;\eta)
+
{\cal O}
\!\left(
\frac{\nabla^2}{(aH)^2}
\right),
\end{equation}
where the omitted terms are suppressed by the same gradient expansion
that underlies the separate-universe approximation. The resulting local
operators can then be expressed in terms of the coarse-grained
variables used to define the effective theory.
In the case we are interested the diffusion term is generated by the composite operator
\begin{equation}
{\cal B}_{\rm diff}
=
\varphi_a'G_S^K\varphi_a' .\label{eq:BilocalDiff}
\end{equation}
where the Keldysh propagator is  part of the operator generated by integrating out the short modes. The role of the profile
$\sigma_\Lambda(k,\eta)$ is different: it specifies how this operator
is weighted by the RG step. The locality analysis should therefore be
performed on the smearing induced by $\sigma_\Lambda$, while keeping
the composite operator ${\cal B}_{\rm diff}$ fixed.

The separate-universe approximation implies that the interaction
generated by \eqref{eq:BilocalDiff} admits a local representation in
terms of the long-wavelength fields. At leading order in the gradient
expansion, the bilocal operator generated by the RG step cannot resolve
the spatial variation of the long modes and therefore reduces to a
local operator, up to corrections suppressed by powers of
$k_L/(aH)$.

Once locality has been established, a second and independent step is
required. The local separate-universe description is formulated in
terms of coarse-grained patch operators defined at a resolution scale
$\mu$. In particular, we represent the local diffusion operator using
the smeared operator
\begin{equation}
[{\cal B}_{\rm diff}]_\Omega(X,\eta)
=
\int_{\bf k}
R(k/\mu)\,
{\cal B}_{\rm diff}({\bf k},\eta)
e^{i{\bf k}\cdot X},
\end{equation}
where $R(k/\mu)$ specifies the coarse-graining profile.

The role of $R$ should not be confused with the emergence of locality.
Locality follows from the separate-universe expansion, whereas
$R(k/\mu)$ specifies how local operators are represented in the
coarse-grained patch description. The operator generated by the RG step
is weighted by the shell profile $\sigma_\Lambda(k,\eta)$, while the
local separate-universe description is written in terms of operators
weighted by $R(k/\mu)$. The purpose of the matching procedure is
therefore to express the local operator generated by the RG step in the
basis of coarse-grained patch operators.

To do so, we decompose the shell profile as
\begin{equation}
\sigma_\Lambda(k,\eta)
=
D_{\rm sp}(\eta)\,R(k/\mu)
+
\delta\sigma_\Lambda(k,\eta),
\end{equation}
where the first term denotes the component represented in the local
patch description and the remainder contains all contributions not
captured by the chosen profile $R$. The coefficient $D_{\rm sp}$ is
therefore the projection coefficient relating the RG-generated profile
to the profile used to define the local patch operator.

To isolate this component we require that the remainder has vanishing
overlap with the profile $R$,
\begin{equation}
\int_{\bf k}
\delta\sigma_\Lambda(k,\eta)\,
R(k/\mu)
=
0.
\end{equation}
This condition is merely a matching prescription. It does not assume
that $R$ belongs to a complete orthonormal basis. It simply fixes the
coefficient of the operator represented in the local patch
description.

Multiplying the decomposition by $R(k/\mu)$ and integrating over
momentum gives
\begin{equation}
\int_{\bf k}
\sigma_\Lambda(k,\eta)
R(k/\mu)
=
D_{\rm sp}(\eta)
\int_{\bf k}
R^2(k/\mu),
\end{equation}
from which it follows that
\begin{equation}
D_{\rm sp}(\eta)
=
\frac{
\int_{\bf k}
\sigma_\Lambda(k,\eta)R(k/\mu)
}{
\int_{\bf k}
R^2(k/\mu)
}.
\label{eq:Dsp_correct_matching}
\end{equation}

With this definition, the local operator generated by the RG step may
be written in terms of the coarse-grained patch operator as
\begin{equation}
\int_{\bf k}
\sigma_\Lambda(k,\eta)
{\cal B}_{\rm diff}({\bf k},\eta)
e^{i{\bf k}\cdot X}
=
D_{\rm sp}(\eta)
[{\cal B}_{\rm diff}]_\Omega(X,\eta)
+\cdots ,
\label{eq:Bdiff_matching}
\end{equation}
where the ellipsis denotes gradient corrections suppressed by
$k_L/(aH)$ together with contributions associated with the component of
the shell profile that is not retained in the local patch
description.

To make the scaling explicit it is convenient to consider first the
idealized limit in which the shell profile is sharply localized at the
crossing scale,
\begin{equation}
\sigma_\Lambda(k,\eta)=\delta(k-\Lambda).
\end{equation}
In this case the projection coefficient \eqref{eq:Dsp_correct_matching}
can be evaluated straightforwardly. The numerator samples the profile
$R(k/\mu)$ at the crossing momentum $k=\Lambda$, while the denominator
provides the normalization associated with the chosen coarse-graining
prescription. One finds
\begin{equation}
D_{\rm sp}(\eta)
=
\frac{\Lambda^2}{\mu^3}
\frac{
R(\Lambda/\mu)
}{
\int_0^\infty dq\,q^2R^2(q)
}.
\label{eq:Dsp_thin_correct}
\end{equation}

The important feature of this result is not the precise normalization,
which depends on the choice of profile $R$, but its scaling with the
physical coarse-graining scales. Since the local operator is defined at
a resolution $\mu\sim aH$ and the shell crossing occurs at
$\Lambda=\epsilon aH$, one obtains
\begin{equation}
D_{\rm sp}
\sim
\frac{\epsilon^2}{aH}.
\end{equation}
Thus the projection coefficient carries one inverse power of the Hubble
scale and two powers of the stochastic parameter $\epsilon$.

Using the matching relation
\eqref{eq:Bdiff_matching}, the diffusion term can now be written as
\begin{equation}
S_{\mathrm{IF}}^{\rm diff}
=
i
\int \dd \eta\,\dd^3X\,
a^4(\eta)\Lambda'(\eta)
D_{\rm sp}(\eta)
[{\cal B}_{\rm diff}]_\Omega(X,\eta)
+\cdots .
\label{eq:diff_patch_B}
\end{equation}
The remaining task is to determine the coefficient multiplying the local
operator.

For a sufficiently narrow shell the Keldysh propagator varies slowly
across the support of $\sigma_\Lambda$ and may therefore be evaluated at
the crossing scale. The composite operator then reduces to
\begin{equation}
[{\cal B}_{\rm diff}]_\Omega
=
G_S^K(\Lambda,\eta,\eta)
[(\varphi_a')^2]_\Omega
+\cdots .
\label{eq:Bdiff_narrow}
\end{equation}
For a massless scalar the equal-time Keldysh propagator behaves as
$G_S^K\propto H^2/\Lambda^3$, so that the diffusion coefficient
contains three distinct ingredients:

\begin{enumerate}
\item the shell velocity $\Lambda'\propto aH\Lambda$,

\item the projection factor
$D_{\rm sp}\propto \Lambda^2/(aH)^3$,

\item the fluctuation amplitude
$G_S^K(\Lambda)\propto H^2/\Lambda^3$.
\end{enumerate}

Taken separately each factor carries a non-trivial dependence on the
crossing scale. However, once they are multiplied together all powers of
$\Lambda$ cancel. The result is
\begin{equation}
\Lambda'
D_{\rm sp}
G_S^K(\Lambda,\eta,\eta)
=
\frac{1+\epsilon^2}{2}
\frac{1}{a^2}
\frac{
R(\Lambda/\mu)
}{
\int_0^\infty dq\,q^2R^2(q)
}.
\label{eq:diff_scaling_final}
\end{equation}

This cancellation is the central result of the matching procedure. The
shell velocity contributes one power of $\Lambda$, the projection onto
the local operator basis contributes $\Lambda^2$, and the equal-time
fluctuation amplitude contributes $\Lambda^{-3}$. The net result is
independent of the crossing scale and leaves only the expected factor
$a^{-2}$ required by dimensional analysis.

Since $\Lambda/\mu\simeq\epsilon$, the residual dependence on
$\epsilon$ appears only through the normalization convention chosen for
the coarse-grained operator. In particular, there is no singular
behaviour as $\epsilon\rightarrow0$. The dependence on the stochastic
crossing scale cancels completely, leaving a finite local diffusion
operator whose normalization is determined only by the choice of
coarse-graining scheme.

Substituting \eqref{eq:diff_scaling_final} into
\eqref{eq:diff_patch_B} gives
\begin{equation}
S_{\mathrm{IF}}^{\rm diff,loc}
=
\frac{i}{2}
{\cal N}_R(\epsilon)
\left[
1+{\cal O}(\epsilon^2)
\right]
\int \dd \eta\,\dd^3X\,
a^2(\eta)
[(\varphi_a')^2]_\Omega(X,\eta)
+\cdots ,
\label{eq:diff_local_conformal}
\end{equation}
with
\begin{equation}
{\cal N}_R(\epsilon)
=
\frac{
R(\epsilon)
}{
\int_0^\infty dq\,q^2R^2(q)
}.
\label{eq:NR_definition}
\end{equation}
The factor ${\cal N}_R$ is not a physical enhancement. It records the
normalization convention used to define the smeared local operator
$[{\cal B}_{\rm diff}]_\Omega$. If the coarse-grained operator is
defined with a normalized profile, one may choose this convention so
that ${\cal N}_R(\epsilon)=1$ at leading order. Otherwise the same factor
must be kept explicitly in the Wilson coefficient. In either convention,
the important point is that the combination generated by the RG step is
\begin{equation}
\Lambda' D_{\rm sp} G_S^K(\Lambda,\eta,\eta)
\propto
\frac{1}{a^2},
\end{equation}
so that the local diffusion term has precisely the conformal-time
measure $a^2 \dd \eta$ appropriate for $(\varphi_a')^2$.

Passing to proper time, using
\begin{equation}
\dd \eta=\frac{dt}{a},
\qquad
\varphi_a'=a\dot\varphi_a,
\end{equation}
one finds
\begin{equation}
\int \dd \eta\,a^2(\varphi_a')^2
=
\int dt\,a^3(\dot\varphi_a)^2 .
\end{equation}
Therefore the local form of the diffusion term is
\begin{equation}
S_{\mathrm{IF}}^{\rm diff,loc}
=
\frac{i}{2}
{\cal N}_R(\epsilon)
\left[
1+{\cal O}(\epsilon^2)
\right]
\int dt\,\dd^3X\,
a^3(t)
[(\dot\varphi_a)^2]_\Omega(X,t)
+\cdots .
\label{eq:diff_local_proper}
\end{equation}
This is the expected separate-universe result. The RG shell fixes the
coefficient of the local noise operator, while the factor $a^3$ is fixed
by locality in physical volume. The dependence on the stochastic
crossing scale $\Lambda=\epsilon aH$ cancels between the shell velocity,
the projection factor, and the equal-time Keldysh propagator, leaving
only the scheme-dependent normalization of the chosen patch operator and corrections suppressed by $\epsilon^2$ and by spatial gradients.

%%%%%%%%%%%%%%%%%%%%%%%%%%%%%%%%%%%%%%%%%%%%%%%%%%%%%%%%%%%%%%%%%%%%%%%%%%%%%%%%%%%%%%%%%%%%%%%%%%%%%%%%%%%%%%%%%%%%%%%%%%%%%%%%%%%%%%%%%%%%%%%%%%%%%%%%%%%%%%%%%%%%%%%%%%%%%%%%%%%%%%%%%%%%%%%%%%%%%%%%%%%%%%%%%%%%%%%%%%%%%%%%%%%%%%%%%%%%%%%%%%%

\section{Diffusive operators at order \texorpdfstring{$\lambda$}{lambda}}
\label{app:ope}

In this appendix we determine the diffusive operators generated at order $\lambda$ by combining the quartic interaction vertices
\begin{align}
S^{(1)}_{\mathrm{int}}
&=
- \frac{\lambda}{3!} \int \dd^4 x\sqrt{-g}\,
\bigg[
(\varphi^L_r)^3 \varphi^S_a
+ 3 (\varphi^L_r)^2 \varphi^L_a \varphi^S_r
+ \frac{3}{4} \varphi^L_r  (\varphi^L_a)^2 \varphi^S_a
+ \frac{1}{4} (\varphi^L_a)^3 \varphi^S_r
\bigg],
\\
S^{(2)}_{\mathrm{int}}
&=
- \frac{\lambda}{3!} \int \dd^4 x\sqrt{-g}\,
\bigg[
3 (\varphi^L_r)^2 \varphi^S_r \varphi_a^S
+ 3 \varphi^L_r \varphi^L_a (\varphi^S_r)^2
+ \frac{3}{4} (\varphi^L_a)^2 \varphi^S_a \varphi^S_r
+ \frac{3}{4} \varphi^L_a \varphi^L_r (\varphi^S_a)^2
\bigg],
\\
S^{(3)}_{\mathrm{int}}
&=
- \frac{\lambda}{3!} \int \dd^4 x\sqrt{-g}\,
\bigg[
\varphi^L_a (\varphi^S_r)^3
+ 3 \varphi^L_r (\varphi^S_r)^2 \varphi^S_a
+ \frac{3}{4} \varphi^L_a \varphi^S_r  (\varphi^S_a)^2
+ \frac{1}{4} \varphi^L_r (\varphi^S_a)^3
\bigg],
\\
S^{(4)}_{\mathrm{int}}
&=
- \frac{\lambda}{3!} \int \dd^4 x\sqrt{-g}\,
\bigg[
(\varphi^S_r)^3 \varphi^S_a
+ \frac{1}{4} \varphi^S_r (\varphi^S_a)^3
\bigg],
\end{align}
together with the long--short mixing term
\begin{align}
S_{\mathrm{cross}}
=
\int_{\bm{k}}\int \dd \eta \, a^2(\eta)\Omega_\Lambda^\prime(k)\bigg[
&-\varphi_r^{\prime L}(\bmk, \eta)\varphi_a^S(-\bmk, \eta)
-\varphi_a^{\prime L}(\bmk, \eta)\varphi_r^S(-\bmk, \eta)
\nonumber\\
&+\varphi_a^L(\bmk, \eta)\varphi_r^{\prime S}(-\bmk, \eta)
+\varphi_r^L(\bmk, \eta)\varphi_a^{\prime S}(-\bmk, \eta)
\bigg].
\end{align}

\paragraph{Hierarchy of contributions.}

Two logically distinct ingredients control the structure of the resulting EFT.

First, there is a hierarchy associated with the short propagators. Contractions involving short advanced fields $\varphi_a^S$ proceed through retarded or advanced kernels and are suppressed on super-Hubble scales. At leading order, only vertices containing short response fields $\varphi_r^S$ contribute.

Second, once a short response field is converted through $S_{\mathrm{cross}}$, the resulting local operator involves either $\varphi_a^{\prime L}$ or $\varphi_a^L$. The relative size of these two structures is controlled by the super-Hubble parameter
\begin{align}
\epsilon_k \equiv -k\eta = \frac{k}{aH}.
\end{align}
Since the shell is centred at $k=\Lambda$, one has $\epsilon_k=\epsilon+\mathcal O(\Delta\Lambda/\Lambda)$ with $\epsilon\ll1$. As in the quadratic diffusion operator, the local term proportional to $\varphi_a^{\prime L}$ is leading, while the accompanying term proportional to $\varphi_a^L$ is suppressed by $\epsilon^2$. This hierarchy is inherited by the higher operators obtained through repeated application of the same conversion rule.

\paragraph{Conversion of a short response leg.}

Consider a term containing a single short response field,
\begin{align}
\int \dd \eta\,a^4(\eta)\int \dd^3 x\,
\mathcal{O}(\varphi_r^L,\varphi_a^L)\,\varphi_r^S.
\end{align}
Contracting $\varphi_r^S$ with $S_{\mathrm{cross}}$ and using the equal-time shell localisation and local spatial limit derived in \App{app:local}, one obtains the local replacement
\begin{align}
\varphi_r^S
\;\longrightarrow\;
\frac{i}{aH} 
\Big[
(1+\epsilon^2)\,\varphi_a^{\prime L}
+\epsilon^2 aH\,\varphi_a^L
\Big],
\label{eq:conversion_precise}
\end{align}
up to corrections suppressed by gradients or by the finite width of the shell.
This expression makes the $\epsilon$-hierarchy manifest: each conversion produces both leading and subleading local structures.

\paragraph{Operators generated from \texorpdfstring{$S_{\mathrm{int}}^{(1)}$}{S1}.}

We now compute explicitly all diffusive operators generated from $S_{\mathrm{int}}^{(1)}$. The relevant terms are those containing one $\varphi_r^S$,
\begin{align}
S_{\mathrm{int}}^{(1)}
\supset
-\frac{\lambda}{2}\int \dd^4x\,\sqrt{-g}\,
(\varphi_r^L)^2\varphi_a^L\varphi_r^S
-\frac{\lambda}{24}\int \dd^4x\,\sqrt{-g}\,
(\varphi_a^L)^3\varphi_r^S .
\end{align}
Applying the conversion rule \eqref{eq:conversion_precise}, one obtains
\begin{align}
S_{\mathrm{IF}}^{(1)}
\supset\;
&- \frac{i\lambda }{2H} 
\int \dd \eta\,a^3(\eta)\int \dd^3 x\,
(1+\epsilon^2)\,
(\varphi_r^L)^2\varphi_a^L\varphi_a^{\prime L}
\nonumber\\
&-\frac{i\lambda}{2} 
\int \dd \eta\,a^2(\eta)\int \dd^3 x\,
\epsilon^2\,
(\varphi_r^L)^2(\varphi_a^L)^2
\nonumber\\
&-\frac{i\lambda}{24 H} 
\int \dd \eta\,a^3(\eta)\int \dd^3 x\,
(1+\epsilon^2)\,
(\varphi_a^L)^3\varphi_a^{\prime L}
\nonumber\\
&-\frac{i\lambda}{24} 
\int \dd \eta\,a^4(\eta)\int \dd^3 x\,
\epsilon^2\,
(\varphi_a^L)^4 .
\label{eq:S1full}
\end{align}
The hierarchy is therefore explicit within a single vertex: operators containing $\varphi_a^{\prime L}$ are leading, while those with additional undifferentiated $\varphi_a^L$ fields are suppressed by powers of $\epsilon^2$.

\paragraph{Operators from higher vertices.}

The same conversion rule may be applied repeatedly to the terms in $S_{\mathrm{int}}^{(2)}$ and $S_{\mathrm{int}}^{(3)}$ that contain only short response fields.

For $S_{\mathrm{int}}^{(2)}$, the relevant term is
\begin{align}
S_{\mathrm{int}}^{(2)}
\supset
-\frac{\lambda}{2}\int \dd^4x\,\sqrt{-g}\,
\varphi_r^L\varphi_a^L(\varphi_r^S)^2 .
\end{align}
Applying the conversion rule twice generates the family
\begin{align}
\varphi_r^L\varphi_a^L(\varphi_a^{\prime L})^2,\qquad
\varphi_r^L(\varphi_a^L)^2\varphi_a^{\prime L},\qquad
\varphi_r^L(\varphi_a^L)^3,
\end{align}
with successive suppression by $\epsilon^2$.

For $S_{\mathrm{int}}^{(3)}$, the relevant term is
\begin{align}
S_{\mathrm{int}}^{(3)}
\supset
-\frac{\lambda}{6}\int \dd^4x\,\sqrt{-g}\,
\varphi_a^L(\varphi_r^S)^3 .
\end{align}
Applying the conversion rule three times generates
\begin{align}
\varphi_a^L(\varphi_a^{\prime L})^3,\qquad
(\varphi_a^L)^2(\varphi_a^{\prime L})^2,\qquad
(\varphi_a^L)^3\varphi_a^{\prime L},\qquad
(\varphi_a^L)^4,
\end{align}
again organised by increasing suppression in $\epsilon^2$.

The local diffusive operators generated at order $\lambda$ are shown in Table~\ref{tab:ope}.
\begin{table}[t]
\centering
\vspace{2mm}
\begin{tabular}{c|c}
\toprule[1.5pt]
Source & Operator structures \\ \hline
$S^{(1)}_{\mathrm{int}}$
& $(\varphi_r^L)^2 \varphi_a^L \varphi_a^{\prime L}\;(\epsilon^0)\,;\quad
   (\varphi_r^L)^2 (\varphi_a^L)^2\;(\epsilon^2)\,;\quad
   (\varphi_a^L)^3 \varphi_a^{\prime L}\;(\epsilon^0)\,;\quad
   (\varphi_a^L)^4\;(\epsilon^2)$
\\ \hline
$S^{(2)}_{\mathrm{int}}$
& $\varphi_r^L\varphi_a^L(\varphi_a^{\prime L})^2\;(\epsilon^0)\,;\quad
   \varphi_r^L(\varphi_a^L)^2\varphi_a^{\prime L}\;(\epsilon^2)\,;\quad
   \varphi_r^L(\varphi_a^L)^3\;(\epsilon^4)$
\\ \hline
$S^{(3)}_{\mathrm{int}}$
& $\varphi_a^L(\varphi_a^{\prime L})^3\;(\epsilon^0)\,;\quad
   (\varphi_a^L)^2(\varphi_a^{\prime L})^2\;(\epsilon^2)\,;\quad
   (\varphi_a^L)^3\varphi_a^{\prime L}\;(\epsilon^4)\,;\quad
   (\varphi_a^L)^4\;(\epsilon^6)$
\\
\bottomrule[1.5pt]
\end{tabular}
\caption{
Local diffusive operators generated by converting short response legs. The indicated powers of $\epsilon$ refer to the suppression relative to the operator with the largest number of time derivatives in each row, obtained by repeated use of the single-leg conversion rule \eqref{eq:conversion_precise}. Terms involving short advanced legs $\varphi_a^S$ are omitted, since they are suppressed at the level of the short propagators.
}
\label{tab:ope}
\end{table}
The structure of the table shows that the super-Hubble expansion acts within each vertex: each interaction generates a family of local operators, organised by the number of time derivatives on the long response fields. The dominant contributions are those involving $\varphi_a^{\prime L}$, while operators with additional undifferentiated $\varphi_a^L$ fields are systematically suppressed by powers of $\epsilon^2$.

Finally, higher orders in $\lambda$ do not introduce new operator structures. They reproduce the same basis and renormalise the corresponding coefficients.

%%%%%%%%%%%%%%%%%%%%%%%%%%%%%%%%%%%%%%%%%%%%%%%%%%%%%%%%%%%%%%%%%%%%%%%%%%%%%%%%%%%%%%%%%%%%%%%%%%%%%%%%%%%%%%%%%%%%%%%%%%%%%%%%%%%%%%%%%%%%%%%%%%%%%%%%%%%%%%%%%%%%%%%%%%

\section{Non-local Langevin equation and its local limit}
\label{sec:nonlocal_langevin}

In this appendix we derive the Langevin equation associated with the non-local influence functional obtained after integrating out the shell modes. We then show how the local stochastic description used in the main text emerges as the leading term in a systematic expansion around the separate-universe limit.
We start from the spatially non-local, but time-local, Keldysh vertex
generated by the shell modes. The leading diffusive term is written as
\begin{align}
S_{\rm IF}^{K}
=
\frac{i}{2}
\int \dd\eta\,\dd^3x\,\dd^3y\,
\varphi_a'({\bf x},\eta)\,
N_\eta({\bf x},{\bf y})\,
\varphi_a'({\bf y},\eta).
\label{eq:nonlocal_K_vertex}
\end{align}
For the shell contribution,
\begin{align}
N_\eta({\bf x},{\bf y})
=
\frac{H^3}{4\pi^2}\,
a^5(\eta)\,
j_0\!\left(
\epsilon aH|{\bf x}-{\bf y}|
\right).
\label{eq:N_eta_kernel}
\end{align}
The kernel is local in time but non-local in space. Thus the locality used here is Markovian locality only; no spatial delta-function limit has been taken.

\paragraph{From the influence functional to a stochastic source.}
The contribution of \eqref{eq:nonlocal_K_vertex} to the path-integral
weight is
\begin{align}
\exp(iS_{\rm IF}^{K})
=
\exp\left[
-\frac12
\int \dd \eta\,\dd^3x\,\dd^3y\,
\varphi_a'({\bf x},\eta)\,
N_\eta({\bf x},{\bf y})\,
\varphi_a'({\bf y},\eta)
\right].
\end{align}
The inverse kernel is defined by
\begin{align}
\int \dd^3z\,
N_\eta({\bf x},{\bf z})
N_\eta^{-1}({\bf z},{\bf y})
=
\delta^{(3)}({\bf x}-{\bf y}) .
\label{eq:N_inverse_def}
\end{align}
The Gaussian may then be linearised as
\begin{align}
&\exp\left[
-\frac12
\int \dd \eta\,\dd^3x\,\dd^3y\,
\varphi_a'({\bf x},\eta)\,
N_\eta({\bf x},{\bf y})\,
\varphi_a'({\bf y},\eta)
\right]
\nonumber\\
&\qquad =
{\cal N}
\int{\cal D}\xi\,
\exp\left[
-\frac12
\int \dd \eta\,\dd^3x\,\dd^3y\,
\xi({\bf x},\eta)\,
N_\eta^{-1}({\bf x},{\bf y})\,
\xi({\bf y},\eta)
\right.
\nonumber\\
&\hspace{4.5cm}
\left.
-i
\int \dd \eta\,\dd^3x\,
\xi({\bf x},\eta)
\varphi_a'({\bf x},\eta)
\right].
\label{eq:HS_nonlocal_kernel}
\end{align}
This is a non-local Hubbard--Stratonovich representation. The linear
coupling to $\varphi_a'$ is local, but the Gaussian measure for
$\xi$ is non-local. This is where all the spatial non-locality is kept.\footnote{The inverse kernel in Eq.~\eqref{eq:HS_nonlocal_kernel} should be
understood as the inverse operator on the space of spatial functions.
It is not the pointwise reciprocal of $N_\eta({\bf x},{\bf y})$.
Instead it is defined by
\begin{align}
\int \dd^3z\,
N_\eta({\bf x},{\bf z})\,
N_\eta^{-1}({\bf z},{\bf y})
=
\delta^{(3)}({\bf x}-{\bf y}) .
\end{align}
Thus $N_\eta^{-1}$ is itself a distributional kernel. In particular,
the identity kernel needed in the Hubbard--Stratonovich transformation
is contained in this inverse relation.

This also fixes the dimensional counting. Since
\begin{align}
[\dd^3z]=-3,
\qquad
[\delta^{(3)}({\bf x}-{\bf y})]=3,
\end{align}
Eq.~\eqref{eq:N_inverse_def} implies $[N_\eta]+[N_\eta^{-1}]-3=3 .$
For the shell kernel, $[N_\eta]=3,$ and therefore $[N_\eta^{-1}]=3 $. Consequently, if the auxiliary field has $[\xi]=2$, then both terms in the Hubbard--Stratonovich representation are
dimensionless:
\begin{align}
\left[
\dd \eta\,\dd^3x\,\dd^3y\,
\xi({\bf x})
N_\eta^{-1}({\bf x},{\bf y})
\xi({\bf y})
\right]
&=
-7+2+3+2
=0,
\\
\left[
\dd \eta\,\dd^3x\,
\xi({\bf x})
\varphi_a'({\bf x})
\right]
&=
-4+2+2
=0.
\end{align}
}

The auxiliary field has covariance
\begin{align}
\left\langle
\xi({\bf x},\eta)
\xi({\bf y},\eta')
\right\rangle
=
N_\eta({\bf x},{\bf y})\,
\delta(\eta-\eta') .
\label{eq:xi_eta_corr}
\end{align}
This follows directly from \eqref{eq:HS_nonlocal_kernel}. Notice that
\eqref{eq:xi_eta_corr} is not a spatial white-noise correlator.

\paragraph{Overdamped limit and first order formalism.}
For fixed $\xi$, combining the linearised Keldysh term with the local
infrared action gives
\begin{align}
S_\xi
=
\int \dd \eta\,\dd^3x
\left[
a^2\varphi_r'\varphi_a'
-a^2\partial_i\varphi_r\partial_i\varphi_a
-a^4V_{,\varphi}(\varphi_r)\varphi_a
-\xi\varphi_a'
\right].
\label{eq:S_xi_second_order}
\end{align}
This action is local for each realization of $\xi$.
We now introduce a phase-space field
\begin{align}
u({\bf x},\eta)
=
\varphi_r'({\bf x},\eta),
\end{align}
with a Lagrange multiplier $\chi_a$ imposing this relation. The action
becomes
\begin{align}
S_\xi
=
\int \dd \eta\,\dd^3x
\left[
a^2u\varphi_a'
-a^2\partial_i\varphi_r\partial_i\varphi_a
-a^4V_{,\varphi}(\varphi_r)\varphi_a
-\xi\varphi_a'
+\chi_a(u-\varphi_r')
\right].
\label{eq:S_xi_phase_raw}
\end{align}
Integrating by parts in the terms containing $\varphi_a'$ gives
\begin{align}
a^2u\varphi_a'
-\xi\varphi_a'
=
-\varphi_a
\left[
(a^2u)'-\xi'
\right],
\end{align}
and the spatial-gradient term gives
\begin{align}
-a^2\partial_i\varphi_r\partial_i\varphi_a
=
a^2(\partial^2\varphi_r)\varphi_a .
\end{align}
Therefore
\begin{align}
S_\xi
=
\int \dd \eta\,\dd^3x
\left\{
-\varphi_a
\left[
(a^2u)'
-\xi'
-a^2\partial^2\varphi_r
+a^4V_{,\varphi}(\varphi_r)
\right]
+
\chi_a(u-\varphi_r')
\right\}.
\label{eq:S_xi_phase_ibp}
\end{align}
The derivative of the stochastic field can be absorbed into the velocity
variable by defining
\begin{align}
X
=
u-\frac{\xi}{a^2}.
\label{eq:X_def}
\end{align}
Then
\begin{align}
(a^2u)'-\xi'
=
(a^2X)' ,
\end{align}
and the phase-space action becomes
\begin{align}
S_\xi
=
\int \dd \eta\,\dd^3x
\left\{
-\varphi_a
\left[
(a^2X)'
-a^2\partial^2\varphi_r
+a^4V_{,\varphi}(\varphi_r)
\right]
+
\chi_a
\left[
X-\varphi_r'
+\frac{\xi}{a^2}
\right]
\right\}.
\label{eq:S_xi_shifted}
\end{align}
Equation \eqref{eq:S_xi_shifted} is the useful form of the path integral. The measure for $\xi$ remains spatially non-local through \eqref{eq:HS_nonlocal_kernel}, but the action at fixed $\xi$ is local. The locality of the action should be distinguished from the locality of the stochastic variable. For a fixed realization of $\xi$, the dynamics is governed by a local action. However, the probability distribution of $\xi$ retains the spatial non-locality inherited from the coarse-graining procedure. The resulting Langevin description is therefore local in time and in its deterministic evolution, while its noise correlations can remain spatially non-local.

The field $\varphi_a$ imposes the following equation for the fast variable
\begin{align}
(a^2X)'
=
a^2\partial^2\varphi_r
-
a^4V_{,\varphi}(\varphi_r).
\label{eq:X_fast_conformal}
\end{align}
In cosmic time, we have $X=aX_t$, such that \eqref{eq:X_fast_conformal} becomes
\begin{align}
(\partial_t+3H)X_t
=
\frac{\partial^2}{a^2}\varphi_r
-
V_{,\varphi}(\varphi_r).
\label{eq:X_fast_cosmic}
\end{align}
The variable $X_t$ describes the fast velocity sector of the theory. Indeed, in the absence of the source term, its evolution is governed by
\begin{align}
(\partial_t+3H)X_t=0 ,
\end{align}
with solution
\begin{align}
X_t(t)=X_t(t_0)e^{-3H(t-t_0)} .
\end{align}
Therefore, the homogeneous component of $X_t$ decays exponentially, and the velocity sector relaxes on a time scale
\begin{align}
\tau_{\rm rel}\sim H^{-1}.
\end{align}
The overdamped approximation amounts to integrating out this rapidly relaxing mode and keeping only its quasi-static response to the slow infrared dynamics of $\varphi_r$.

The exact retarded solution for $X_t$ contains a memory kernel with width set by the relaxation time $\tau_{\rm rel}\sim H^{-1}$. When the infrared field evolves on a much longer time scale, $\omega\ll H$, the memory integral can be expanded locally in time.  Equivalently, one may expand the inverse differential operator as
\begin{align}
(\partial_t+3H)^{-1}
=
\frac{1}{3H}
\left[
1-\frac{\partial_t}{3H}+\cdots
\right],
\end{align}
so that
\begin{align}
X_t
=
\frac{1}{3H}
\left[
\frac{\partial^2}{a^2}\varphi_r
-
V_{,\varphi}(\varphi_r)
\right]
+\cdots .
\label{eq:X_overdamped}
\end{align}
The remaining constraint in \eqref{eq:S_xi_shifted} gives
\begin{align}
\varphi_r'
=
X+\frac{\xi}{a^2}.
\end{align}
Thus, at leading order in the overdamped expansion, the fast variable $X_t$ is no longer an independent degree of freedom but is fixed algebraically by the infrared configuration $\varphi_r$.
Dividing by $a$ and using $X=aX_t$, this becomes
\begin{align}
\dot\varphi_r
=
X_t+\frac{\xi}{a^3}.
\end{align}
Using \eqref{eq:X_overdamped}, the first-order Langevin equation is
\begin{align}
\dot\varphi_r({\bf x},t)
=
\frac{1}{3H}
\frac{\partial^2}{a^2}\varphi_r({\bf x},t)
-
\frac{1}{3H}
V_{,\varphi}(\varphi_r({\bf x},t))
+
f({\bf x},t)
+\cdots ,
\label{eq:first_order_langevin_nonlocal}
\end{align}
where
\begin{align}
f({\bf x},t)
=
\frac{\xi({\bf x},\eta)}{a^3(t)} .
\label{eq:f_xi_relation}
\end{align}
Equivalently, substituting the overdamped solution back into the phase-space action gives the first-order Martin--Siggia--Rose form of the infrared theory,
\begin{align}
S_{\rm red}
=
\int \dd t\,\dd^3x\,a^3(t)
\chi_a
\left[
\dot\varphi_r
-
\frac{1}{3H}
\left(
\frac{\partial^2}{a^2}\varphi_r
-
V_{,\varphi}(\varphi_r)
\right)
-f
\right]
+\cdots .
\end{align}
This makes explicit that the first-order stochastic description is obtained by integrating out the rapidly relaxing velocity sector of the original second-order Schwinger--Keldysh theory.

\paragraph{Noise correlations.}

The correlation function of $f$ follows from
\eqref{eq:xi_eta_corr}. Since
\begin{align}
\delta(\eta-\eta')
=
a(t)\delta(t-t'),
\end{align}
one obtains
\begin{align}
\left\langle
f({\bf x},t)
f({\bf y},t')
\right\rangle
&=
\frac{1}{a^3(t)a^3(t')}
N_\eta({\bf x},{\bf y})
\delta(\eta-\eta')
\nonumber\\
&=
\frac{1}{a^5(t)}
N_\eta({\bf x},{\bf y})
\delta(t-t') .
\label{eq:f_corr_general}
\end{align}
Substituting \eqref{eq:N_eta_kernel} gives
\begin{align}
\left\langle
f({\bf x},t)
f({\bf y},t')
\right\rangle
=
\frac{H^3}{4\pi^2}
j_0\!\left(
\epsilon aH|{\bf x}-{\bf y}|
\right)
\delta(t-t') .
\label{eq:f_corr_nonlocal}
\end{align}
Thus the stochastic source in the first-order Langevin equation is
spatially correlated. No spatial white-noise limit has been taken.

\paragraph{Reduced stochastic path integral.} The first-order action is obtained by substituting the overdamped
solution for $X_t$ back into \eqref{eq:S_xi_shifted}. The term
proportional to $\varphi_a$ vanishes because it multiplies the fast
constraint. After converting to cosmic time and writing the response
field in cosmic-time normalization, one obtains
\begin{align}
S_{\rm red}[f]
=
\int dt\,\dd^3x\,a^3(t)\,
\chi_a
\left[
-\dot\varphi_r
+
\frac{1}{3H}
\frac{\partial^2}{a^2}\varphi_r
-
\frac{1}{3H}
V_{,\varphi}(\varphi_r)
+
f({\bf x},t)
\right]
+\cdots .
\label{eq:first_order_action_f}
\end{align}
The full first-order path integral is therefore
\begin{align}
Z
=
\int{\cal D}f\,
\exp\left[
-\frac12
\int dt\,\dd^3x\,\dd^3y\,
f({\bf x},t)
{\cal N}_t^{-1}({\bf x},{\bf y})
f({\bf y},t)
\right]
\int{\cal D}\varphi_r{\cal D}\chi_a\,
\exp\left[
iS_{\rm red}[f]
\right],
\label{eq:first_order_path_integral}
\end{align}
where
\begin{align}
{\cal N}_t({\bf x},{\bf y})
=
\frac{H^3}{4\pi^2}j_0\!\left(
\epsilon aH|{\bf x}-{\bf y}|
\right).
\label{eq:Nt_kernel}
\end{align}
Equivalently,
\begin{align}
\left\langle
f({\bf x},t)
f({\bf y},t')
\right\rangle
=
{\cal N}_t({\bf x},{\bf y})
\delta(t-t') .
\end{align}

\paragraph{Separate-universe local limit.}

The separate-universe limit is taken only after deriving the non-local stochastic theory. It corresponds to evaluating the spatial kernel within a local Hubble patch and should not be confused with a spatial white-noise limit. For points satisfying
\begin{align}
\epsilon aH|{\bf x}-{\bf y}|\ll1,
\end{align}
the kernel admits the expansion
\begin{align}
j_0\!\left(
\epsilon aH|{\bf x}-{\bf y}|
\right)
=
1
+
{\cal O}\!\left(
\epsilon^2a^2H^2|{\bf x}-{\bf y}|^2
\right),
\end{align}
so that the local stochastic variable is obtained by taking the coincident limit of the non-local kernel within the same patch, rather than by replacing it with a spatial delta function.

This distinction becomes apparent when constructing the Martin--Siggia--Rose representation of the local diffusive action,
\begin{align}
S_K^{\rm loc}
=
\frac{i}{2}
\int dt\,\dd^3X\,a^3(t)\,
(\dot\varphi_a)_\Omega^2 .
\label{eq:local_diffusive_action}
\end{align}
If this operator were incorrectly interpreted as a pointwise quadratic form, one would formally write
\begin{align}
\exp(iS_K^{\rm loc})
=
{\cal N}
\int{\cal D}\eta\,
\exp\left[
-\frac12
\int dt\,\dd^3X\,a^3(t)\eta^2
-i
\int dt\,\dd^3X\,a^3(t)\eta\dot\varphi_a
\right],
\label{eq:HS_local_diffusion_wrong}
\end{align}
which would imply the spatial white-noise correlator
\begin{align}
\left\langle
\eta(t,\mathbf X)
\eta(t',\mathbf Y)
\right\rangle
=
\frac{1}{a^3(t)}
\delta(t-t')
\delta^{(3)}(\mathbf X-\mathbf Y).
\label{eq:eta_local_noise_wrong}
\end{align}
However, this is not the correct local limit. The operator $(\dot\varphi_a)_\Omega$ already represents a field averaged over a local patch, so there is no remaining independent spatial point whose correlations should be described by a Dirac delta. Instead, the local noise is inherited from the non-local kernel \eqref{eq:N_eta_kernel} by taking its coincident limit,
\begin{align}
\left\langle
\eta(t,\mathbf X)
\eta(t',\mathbf X)
\right\rangle
&=
\lim_{\mathbf Y\rightarrow\mathbf X}
\frac{1}{a^3(t)}
\frac{H^3}{4\pi^2}
j_0\!\left(
\Lambda|\mathbf X-\mathbf Y|
\right)
\delta(t-t')
\nonumber\\
&=
\frac{H^3}{4\pi^2a^3(t)}
\delta(t-t').
\label{eq:eta_local_noise_correct}
\end{align}
This reproduces the familiar Starobinsky--Yokoyama noise correlator and makes explicit that the local stochastic description is obtained by evaluating the non-local kernel within a Hubble patch, rather than by introducing an independent spatial white-noise field.

%%%%%%%%%%%%%%%%%%%%%%%%%%%%%%%%%%%%%%%%%%%%%%%%%%%%%%%%%%%%%%%%%%

\paragraph{Non-Gaussian noise correction.}

The overdamped reduction developed above also allows one to evaluate the leading non-Gaussian noise correction to the stochastic dynamics. While the quadratic Keldysh sector generates Gaussian noise, higher powers of the response field induce higher cumulants of the stochastic force. We
illustrate this mechanism for the cubic response vertex already present in the local Schwinger--Keldysh action,
\begin{align}
S_{\rm red}^{(3)}
=
-\frac{\lambda}{4!}
\int \dd^3x\int \dd t\, a^3(t)\,
\varphi_r^L(\varphi_a^L)^3 ,
\end{align}
and project it onto the overdamped sector. Using the same overdamped expansion as above,
\begin{align}
\varphi_a^L
=
-\frac{1}{3H}\chi_a^L
+\mathcal O\!\left(\frac{\omega}{H}\right).
\end{align}
The reduced action therefore contains the cubic response operator
\begin{align}
S_{\rm red}^{(3)}
\supset \frac{\lambda}{4!(3H)^3}
\int \dd^3x\int \dd t\, a^3(t)\,
\varphi_r^L(\chi_a^L)^3
+\cdots .
\label{eq:S3_firstorder_trim}
\end{align}
The Gaussian dynamics is still governed by the reduced action
\eqref{eq:first_order_action_f}, which implies the Langevin equation of the form
\begin{align}
\dot\varphi_r^L
=
- A[\varphi_r^L]
+\eta_\xi \,. 
\end{align}
The cubic operator \eqref{eq:S3_firstorder_trim} does not
modify this drift term. Instead, it changes the statistics of the stochastic force. While Gaussian noise is completely characterised by its two-point function, the cubic response vertex generates a non-vanishing third connected cumulant,
\begin{align}
\langle
\eta_\xi(x_1)\eta_\xi(x_2)\eta_\xi(x_3)
\rangle
\neq 0 \,.
\end{align}
The explicit expression for this third cumulant is encoded in the coefficient $N_3$, derived below in \Eq{eq:N3_momentum_trim}, and is obtained by matching to the microscopic Schwinger--Keldysh theory \cite{Salcedo:2024smn}.

Instead, it affects the statistics of the noise, $\eta_\xi$, which is not anymore fully characterised by its two-point function but instead has non-trivial three-point statistics \cite{Salcedo:2024smn}.

Besides the Langevin equation, non-Gaussian noise also modifies the Fokker--Planck equation governing the time evolution of the probability functional $P[\varphi^L]$.
As discussed in the main text, powers of the response field generate higher functional derivatives in the evolution equation for the probability functional. Consequently, the operator $(\chi_a)^3$ produces the leading correction to the Kramers--Moyal expansion,
\begin{align}
\frac{\partial P[\varphi^L]}{\partial t}
\supset
\int \dd^3x\;
N_3(\varphi^L(x))
\,
\frac{\delta^3 P[\varphi^L]}
{\delta \varphi^L(x)^3}.
\label{eq:KM3_appendix}
\end{align}
This is the third Kramers--Moyal coefficient and represents the leading
non-Gaussian correction to the stochastic dynamics.
From the microscopic point of view, the coefficient $N_3$ arises by
combining the same shell-localised Keldysh contraction responsible for
Gaussian diffusion with a single insertion of the cubic response
vertex. Each insertion of $\varphi_a^L$ produces a retarded propagator,
so the resulting kernel involves one $G_K$ and three $G_R$'s. In
momentum space this takes the form
\begin{align}
&N_3(\bmk_1,\bmk_2,\bmk_3;t_0)
=
\lambda
\int_{\bmk}
(2\pi)^3\delta^{(3)}(\bmk+\bmk_1+\bmk_2+\bmk_3)\,
\phi^L(\bmk,t_0)\,
\dot\Omega_\Lambda(t_0)\,
f_\varphi^2(k,t_0)
\nonumber\\
&\qquad\times
\int^{t_0}\dd t\,a^3(t)\,
G_K(k;t,t_0)\,
G_R(k_1;t,t_0)\,
G_R(k_2;t,t_0)\,
G_R(k_3;t,t_0).
\label{eq:N3_momentum_trim}
\end{align}
This makes explicit how the third cumulant is matched to the microscopic SK theory.
The local coefficient appearing in
\eqref{eq:KM3_appendix} is obtained from this expression by performing
the spatial integrals and taking the coincident-point limit. The
presence of a non-vanishing third cumulant shows that the noise is no longer Gaussian. Hence, just as the quadratic Keldysh kernel determines the Gaussian noise correlator, the cubic response vertex determines the leading non-Gaussian cumulant of the stochastic force after projection onto the
first-order theory.

%%%%%%%%%%%%%%%%%%%%%%%%%%%%%%%%%%%%%%%%%%%%%%%%%%%%%%%%%%%%%%%%%%%%%%%%%%%%%%%%%%%%%%%%%%%%%%%%%%%%%%%%%%%%%%%%%%%%%%%%%%%%%%%%%%%%%%%%

\section{From renormalization group flow to open effective field theory}\label{app:derivation_polchinski}

In this appendix, we derive the exact Schwinger--Keldysh renormalization group identity used in Sec.~\ref{sec:RGPolchinski} and explain how it gives rise to the local open effective field theory. Starting from the exact RG flow, we show how the quadratic $aa$ sector is reduced to a local-time form, allowing the bilocal kernel to be interpreted as the stochastic noise of the infrared effective theory.

\subsection*{Polchinski equation in the Schwinger--Keldysh contour}

In this appendix, we derive the RG identity used in Sec.~\ref{sec:RGPolchinski}, starting from
\begin{align}
\frac{\dd}{\dd\log\Lambda}Z_\Lambda[J_r,J_a]=0\, .
\label{eq:appZRG0}
\end{align}

We use condensed Schwinger--Keldysh notation,
\begin{align}
A_\alpha\circ B^{\alpha\beta}\circ C_\beta
=
\int_{\mathbf k}\int \dd\eta\,\dd\tilde\eta\;
A_\alpha(\mathbf k,\eta)\,
B^{\alpha\beta}(k;\eta,\tilde\eta)\,
C_\beta(-\mathbf k,\tilde\eta)\, ,
\end{align}
and write
\begin{align}
Z_\Lambda
=
\int \mathcal D\varphi_r\,\mathcal D\varphi_a\;
\exp\!\left\{
iS_{\mathrm{eff}}^\Lambda+iS_J
\right\},\qquad
S_{\mathrm{eff}}^\Lambda=S_0^\Lambda+S_{\mathrm{int}}^\Lambda,
\end{align}
with
\begin{align}
S_0^\Lambda
=
-\frac12\,\varphi_\alpha\circ \widehat D_\Lambda^{\alpha\beta}\circ \varphi_\beta,
\qquad
\widehat D_\Lambda\circ G_\Lambda=\mathbf 1\, .
\end{align}

\paragraph{Key identity.}

Differentiating the inverse relation gives
\begin{align}
\partial_{\log\Lambda}\widehat D_\Lambda
=
-\widehat D_\Lambda\circ
(\partial_{\log\Lambda}G_\Lambda)
\circ \widehat D_\Lambda\, .
\end{align}
This implies
\begin{align}
\partial_{\log\Lambda}S_0^\Lambda
=
\frac12
(\widehat D_\Lambda\circ\varphi)_\alpha
(\partial_{\log\Lambda}G_\Lambda^{\alpha\beta})
(\widehat D_\Lambda\circ\varphi)_\beta .
\label{eq:appdS0_short}
\end{align}

\paragraph{Flow for the interaction.}

Using \eqref{eq:appZRG0} and \eqref{eq:appdS0_short}, one obtains
\begin{align}
0
=
\int \mathcal D\varphi\;
e^{iS_{\mathrm{eff}}^\Lambda+iS_J}
\left[
\frac{i}{2}
(\widehat D_\Lambda\circ\varphi)_\alpha
(\partial_{\log\Lambda}G_\Lambda^{\alpha\beta})
(\widehat D_\Lambda\circ\varphi)_\beta
+
i\,\partial_{\log\Lambda}S_{\mathrm{int}}^\Lambda
\right].
\end{align}
Requiring this to be reproduced by a functional equation for
$e^{iS_{\mathrm{int}}^\Lambda}$ fixes uniquely
\begin{align}
\boxed{
\partial_{\log\Lambda}e^{iS_{\mathrm{int}}^\Lambda}
=
\frac{i}{2}\,
(\partial_{\log\Lambda}G_\Lambda^{\alpha\beta})
\frac{\delta^2 e^{iS_{\mathrm{int}}^\Lambda}}
{\delta\varphi_\beta\,\delta\varphi_\alpha}
}
\label{eq:appPolchInt_short}
\end{align}
up to field-independent normalization terms.

\paragraph{Flow for the full effective action.}

Using
\begin{align}
e^{iS_{\mathrm{eff}}^\Lambda}
=
e^{iS_0^\Lambda}e^{iS_{\mathrm{int}}^\Lambda},
\end{align}
and rewriting the second functional derivative in terms of
$e^{iS_{\mathrm{eff}}^\Lambda}$, one finds after a direct expansion
\begin{align}
&e^{iS_0^\Lambda}
\frac{\delta^2 e^{iS_{\mathrm{int}}^\Lambda}}
{\delta\varphi_\beta\,\delta\varphi_\alpha}
=
\frac{\delta^2 e^{iS_{\mathrm{eff}}^\Lambda}}
{\delta\varphi_\beta\,\delta\varphi_\alpha}
+
i\,\widehat D_{\Lambda\,\alpha\beta}\,e^{iS_{\mathrm{eff}}^\Lambda}
\nonumber\\
&\quad
+
i(\widehat D_\Lambda\circ\varphi)_\alpha
\frac{\delta e^{iS_{\mathrm{eff}}^\Lambda}}{\delta\varphi_\beta}
+
i(\widehat D_\Lambda\circ\varphi)_\beta
\frac{\delta e^{iS_{\mathrm{eff}}^\Lambda}}{\delta\varphi_\alpha}
-
(\widehat D_\Lambda\circ\varphi)_\alpha
(\widehat D_\Lambda\circ\varphi)_\beta
e^{iS_{\mathrm{eff}}^\Lambda}.
\end{align}
The last term cancels against the variation of $S_0^\Lambda$.
Using symmetry in $\alpha,\beta$, the flow becomes
\begin{align}
\partial_{\log\Lambda}e^{iS_{\mathrm{eff}}^\Lambda}
&=
\frac{i}{2}\,
\frac{\partial G_\Lambda^{\alpha\beta}}{\partial\log\Lambda}
\Bigg[
\frac{\delta^2 e^{iS_{\mathrm{eff}}^\Lambda}}
{\delta\varphi_\beta\,\delta\varphi_\alpha}
+
i\,\widehat D_{\Lambda\,\alpha\beta}\,e^{iS_{\mathrm{eff}}^\Lambda}
+
2i(\widehat D_\Lambda\circ\varphi)_\beta
\frac{\delta e^{iS_{\mathrm{eff}}^\Lambda}}{\delta\varphi_\alpha}
\Bigg].
\end{align}
At last, using
\begin{align}
\frac{\delta S_0^\Lambda}{\delta\varphi_\beta}
=
-(\widehat D_\Lambda\circ\varphi)_\beta,
\end{align}
this can be written in the compact form used in the main text,
\begin{align}
\boxed{
\frac{\partial e^{iS^\Lambda_{\mathrm{eff}}}}{\partial \log \Lambda}
=
-\frac{i}{2}\,
\frac{\partial G^{\alpha \beta}_\Lambda}{\partial \log \Lambda} \circ
\left[
- \frac{\delta^2 e^{i S^\Lambda_{\mathrm{eff}}}}{\delta \varphi_{\beta}\delta \varphi_{\alpha}}
+2i\frac{\delta}{\delta\varphi_\alpha}
\left(
\frac{\delta S^\Lambda_{0}}{\delta \varphi_{\beta}}
\,e^{iS^\Lambda_{\mathrm{eff}}}
\right)
\right]
}
\label{eq:appRGdiff_short}
\end{align}
up to field-independent normalization terms.

\paragraph{Short vs long propagator.}

Writing the flow in terms of the complementary propagator
$\bar G_\Lambda$,
\begin{align}
G_\Lambda+\bar G_\Lambda=G_{\rm full},
\qquad
\partial_{\log\Lambda}G_\Lambda
=
-\partial_{\log\Lambda}\bar G_\Lambda,
\end{align}
simply flips the overall sign of the kernel.

%%%%%%%%%%%%%%%%%%%%%%%%%%%%%%%%%%%%%%%%%%%%%%%%%%%%%%%%%%%%%%%%%%%%%%%%%%%%%%%%%%%%%%

\subsection*{Reduction to the local open EFT and emergence of the noise sector}

In this appendix we clarify how the exact quadratic $aa$ sector of the
RG flow reduces to the local noise sector of the open EFT. The key point
is that operator independence can only be discussed after the bilocal
kernel has been reduced to a local-time form.

\paragraph{Exact bilocal kernel.}

The exact quadratic contribution to the flow is
\begin{align}
\frac{\partial S_{\mathrm{eff}}^\Lambda}{\partial\log\Lambda}\Big|_{aa}
&=
-\varphi_a^L\circ \widehat D_\Lambda^A \circ
\frac{\partial G_\Lambda^K}{\partial\log\Lambda}
\circ \widehat D_\Lambda^R \circ \varphi_a^L
+\mathrm{B.T.}\, .
\label{eq:app2-exact-aa_trim}
\end{align}
Writing
\begin{align}
G_{S,\Lambda}^K(k,\eta,\tilde\eta)
=
\bar\Omega_\Lambda(k,\eta)\,
\bar\Omega_\Lambda(k,\tilde\eta)\,
G^K(k,\eta,\tilde\eta),
\end{align}
the corresponding bilocal kernel is
\begin{align}
\mathcal M_\Lambda(\eta,\tilde\eta;k)
=
\widehat D_\Lambda^A(\eta)\,
\frac{\partial G_{S,\Lambda}^K(k,\eta,\tilde\eta)}{\partial\log\Lambda}\,
\widehat D_\Lambda^R(\tilde\eta),
\end{align}
so that
\begin{align}
\frac{\partial S_{\mathrm{eff}}^\Lambda}{\partial\log\Lambda}\Big|_{aa}
=
-\int_{\mathbf k}\int \dd\eta\,\dd\tilde\eta\;
\varphi_a^L(\mathbf k,\eta)\,
\mathcal M_\Lambda(\eta,\tilde\eta;k)\,
\varphi_a^L(-\mathbf k,\tilde\eta)
+\mathrm{B.T.}\, .
\end{align}

Before local reduction, the most general structure is bilocal and takes
the schematic form
\begin{align}
\varphi_a^{(m)}(\eta)\,
\mathcal C_{mn}(\eta,\tilde\eta;k)\,
\varphi_a^{(n)}(\tilde\eta),
\qquad
m,n\in\{0,1,2\}.
\label{eq:app2-biloc-general_trim}
\end{align}
At this stage there is no useful notion of operator redundancy, because
these are kernels rather than local operators.

\paragraph{Local-time reduction.}

In the shell-crossing regime the derivative of the dressed propagator
localises on equal shell,
\begin{align}
\frac{\partial G_{S,\Lambda}^K(k,\eta,\tilde\eta)}{\partial\log\Lambda}
\;\longrightarrow\;
\mathcal F_\Lambda(k;\eta,\tilde\eta)\,
\delta\!\big(\Lambda(\eta)-\Lambda(\tilde\eta)\big).
\end{align}
For monotonic $\Lambda(\eta)$ this implies
\begin{align}
\delta\!\big(\Lambda(\eta)-\Lambda(\tilde\eta)\big)
=
\frac{\delta(\eta-\tilde\eta)}
{|\Lambda'(\eta)|},
\end{align}
so the bilocal kernel reduces to a local distribution,
\begin{align}
\mathcal M_\Lambda(\eta,\tilde\eta;k)
\;\longrightarrow\;
\sum_{m,n=0}^2
f_{mn}(\eta,k)\,
\partial_\eta^m\partial_{\tilde\eta}^n
\delta(\eta-\tilde\eta).
\end{align}
The quadratic action then becomes
\begin{align}
\frac{\partial S_{\mathrm{eff}}^\Lambda}{\partial\log\Lambda}\Big|_{aa,\mathrm{loc}}
=
\int_{\mathbf k}\int \dd\eta\;
\Big[
c_2(\eta,k)\,(\varphi_a^{\prime L})^2
+c_1(\eta,k)\,\varphi_a^{\prime L}\varphi_a^L
+c_0(\eta,k)\,(\varphi_a^L)^2
\Big]
+\mathrm{B.T.}\, .
\label{eq:app2-local-general_trim}
\end{align}
Only at this stage does the question of operator independence become
well posed. The mixed term is redundant up to a boundary term,
\begin{align}
\int \dd\eta\; c_1(\eta,k)\,\varphi_a^{\prime L}\varphi_a^L
=
-\frac12\int \dd\eta\; c_1'(\eta,k)\,(\varphi_a^L)^2
+\mathrm{B.T.}\, ,
\end{align}
so the local quadratic $aa$ sector contains only two independent bulk
operators,
\begin{align}
(\varphi_a^{\prime L})^2,
\qquad
(\varphi_a^L)^2.
\label{eq:app2-twoops_trim}
\end{align}
No further redundancy follows from local field redefinitions, since such
redefinitions remove only terms proportional to equations of motion,
whereas the quadratic $aa$ sector belongs to the noise sector.

\paragraph{Massless scalar in de Sitter.}

For a free massless scalar in de Sitter,
\begin{align}
a(\eta)=-\frac{1}{H\eta},
\qquad
\widehat D
=
a^2(\eta)
\left(
\partial_\eta^2
-\frac{2}{\eta}\partial_\eta
+k^2
\right).
\end{align}
With the moving cut-off $\Lambda(\eta)=-\epsilon/\eta$, one has $\Lambda'(\eta)=-\Lambda(\eta)/\eta$ together with
\begin{align}
\partial_{\log\Lambda}\bar\Omega_\Lambda(k,\eta)
\sim
\Lambda(\eta)\,\delta\!\big(k-\Lambda(\eta)\big),
\qquad
\partial_\eta \bar\Omega_\Lambda(k,\eta)
\sim
-\frac{\Lambda(\eta)}{\eta}\,
\delta\!\big(k-\Lambda(\eta)\big).
\label{eq:app2-shellpieceDS_trim}
\end{align}
These are the shell-crossing factors.

The Keldysh propagator is smooth on super-Hubble scales,
\begin{align}
G^K(k,\eta,\tilde\eta)
\sim
\frac{H^2}{2k^3}
\Big[
1+O(k^2\eta^2,k^2\tilde\eta^2)
\Big],
\end{align}
and its time derivatives are correspondingly suppressed by powers of
$k\eta\sim\epsilon$. Therefore the dominant local contribution comes from
the part of the differential operators in which one derivative on each
leg acts on the moving filter. Under these assumptions, the exact kernel
reduces to the shell-crossing structure
\begin{align}
\frac{\partial S_{\mathrm{eff}}^\Lambda}{\partial\log\Lambda}\Big|_{aa}^{\mathrm{shell}}
&=
\int_{\mathbf{k}}\int \dd \eta\,\dd\tilde\eta\;
a^2(\eta)\bar\Omega'_\Lambda(k,\eta)\,
a^2(\tilde\eta)\bar\Omega'_{\tilde\Lambda}(k,\tilde\eta)\,
\mathcal B(\eta,\tilde\eta;k)
+\mathrm{B.T.}\, ,
\end{align}
with
\begin{align}
\mathcal B(\eta,\tilde\eta;k)
&=
\varphi_a^{\prime L}(\mathbf k,\eta)\,
G^K(k,\eta,\tilde\eta)\,
\varphi_a^{\prime L}(-\mathbf k,\tilde\eta)
\nonumber\\
&\quad
-2\,\varphi_a^{\prime L}(\mathbf k,\eta)\,
\partial_{\tilde\eta}G^K(k,\eta,\tilde\eta)\,
\varphi_a^L(-\mathbf k,\tilde\eta)
+\varphi_a^L(\mathbf k,\eta)\,
\partial_\eta\partial_{\tilde\eta}G^K(k,\eta,\tilde\eta)\,
\varphi_a^L(-\mathbf k,\tilde\eta).
\label{eq:app2-Bfinal_trim}
\end{align}

The extra terms beyond the shell-crossing contribution can only be
classified after the local reduction. If they reduce to the mixed
structure $\varphi_a^{\prime L}\varphi_a^L$, they are redundant in the
bulk. If they contribute to $(\varphi_a^{\prime L})^2$ or
$(\varphi_a^L)^2$, they are not redundant, but for the massless de
Sitter shell-crossing regime they are parametrically suppressed because
they do not contain the sharply localised derivatives of the moving
filter.

Thus redundancy is a statement about the local operator basis, whereas
suppression is a statement about the size of the corresponding
coefficients once the de Sitter shell-crossing regime has been
specified.

%%%%%%%%%%%%%%%%%%%%%%%%%%%%%%%%%%%%%%%%%%%%%%%%%%%%%%%%%%%%%%%%%%%%%%%%%%%%%%%%%%%%%%%%%%%%%%%%

\section{Infrared scaling of the reduced first-order stochastic theory}
\label{app:sh-ir-scaling}
%%%%%%%%%%%%%%%%%%%%%%%%%%%%%%%%%%%%%%%%%%%%%%%%%%%%%%%%%%%%%%%%%%%%%%%%%%%%%%%%%%%%%%

Once the local super-Hubble EFT has been reduced to the first-order stochastic theory, one may ask which projected operators are relevant at its infrared fixed point. In this appendix, we perform a canonical scaling analysis of the reduced stochastic theory.

To define the infrared scaling, one must retain the leading Laplacian in
the reduced action,
\begin{align}
S_{\rm red}
=
\int \dd t\,\dd^d x\, a^d(t)\,
\chi_a^L
\left[
\eta_\xi
-\partial_t\varphi_r^L
-\kappa\,\nabla_{\rm phys}^2\varphi_r^L
+\mu\,(\varphi_r^L)^3
\right]
+\cdots ,
\end{align}
where
\begin{align}
\nabla_{\rm phys}^2\equiv\frac{\partial^2}{a^2},
\qquad
\kappa=\frac{1}{3H},
\qquad
\mu=\frac{\lambda}{18H}.
\end{align}
Introducing the rescaled response field
\begin{align}
\widehat\chi_a^L
\equiv
a^d(t)\chi_a^L,
\end{align}
and integrating out the Gaussian noise using the correlator obtained in
Appendix~\ref{sec:nonlocal_langevin}, one finds
\begin{align}
-\frac12
\int
\dd t\,\dd^d x\,
\dd t'\,\dd^d y\,
\widehat\chi_a^L(t,\bmx)\,
\langle
\eta_\xi(t,\bmx)
\eta_\xi(t',\bmy)
\rangle
\widehat\chi_a^L(t',\bmy)
=
-iD_0
\int
\dd t\,\dd^d x\,
(\widehat\chi_a^L)^2.
\end{align}
Here $D_0$ denotes the local diffusion coefficient. Since the canonical
scaling analysis is performed locally in time, the slowly varying
background factor is absorbed into its normalization. The reduced action
therefore takes the Martin--Siggia--Rose form
\begin{align}
S_{\rm MSR}^{(0)}
=
\int
\dd t\,\dd^d x
\left[
\widehat\chi_a^L
\left(
\partial_t\varphi_r^L
-\kappa\nabla_{\rm phys}^2\varphi_r^L
+\mu(\varphi_r^L)^3
\right)
-iD_0(\widehat\chi_a^L)^2
\right]
+\cdots .
\label{eq:MSR0_trim}
\end{align}

The Gaussian fixed point is diffusive. Requiring the time-derivative and
Laplacian terms to scale identically, while treating the background
scale factor as fixed under the scaling transformation, gives
\begin{align}
\bmx\rightarrow b\,\bmx,
\qquad
t\rightarrow b^z t,
\qquad
z=2.
\end{align}
Hence
\begin{align}
[\bmx]=-1,
\qquad
[t]=-2,
\qquad
[\partial_t]=2,
\qquad
[\nabla_{\rm phys}]=1.
\end{align}

The field dimensions follow from invariance of the Gaussian action. The
term
$\widehat\chi_a^L\partial_t\varphi_r^L$
implies
\begin{align}
[\widehat\chi_a^L]
+
[\varphi_r^L]
=
d,
\end{align}
while the noise term requires
\begin{align}
2[\widehat\chi_a^L]
=
d+2.
\end{align}
Therefore
\begin{align}
[\widehat\chi_a^L]
=
\frac{d+2}{2},
\qquad
[\varphi_r^L]
=
\frac{d-2}{2}.
\end{align}
In three spatial dimensions,
\begin{align}
[\widehat\chi_a^L]
=
\frac52,
\qquad
[\varphi_r^L]
=
\frac12.
\end{align}

The quartic drift operator
\begin{align}
S_{1,3}
=
g_{1,3}
\int
\dd t\,\dd^d x\,
\widehat\chi_a^L
(\varphi_r^L)^3
\end{align}
has coupling dimension
\begin{align}
[g_{1,3}]
=
4-d,
\end{align}
and is therefore relevant for $d<4$.

The field-dependent diffusion operator
\begin{align}
S_{2,2}
=
g_{2,2}
\int
\dd t\,\dd^d x\,
(\widehat\chi_a^L)^2
(\varphi_r^L)^2
\end{align}
has
\begin{align}
[g_{2,2}]
=
2-d,
\end{align}
and is already irrelevant in $d=3$.

Likewise, the projected higher-response interaction
\begin{align}
S_{3,1}
=
g_{3,1}
\int
\dd t\,\dd^d x\,
(\widehat\chi_a^L)^3
\varphi_r^L
\end{align}
has
\begin{align}
[g_{3,1}]
=
-d,
\end{align}
and is strongly irrelevant for all $d>0$.

Derivative interactions are even more irrelevant. After the overdamped
reduction every occurrence of $\dot\varphi_a^L$ is replaced by the
response field $\widehat\chi_a^L$, while derivatives acting on the
physical field contribute positive scaling dimension according to
\begin{align}
[\partial_t]=2,
\qquad
[\nabla_{\rm phys}]=1.
\end{align}
Consequently every additional time derivative, or equivalently every
additional pair of spatial derivatives, lowers the canonical dimension
of the corresponding coupling by two units. Higher-derivative operators
generated in the local super-Hubble EFT therefore constitute
increasingly irrelevant perturbations of the diffusive fixed point.

The resulting hierarchy is
\begin{align}
\widehat\chi_a^L
(\varphi_r^L)^3
&:
\qquad
\text{relevant for } d<4,
\\
(\widehat\chi_a^L)^2
(\varphi_r^L)^2
&:
\qquad
\text{irrelevant for } d>2,
\\
(\widehat\chi_a^L)^3
\varphi_r^L
&:
\qquad
\text{strongly irrelevant for } d>0,
\end{align}
with higher-derivative operators becoming progressively more
irrelevant.

In particular, in $d=3$ the quartic drift operator is the unique leading
infrared interaction. This ordering is distinct from the local
super-Hubble hierarchy discussed in
Appendix~\ref{sec:nonlocal_langevin}: operators that appear at the same order
in the local matching need not have the same relevance once the reduced
stochastic theory is viewed as an infrared field theory.

%%%%%%%%%%%%%%%%%%%%%%%%%%%%%%%%%%%%%%%%%%%%%%%%%%%%%%%%%%%%%%%%%%%%%%%%%%%%%%%%%%%%%%%%%%%%%%%%%%%%%%%%%%%%%%%%%%%%%%%%%%%%%%%%%%%%%%%%%%%%%%%%%%%

\section{Dynamical Kubo–Martin–Schwinger symmetry}\label{app:dKMS}

The equilibrium properties of the infrared theory can be identified directly at the level of the local Schwinger--Keldysh action. In this appendix, we show that it exhibits an emergent dynamical Kubo--Martin--Schwinger (KMS) symmetry, which implies the fluctuation-dissipation relation among the EFT coefficients.
We consider the dynamical KMS transformation of the local reduced
action. Neglecting spatial gradients, the relevant action is
\begin{equation}
S_{\rm red}^{\rm loc}
=
\int \dd t\,\dd^3\bmx\,
\left\{
-\dot\varphi_a^L
\left[
\dot\varphi_r^L
+
\Gamma V'(\varphi_r^L)
\right]
+
\frac{i}{2}
\left(\dot\varphi_a^L\right)^2
\right\},
\label{eq:local_reduced_action_kms}
\end{equation}
with $\Gamma=1/(3H)$. The dynamical KMS transformation \cite{Crossley:2015evo, Liu:2018kfw, Hongo:2018ant} follows by requiring invariance of
Eq.~\eqref{eq:local_reduced_action_kms} under time reversal accompanied by a shift of the response field.
Consider the transformation
\begin{align}
\varphi_r^L(t,\bmx)
&\rightarrow
\varphi_r^L(-t,\bmx),
\nonumber\\
\dot\varphi_a^L(t,\bmx)
&\rightarrow
\dot\varphi_a^L(-t,\bmx)
+
i\alpha\,
\dot\varphi_r^L(-t,\bmx).
\label{eq:dkms_transformation_local}
\end{align}
After changing the integration variable $t\rightarrow -t$, the
transformed action becomes
\begin{align}
\widetilde S_{\rm red}^{\rm loc}
=
\int \dd t\,\dd^3\bmx\,
\bigg[
-&
\left(
\dot\varphi_a^L
+
i\alpha\dot\varphi_r^L
\right)
\left(
-\dot\varphi_r^L
+
\Gamma V'
\right)+
\frac{i}{2}
\left(
\dot\varphi_a^L
+
i\alpha\dot\varphi_r^L
\right)^2
\bigg].
\end{align}
Expanding the transformed action, the coefficient of
$\dot\varphi_a^L\dot\varphi_r^L$ agrees with the original action
provided $1-\alpha=-1$, 
which fixes $\alpha=2$.
For this value the terms proportional to
$(\dot\varphi_r^L)^2$ cancel identically. The remaining variation is
\begin{equation}
\widetilde S_{\rm red}^{\rm loc}
-
S_{\rm red}^{\rm loc}
=
-2i\Gamma
\int \dd t\,\dd^3\bmx\,
\dot\varphi_r^L
V'(\varphi_r^L).
\end{equation}
Therefore, 
\begin{equation}
\widetilde S_{\rm red}^{\rm loc}
-
S_{\rm red}^{\rm loc}
=
-i\beta_{\rm loc}
\left[
\int \dd^3\bmx\,
V(\varphi_r^L) \right]^{t_f}_{t_i},
\end{equation}
with $\beta_{\rm loc} = 2\Gamma = 2/(3H)$.
Since $[\beta_{\rm loc}]=M^{-1},$
the parameter appearing in the dynamical KMS transformation has the
dimensions of an inverse temperature. Equivalently, the local
non-relativistic theory defines the temperature scale
\begin{equation}
T_{\rm loc}
=
\beta_{\rm loc}^{-1}
=
\frac{3H}{2}.
\label{eq:local_kms_temperature}
\end{equation}

This KMS temperature should be distinguished from the coefficient
appearing in the equilibrium Fokker--Planck distribution. The latter also depends on the physical normalization of the noise inherited from the non-local diffusion kernel. From the coincident noise correlator
\begin{equation}
\left\langle
\eta(t,\bmx)\eta(t',\bmx)
\right\rangle
=
\frac{H^3}{4\pi^2}
\delta(t-t')\,
\label{eq:local_noise_inherited}
\end{equation}
the Starobinsky--Yokoyama (SY) diffusion
coefficient is $D_{\rm SY}
=
H^3/(8\pi^2)$.
Consequently, the coefficient controlling the equilibrium distribution
is
\begin{equation}
\beta_{\rm SY}
=
\frac{\Gamma}{D_{\rm SY}}
=
\frac{8\pi^2}{3H^4},
\label{eq:beta_SY_from_noise}
\end{equation}
and
\begin{equation}
P_{\rm eq}(\varphi)
\propto
\exp\left[
-\beta_{\rm SY}V(\varphi)
\right].
\end{equation}
Thus the dynamical KMS parameter
$\beta_{\rm loc}=2/(3H)$ and the coefficient
$\beta_{\rm SY}=8\pi^2/(3H^4)$ entering the equilibrium
Fokker--Planck measure are different quantities. The former is the
inverse-temperature parameter associated with the dynamical KMS
symmetry of the normalized local non-relativistic action, while the latter additionally contains the physical noise normalization inherited from the non-local theory.

The dKMS symmetry is compatible with the diffusive infrared scaling derived above in \App{app:sh-ir-scaling}. Once performing the first-order reduction, since
\begin{equation}
[\chi_a^L]
=
[\dot\varphi_r^L]
=
\frac{d+2}{2},
\end{equation}
for $\chi_a^L=-(\partial_t+3H)\varphi_a^L$, the two terms entering the KMS transformation have identical scaling dimension. The RG flow must therefore organize interactions into complete dKMS multiplets rather than independent operators. For example, field-dependent noise coefficients generate multiplicative-noise vertices of the form
\begin{equation}
iD(\varphi_r^L)(\chi_a^L)^2,
\end{equation}
whose dKMS completion necessarily contains the corresponding
field-dependent dissipative terms. More generally, vertices with
additional response fields are related by the dKMS transformation to operators containing fewer response fields and additional time
derivatives.

The infrared RG hierarchy and dKMS symmetry therefore impose
two complementary structures on the stochastic effective theory. The RG flow determines which dKMS multiplets are relevant in the infrared, while dKMS relates the noise and dissipative interactions within each multiplet. The stationary Fokker--Planck distribution is then the equilibrium boundary weight associated with this symmetry.

	\bibliographystyle{JHEP}
	\bibliography{biblio}

\end{document}

%% file: newcommands.tex
\newcommand{\ee}{e}

\newcommand{\boldmathsymbol}[1]{{\ensuremath{\boldsymbol{#1}}}}

\newcommand{\bmk}{\boldmathsymbol{k}}

\newcommand{\bmx}{\boldmathsymbol{x}}
\newcommand{\bmy}{\boldmathsymbol{y}}

\newcommand{\teta}{\tilde{\eta}}

\newcommand{\beq}{\begin{equation}}
\newcommand{\eeq}{\end{equation}}
\newcommand{\bea}{\begin{equation}\begin{aligned}}
\newcommand{\eea}{\end{aligned}\end{equation}}
\newlength{\wsingfig}
\newlength{\wdblefig}
\newlength{\wquadfig}
\newlength{\wtriplefig}
\newcommand{\Eq}[1]{Eq.~(\ref{#1})}
\newcommand{\Eqs}[1]{Eqs.~(\ref{#1})}
\newcommand{\Fig}[1]{Fig.~{\ref{#1}}}

\newcommand{\App}[1]{Appendix~\ref{#1}}